\documentclass[reprint,superscriptaddress,nofootinbib,amsmath,amssymb,aps,floatfix]{revtex4-2}
\usepackage{graphicx,bm,amsmath,amssymb,amsthm,booktabs,hyperref}

\newtheorem{theorem}{Theorem}
\newtheorem{proposition}{Proposition}

\newtheorem{corollary}{Corollary}
\newtheorem{conjecture}{Conjecture}
\theoremstyle{definition}
\newcommand{\Z}{\mathbb{Z}}\newcommand{\R}{\mathbb{R}}\newcommand{\F}{\mathbb{F}}
\newcommand{\dX}{\mathrm{d}_X}\newcommand{\dZ}{\mathrm{d}_Z}

\begin{document}
\raggedbottom

\title{Tilting the Three-Dimensional Toric Code for Distance-Optimal Lattices and Their Memory Costs}

\author{Xiang Zou}
\email{xiang.zou@mail.utoronto.ca}
\affiliation{Department of Physics, University of Toronto, 60 St. George Street, Toronto, ON, M5S 1A7, Canada}

\date{November 27, 2026}

\begin{abstract}
Rotating the lattice of a two-dimensional toric or surface code reduces
the number of qubits needed for a given code distance. We study the same
question for the three-dimensional toric code, which supports single-shot
correction of $X$ errors and, on three copies, a constant-depth CCZ gate.
For periodic cubic-lattice codes, the densest lattice packing of
octahedra, found by Minkowski, gives the sharp bound $n\ge19\dZ^3/36$ on
the number $n$ of data qubits at string distance $\dZ$.
The Minkowski codes attain this bound at $\dZ=6m$ for every positive
integer $m$, and we prove that their membrane distance is $\dX=19m^2$.
Distance alone does not set the failure rate: for every $d$ divisible by
six we construct cubic and tilted layouts with equal $n$, $\dZ=d$ and
$\dX=d^2$ whose leading failure coefficients, with perfect checks and
maximum-likelihood decoding, differ by a factor that grows exponentially
with $d$. Our loop model uses the circuit's detector error model to
estimate how often at least one of the three logical $X$ readouts is
wrong, with no parameter fitted to logical-error data.
For a greedy schedule and matching at
$p=(1\text{--}3)\times10^{-4}$, its predictions lie within a factor
of $1.5$ of simulation in all $57$ runs with at least $30$ failures
($24$ codes, $5\le\dZ\le15$). At
$10^{-9}$ failures per round on another schedule, where it
underestimates large tilted codes by up to a third, its central
estimates give cubic codes $1.3$ to
$2.0$ times as many data qubits as body-centred cubic (BCC) codes for
$p\le3\times10^{-4}$, and face-centred cubic codes fewer than BCC at
the three lowest of seven noise levels. For
surface codes on finite volumes, tilting the
rough boundaries raises $\dZ\dX/n$, yet in circuit simulations at
$p=3\times10^{-4}$ the axis-aligned volume reaches $10^{-5}$ failures per
round with the fewest data qubits.
\end{abstract}

\maketitle

\section{Introduction}

Surface-code memories now operate below threshold~\cite{Google1,Google2},
making the qubit cost of reaching a specified logical error rate a
practical design question. Geometry changes that cost even within one
code family. Tilting the square lattice turns the $[[2d^2,2,d]]$ toric
code~\cite{Toriccode1} into the $[[d^2,2,d]]$ rotated
code~\cite{BombinOptimal2007,KovalevPryadko2012}; searches over twisted
tori~\cite{LiangTwisted2025,AHHW2025} and constructions of new
toric codes~\cite{Trinca2023} extend this approach.
Which layout needs the fewest qubits at a given distance, and which
needs the fewest at a given logical failure rate? Distance measures the
weight of the shortest logical error. The number of errors that look
alike to the decoder also affects the failure rate.

Three-dimensional toric codes are useful for this comparison because their
qubit count grows cubically with string distance, while they support
single-shot correction of $X$ errors and non-Clifford logical
gates~\cite{3DToric_code1,3DToric_code2,KubicaYoshidaPastawski2015,Zhu2025}.
Rotated three-dimensional surface-code layouts have also been studied
for biased noise~\cite{Huang2023}.

Minkowski's theorem on the densest lattice packing of regular
octahedra~\cite{Minkowski1904,BetkeHenk2000} suggests how to choose the
periods of a cubic-lattice toric code. The shortest nonzero $\ell_1$
length in the period lattice is the string distance $\dZ$. Octahedra of
$\ell_1$ radius $\dZ/2$ centred at its points therefore have disjoint
interiors. Each has volume $\dZ^3/6$, and a fundamental cell has volume
$n/3$. Minimizing the qubit count at fixed $\dZ$ thus amounts to
maximizing the packing density. Minkowski's density bound of $18/19$
gives $n\ge19\dZ^3/36$. The theory of lattice rules for numerical
integration already contains this bound: Cools and Lyness state it for
the index of a three-dimensional lattice rule of given enhanced degree,
which are the number of vertices and the string distance of our
torus~\cite[Eqs.~(1.4)--(1.5)]{CoolsLyness2001}; see also
Ref.~\cite{Lyness2003} on lattice rules and octahedron packing.

Scaling the centre lattice of the optimal unit octahedra by three gives
integer periods of determinant $38$, hence a $114$-qubit code with
$\dZ=6$. Aasen, Haah, Hastings and Wang found
the same minimum determinant at distance six by exhaustive lattice search
and described the integer rescalings~\cite[Secs.~II\,A--B and X, Table~I]{AHHW2025}.
Jacob, McLauchlan and Browne obtained the same code from a tricycle
construction and reported membrane distance nineteen and a depth-two CCZ
circuit~\cite[Sec.~V\,C and Table~III]{TricycleJMB2025}. We call the
$114$-qubit code and its integer rescalings \emph{Minkowski codes}.
The qubit bound and the codes that attain it are therefore not new.
What we add is the joint bound on $k\,\dZ\dX/n$
(Theorem~\ref{thm:packing}) and the membrane distance of the Minkowski
codes: we prove that it is $\dX=19m^2$ for every integer $m\ge1$.

For the failure-rate question, the competition between distance and
error multiplicity is established in two
dimensions~\cite[Secs.~III--V]{Beverland2019}. Path counting enters
decoding and low-noise estimates~\cite{CrigerAshraf2018,WatsonBarrett2014,
ValentiniForlivesiChiani2023}; Fowler computes circuit-derived low-noise
coefficients without fitting logical-error data~\cite[Table~I]{Fowler2013}.
Beverland \emph{et al.}\ obtain the leading failure coefficient of a
syndrome circuit by counting, for each syndrome, its minimum-weight
errors in each logical class~\cite[Sec.~4.3]{Beverland2025}.
Equal-error circuit resource comparisons are also
established~\cite{ORourkeDevitt2025}. We first compare two layouts with
the same qubit count and both distances. We then model the failure rate
from the number of winding loops, the errors they share and the circuit
noise over successive rounds (Sec.~\ref{sec:entropy}).

We give model extrapolations to $10^{-9}$ failures per round
(Sec.~\ref{sec:design}) and compare forecasts fixed before the
confirmation samples were drawn with simulations of 18 layouts taken
from the lattice families studied here
(Appendix~\ref{app:frozen-forecasts}). We also compare three orientations
of the rough boundaries of a surface code (Sec.~\ref{sec:patches}).

\section{Period lattices and distance bounds}
\label{sec:setup}

Let $\Lambda\subset\Z^D$ be the integer span of $D$ independent period
vectors. Identifying $x$ with $x+v$ for $v\in\Lambda$ gives the cubic
cellulation $T_\Lambda$ of $\R^D/\Lambda$, with $V=\det\Lambda$
vertices. Its edge code has $n=DV$ data qubits and $k=D$ logical qubits.
Vertex $X$ checks and plaquette $Z$ checks have weights $2D$ and $4$
when no unit vector belongs to $\Lambda$; unit periods can cancel
repeated Pauli incidences and lower these weights.

A closed walk lifts to a path from $x$ to $x+v$ in $\Z^D$, with winding
$v\in\Lambda$. Its shortest possible length is
$\|v\|_1=\sum_i|v_i|$. A path attaining that length is
\emph{monotone}. Every step along axis $i$ has the sign of $v_i$.
A nonstraight monotone path is a \emph{staircase}.
Binary string classes are
$H_1(T_\Lambda;\Z_2)\cong\Lambda/2\Lambda$; a winding in $2\Lambda$
is trivial. Membrane classes are
$H^1(T_\Lambda;\Z_2)\cong\mathrm{Hom}(\Lambda,\Z_2)$.
A string of winding $v$ anticommutes with a membrane of class $\alpha$
exactly when $\alpha(v)=1$~\cite{ZouLo2025}.
We write $[[n,k,(\dZ,\dX)]]$ for the data-qubit count, logical-qubit
count, string distance and membrane distance.

For a lattice $\Gamma$, define its $\ell_1$-systole
$\lambda(\Gamma)=\min_{v\in\Gamma\setminus0}\|v\|_1$.
The dual is
$\Lambda^*=\{\varphi\in\R^D:\varphi\cdot v\in\Z\text{ for all }v\in\Lambda\}$.
We write $\varphi(v)=\varphi\cdot v$.
We use $e_i$ for coordinate unit vectors,
$\langle b_1,\ldots,b_D\rangle$ for an integer basis,
$m\Lambda$ for integer scaling, and $\Lambda_N$ for a lattice of
determinant $N$. A cyclic lattice has
$x_1\equiv a_2x_2+\cdots+a_Dx_D\bmod N$.
The setup and the following distance bounds apply in any dimension $D$.

\begin{proposition}[Distance bounds]\label{prop:dist}
\begin{enumerate}
\item[(a)] $\dZ=\lambda(\Lambda)$.
\item[(b)] A class $\alpha\in\mathrm{Hom}(\Lambda,\Z_2)$ contains a membrane of weight at most $V\min\{\|\varphi\|_1:\varphi\in\Lambda^*,\ \varphi(v)\equiv\alpha(v)\bmod2\text{ for all }v\in\Lambda\}$. In particular $\dX\le V\lambda(\Lambda^*)$.
\item[(c)] Let $\nu$ be a probability distribution on the vectors $v\in\Lambda$ with $\alpha(v)=1$, and put $t_\nu=\max_i\mathbb E_\nu|v_i|$. Then every membrane in class $\alpha$ has weight at least $V/t_\nu$.
\end{enumerate}
\end{proposition}

Part~(a) is established~\cite[Secs.~II\,A--B]{AHHW2025}.
Part~(b) takes the edges where the integer function
$\lfloor\varphi\cdot x\rfloor$ changes by an odd amount.
The factor $V$ converts the dual covector's displacement per edge into
the total number of supported edges. Part~(c) packs translated
monotone paths fractionally: a membrane must meet every path with
$\alpha(v)=1$. Appendix~\ref{app:proof-dist} proves both bounds.
For the five families in Table~\ref{tab:lattices} the two bounds
coincide without rounding, which determines their membrane distances at
every integer scale.

In three dimensions we conjecture that the upper bound in part~(b)
is sharp after minimizing over all nonzero membrane classes.

\begin{conjecture}[Membrane distance]\label{conj:exact}
$\dX=V\lambda(\Lambda^*)$ for every $\Lambda\subset\Z^3$.
\end{conjecture}

The evidence is numerical. For every integer lattice of determinant
at most $48$, which makes $76\,060$ lattices with $532\,420$ nonzero
membrane classes, a computed lower bound on the weight of each class is
at least $V\lambda(\Lambda^*)$: a linear program gives it for all but
$84$ classes, and integer programming for those $84$. The solvers work
in floating point, so this is a finite numerical check and not a proof.

\section{Packing bounds}
\label{sec:bounds}

The distance bounds turn a geometric packing constraint into a
constraint on code resources. We apply the classical relation between
packing density and the shortest lattice vector to both the period
lattice and its dual~\cite[Sec.~2]{BetkeHenk2000}.

\begin{theorem}[Code consequences of lattice packing]\label{thm:packing}
Let $\delta_D$ be the density of the densest lattice packing of the cross-polytope $\{\|x\|_1\le1\}\subset\R^D$. Every toric code on a twisted torus $\Z^D/\Lambda$ satisfies
\begin{equation}
n\;\ge\;\frac{D\,\dZ^{\,D}}{D!\,\delta_D},\qquad k\,\dZ\dX\;\le\;(D!\,\delta_D)^{2/D}\,n. \label{eq:bounds}
\end{equation}
\end{theorem}

\begin{proof}
Open $\ell_1$-balls of radius $\lambda(\Lambda)/2$ centred on the points of $\Lambda$ are disjoint, and each has volume $\lambda(\Lambda)^D/D!$. Hence $\lambda(\Lambda)^D\le D!\,\delta_D\det\Lambda$. The same argument applied to $\Lambda^*$, whose determinant is $1/\det\Lambda$, gives $\lambda(\Lambda^*)^D\le D!\,\delta_D/\det\Lambda$. Combine these with Proposition~\ref{prop:dist}(a,b), $n=D\det\Lambda$ and $k=D$.
\end{proof}

The first inequality bounds the qubit cost of a prescribed string
distance. In the second, the determinants of the lattice and its dual
cancel, bounding the joint protection per data qubit. This argument
uses only the membrane upper bound in Proposition~\ref{prop:dist}(b)
and does not assume Conjecture~\ref{conj:exact}.

\subsection{Two and three dimensions}

In two dimensions the $\ell_1$ unit ball tiles the plane, so $\delta_2=1$.
Theorem~\ref{thm:packing} thus recovers the known bound $n\ge d^2$ for
square-lattice toric codes on twisted tori, where $d=\dZ=\dX$. For even
$d$, the rotated toric code attains $n=d^2$~\cite{BombinOptimal2007}.
For odd $d$, the qubit count $n=2\det\Lambda$ is even, so $n\ge d^2+1$;
twisted codes attain this bound~\cite{HomologicalQEC1,KovalevPryadko2012}.

In three dimensions the densest lattice packing of the octahedron
has density $\delta_3=18/19$,
determined by Minkowski~\cite{Minkowski1904} and confirmed by
Betke and Henk~\cite[Sec.~5]{BetkeHenk2000}.
Theorem~\ref{thm:packing} therefore gives
\begin{equation}
n\ge\tfrac{19}{36}\,\dZ^3,\qquad k\,\dZ\dX\le\left(\tfrac{108}{19}\right)^{2/3}n\approx3.185\,n. \label{eq:3d}
\end{equation}
For a three-dimensional lattice rule with $N$ points and enhanced degree
$\delta$, Cools and Lyness give $N\ge\delta^3/(3!\,\theta(3))$ with
$\theta(3)=18/19$, the density of the densest lattice packing of the
octahedron~\cite[Eqs.~(1.4)--(1.5)]{CoolsLyness2001}. The dual lattice
of the rule plays the role of our period lattice: its determinant is
$N$, so $N=V=n/3$, and the enhanced degree is its shortest nonzero
$\ell_1$ length, so
$\delta=\dZ$~\cite[Theorem~2.1 and Eq.~(2.2)]{CoolsLyness2001}.
Substituting gives $n\ge19\dZ^3/36$. The second inequality, which
involves the membrane distance, has no counterpart there.
The Minkowski codes in Sec.~\ref{sec:extremal} attain the first
bound at every string distance divisible by six and have $k\,\dZ\dX/n=3$.
\hypertarget{joint-lattice-bound}{We conjecture} that every full-rank
lattice $\Lambda\subset\R^3$ satisfies $\lambda(\Lambda)\lambda(\Lambda^*)\le3$,
which by Proposition~\ref{prop:dist}(a,b) would make $3$ the sharp upper
bound on $k\,\dZ\dX/n$. As evidence, a numerical maximization of
$\lambda(\Lambda)\lambda(\Lambda^*)$ over real lattice bases, started
from $200$ random bases, returned values within $10^{-12}$ of $3$ in its
ten best runs and none above $3$ beyond rounding error. This is a search
for counterexamples and not a proof.

\subsection{Higher-dimensional edge codes}
\label{sec:higher-dim}

For arbitrary $D\ge2$, we continue to put qubits on edges. Logical
$Z$ operators are strings and logical $X$ operators are dual
$(D-1)$-dimensional membranes. The cubic periods $m\Z^D$ give
$[[Dm^D,D,(m,m^{D-1})]]$ and $k\,\dZ\dX/n=1$.
These parameters follow from straight strings and coordinate
membranes: the translated straight strings give the matching
membrane lower bound in Proposition~\ref{prop:dist}(c).

Let $\mathcal S_D$ be the largest string distance normalized by the
$D$th root of the number of vertices, taking a supremum over
full-rank integer period lattices:
\begin{equation}\label{eq:packing-normalized-systole}
 \mathcal S_D
 =\sup_{\Lambda\subseteq\Z^D}\frac{\dZ}{V^{1/D}}
 =(D!\,\delta_D)^{1/D}.
\end{equation}
The last equality follows from the packing density
$\lambda(\Lambda)^D/(D!\det\Lambda)$. Rational bases can approximate
any real lattice, and clearing their denominators leaves this
normalized ratio unchanged, so the integer restriction does not
change the supremum.
Aasen \emph{et al.}~\cite[Sec.~VIII\,B, Theorem~VIII.1]{AHHW2025}
prove $\mathcal S_D=\Theta(D)$ using a random-lattice construction
for the lower bound. The elementary upper bound $\delta_D\le1$
gives $\mathcal S_D\le(D!)^{1/D}=(1+o(1))D/e$ by Stirling's formula.
Consequently the coefficient $\mathcal S_D^2$ in the joint packing
bound grows quadratically with dimension.

The supremum allows unrestricted integer scales. At a specified finite
distance the best available shape can fall short of it. Uniform scaling
of any one shape gives $n\propto\dZ^D$.

For an explicit comparison, Aasen \emph{et al.}~\cite[Sec.~VIII\,A]{AHHW2025}
use Sylvester Hadamard matrices as period bases in power-of-two dimensions.
More generally, take periods $mh_j$, where the $h_j$ are the rows of any
order-$D$ Hadamard matrix and $m$ is a positive integer. The known lattice
volume and shortest $\ell_1$ length give $V=m^D D^{D/2}$ and
$\dZ=mD$~\cite[Theorem~1]{EtzionVardyYaakobi2013}. Our membrane bounds
give $\dX=V/m$: every nonzero class evaluates to one on some $mh_j$,
whose coordinates all have magnitude $m$, so Proposition~\ref{prop:dist}(c)
gives the lower bound. The dual vector $h_j/(mD)$ has $\ell_1$ norm $1/m$,
giving the matching upper bound by part~(b). Thus $k\,\dZ\dX/n=D$,
compared with one for cubic periods.

\section{Three-dimensional reference lattices}
\label{sec:extremal}

\subsection{Minkowski codes at fixed string distance}

To attain the qubit bound in Eq.~\eqref{eq:3d}, take the packing basis
for the unit octahedron given by Betke and Henk~\cite[Sec.~5]{BetkeHenk2000}.
Clearing its denominator of three and rotating the coordinate axes
gives the period lattice $\Lambda_{38}$ below
(Appendix~\ref{app:proof-minkowski}). This recovers the distance-six
entry of Ref.~\cite[Table~I]{AHHW2025} and the $114$-qubit tricycle code
of Ref.~\cite[Sec.~V\,C and Table~III]{TricycleJMB2025}.
In lattice-rule form it is the rule of index $38$ and enhanced
degree six of Cools and Lyness~\cite[Tables~1 and 3]{CoolsLyness2001}.
To determine the membrane
distance at every scale, we bound the weight in each logical class.

\begin{theorem}[Exact distances of Minkowski codes]\label{thm:minkowski}
Let $\Lambda_{38}=\{x\in\Z^3:x_1+7x_2+11x_3\equiv0\bmod38\}$, with basis $(1,-1,4)$, $(2,2,2)$, $(3,-2,1)$. For every integer $m\ge1$ the toric code on $\Z^3/m\Lambda_{38}$ is a $[[114m^3,3,(6m,19m^2)]]$ code. It meets $n\ge\frac{19}{36}\dZ^3$ with equality and has $k\,\dZ\dX=3n$.
\end{theorem}

\begin{figure*}[tp]\centering
\includegraphics[width=0.8\textwidth]{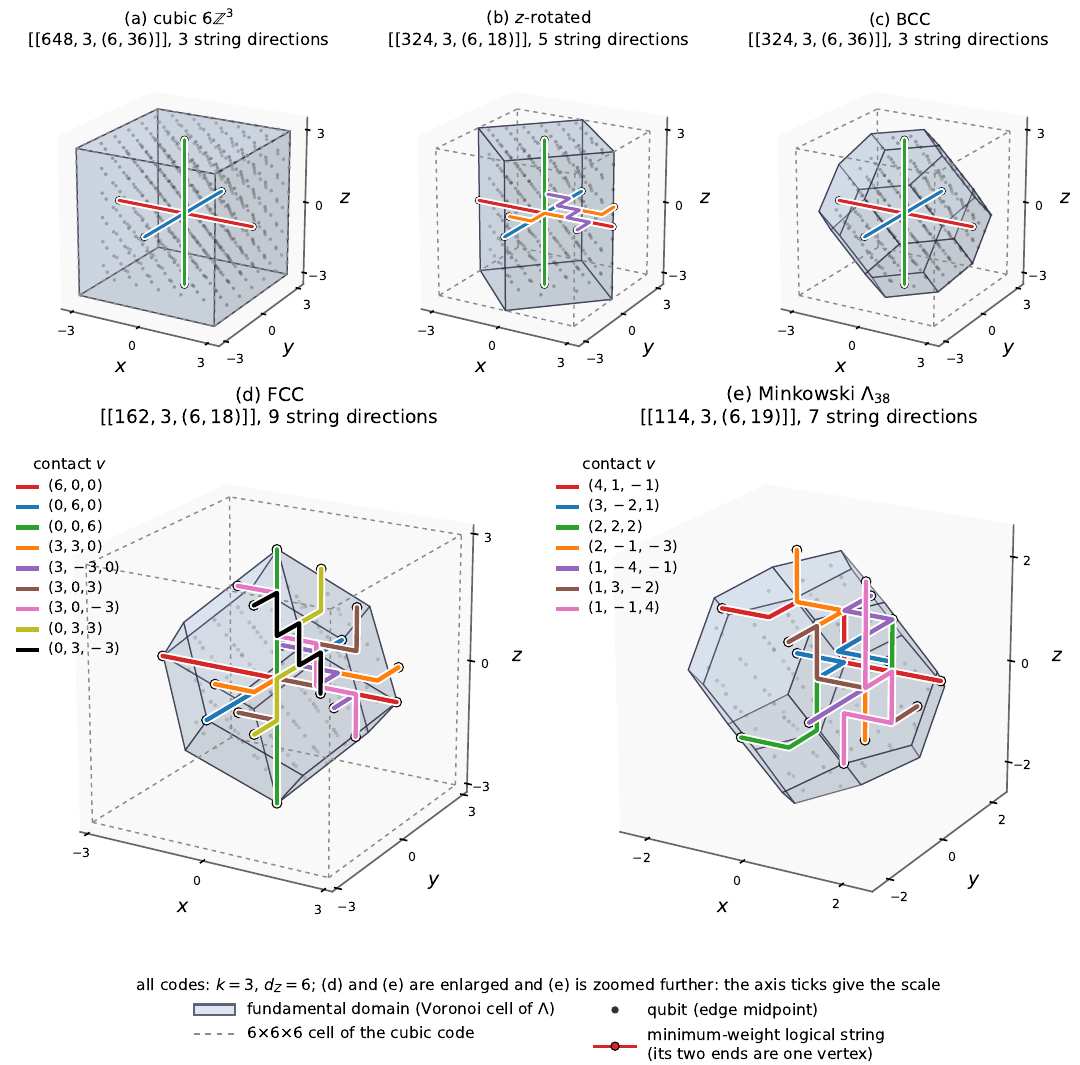}
\caption{\raggedright
\textbf{Reference period lattices at $\dZ=6$.}
Dots mark data qubits in the shaded Euclidean Voronoi cell. The dashed
cube in (a--d) is the cubic reference. We draw one shortest logical
string per contact pair, with endpoints identified on the torus.
In panels (a--d), red, blue and green mark strings along $x,y,z$,
respectively. In the enlarged panels (d,e), keys give the contact of
each string and ticks give the scale; the Minkowski code has no
axis strings, so in panel (e) all seven colours mark staircases.
Panel (e) shows the $114$-qubit Minkowski code on
$\Z^3/\Lambda_{38}$. Table~\ref{tab:lattices} lists distances and string counts.}
\label{fig:cells}
\end{figure*}

Table~\ref{tab:lattices} and Fig.~\ref{fig:cells} compare five reference
families. At equal string distance the Minkowski code uses $19/108$
as many data qubits as the cubic code, the ratio already reported for
the base~\cite[Table~III]{TricycleJMB2025}.
At the same string distance, BCC uses more qubits than the Minkowski
code and has a larger membrane distance.

\begin{table*}[tp]\centering\small\setlength{\tabcolsep}{6pt}
\begin{tabular}{lrrrrcr}
\toprule
period lattice ($\dZ=6$) & $n$ & $\dX$ & $k\dZ\dX/n$ & $n/\dZ^3$ & contacts & $N_{\min}$\\\midrule
cubic $6\Z^3$ & 648 & 36 & 1.00 & 3.00 & $6+0$ & 108\\
$z$-rotated & 324 & 18 & 1.00 & 1.50 & $6+4$ & 774\\
BCC & 324 & 36 & 2.00 & 1.50 & $6+0$ & 54\\
FCC & 162 & 18 & 2.00 & 0.75 & $6+12$ & 1107\\
Minkowski $\Lambda_{38}$ & \textbf{114} & 19 & 3.00 & \textbf{0.528} & $0+14$ & 2280\\
\bottomrule
\end{tabular}
\caption{\raggedright \textbf{Reference toric codes at $\dZ=6$ ($k=3$).}
At even string distance $d$ the period lattices are $d\Z^3$ (cubic),
$\langle\frac d2(1,1,0),\frac d2(1,-1,0),de_3\rangle$ ($z$-rotated),
$d\Z^3+\Z\frac d2(1,1,1)$ (BCC),
$\frac d2\{x\in\Z^3:\sum_ix_i\ \text{even}\}$ (FCC) and
$\frac d6\Lambda_{38}$ (Minkowski, for $d$ divisible by six).
Bold entries mark equality in the qubit bound $n/\dZ^3\ge19/36$,
not a bound on the joint ratio. ``Contacts'' gives the numbers of
axis and staircase contact vectors, counted with both signs; the
octahedral packing density is $\dZ^3/(2n)$. $N_{\min}$ counts
minimum-weight logical strings. Path-packing lower bounds
(Proposition~\ref{prop:dist}(c)) meet the dual-covector upper bounds
(Proposition~\ref{prop:dist}(b)), which establishes the membrane distances at
every integer scale of these five families. Appendix~\ref{app:simulation-inventory}
also gives the membrane distances of the cyclic codes at $\dZ=8,10$.}
\label{tab:lattices}
\end{table*}

\subsection{From shortest vectors to error-chain multiplicity}

The BCC code halves the cubic volume while keeping every shortest
string straight~\cite[Lemma~2.1]{BetkeHenk2000}.

\begin{proposition}[BCC and the classical lattice classification]\label{prop:bcc}
Let $d$ be even and let $\Lambda$ be a period lattice whose only nonzero vectors with $\|v\|_1\le d$ are $\pm de_1,\pm de_2,\pm de_3$; equivalently, $\dZ=d$ and the minimum-weight logical strings are straight and run along all three axes. Then $\det\Lambda\ge d^3/2$, with equality exactly for the body-centred cubic lattice $\Lambda_{\rm BCC}=d\Z^3+\Z\tfrac d2(1,1,1)$. The BCC code is a $[[\tfrac32d^3,3,(d,d^2)]]$ code with $\tfrac32d^2$ minimum-weight logical strings. The standard code is $[[3d^3,3,(d,d^2)]]$ with $3d^2$.
\end{proposition}

After division by $d$, the hypothesis says that the closed unit
octahedron contains only the origin and its six vertices as lattice
points. Appendix~\ref{app:proof-bcc} gives a direct proof of the
determinant bound, membrane distance and string count.

Two packed octahedra touch when their centre displacement $v\in\Lambda$
has $\|v\|_1=\dZ$; call such a vector a \emph{contact}.
The number of step orderings extends the two-dimensional path count
of Ref.~\cite[Sec.~III]{Beverland2019}:

\begin{proposition}[Counting shortest strings]\label{prop:nmin}
The minimum-weight logical strings of the toric code on $\Z^3/\Lambda$ are the monotone lattice paths from a vertex $x$ to $x+v$, closed up on the torus, where $v$ is a contact. Such a string represents the logical class $v\bmod2\Lambda$, and their number is
\begin{equation}
N_{\min}=\frac{V}{\dZ}\sum_{\{\pm v\}}\binom{\dZ}{|v_1|,|v_2|,|v_3|},\label{eq:nmin}
\end{equation}
where the sum runs over the pairs $\pm v$ of contacts and the multinomial coefficient $\dZ!/(|v_1|!\,|v_2|!\,|v_3|!)$ counts the orderings of the steps.
\end{proposition}

At $\dZ=6$, the Minkowski code has over forty times as many shortest
strings as BCC (Table~\ref{tab:lattices}). Under uniform scaling, the
number of step orderings grows exponentially for a staircase and stays
one for a straight string. The loop model uses this distinction.
Appendix~\ref{app:proof-nmin} proves the count, and
Table~\ref{tab:lattices} lists the contacts of the reference families.

\begin{proposition}[Matched distances, different failure coefficients]\label{prop:matched}
For every positive multiple $d$ of six, compare the cubic lattice
$d\Z^3$ with the tilted lattice generated by
\begin{equation}\label{eq:matched-basis}
\begin{aligned}
 b_1&=\tfrac d3(1,1,1),\quad b_2=d(1,-1,0),\\
 b_3&=\tfrac d2(-1,-1,2).
\end{aligned}
\end{equation}
Both codes have parameters $[[3d^3,3,(d,d^2)]]$. With independent
data-$Z$ flips of probability $q$, perfect syndrome measurements and
maximum-likelihood decoding of the logical class, let $P_{\rm fail}$ be
the probability of an error in any string-sector observable. For fixed
$d$ as $q\to 0^+$, write $P_{\rm fail}=A(d)q^{d/2}+O_d(q^{d/2+1})$.
Then
\begin{equation}\label{eq:matched-coeff}
\begin{aligned}
 A_{\rm cubic}(d)&=\tfrac32 d^2\binom{d}{d/2},\\
 A_{\rm tilted}(d)&\ge d^2\binom{d}{d/3,d/3,d/3}.
\end{aligned}
\end{equation}
In particular $A_{\rm tilted}/A_{\rm cubic}=\Omega((3/2)^d/\sqrt d)$.
\end{proposition}

\begin{figure}[tb]\centering
\includegraphics[width=\columnwidth]{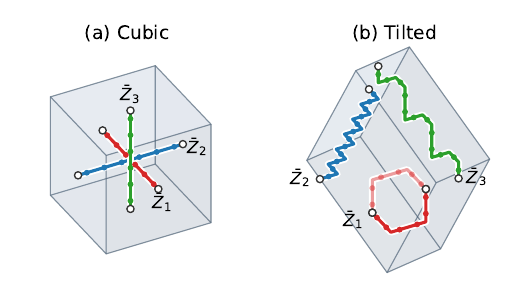}
\caption{\raggedright
\textbf{Logical $Z$ generators on cubic and tilted tori at $d=6$ (same scale).}
Dots mark qubits, and each path's open endpoints coincide on the torus.
(a) The cubic strings $\bar Z_1$ (red), $\bar Z_2$ (blue) and $\bar Z_3$
(green) each follow six edges along one axis.
(b) The red paths show two of the $6!/(2!)^3=90$ six-step routes from
$x$ to $x+b_1$ with $b_1=(2,2,2)$, all realizing $\bar Z_1$.
The blue and green strings each have weight 12, with windings
$b_2=(6,-6,0)$ and $b_3=(-3,-3,6)$, respectively.}
\label{fig:matched-cells}
\end{figure}

Figure~\ref{fig:matched-cells} shows three straight contact pairs on
the cubic torus. The tilted torus has one contact pair, $\pm b_1$, with
many possible step orderings. Alternating
edges of each tilted shortest loop give two errors with the same
syndrome and different logical actions. Neither error in such a pair
belongs to an alternating pair from another loop, so at least one
failure per loop survives even optimal decoding. Deciding each syndrome
by maximum likelihood among all distinct halves gives a fifteenfold
increase in the leading failure coefficient at $d=6$
(Table~\ref{tab:matched} and Appendix~\ref{app:proof-matched}).
This comparison takes $q\to0$ at fixed $d$. Section~\ref{sec:matched-circuit}
tests the ordering with noisy checks.

\begin{table}[tb]\centering
\caption{\raggedright \textbf{The two $[[648,3,(6,36)]]$ codes in Fig.~\ref{fig:matched-cells}, with perfect checks.} A half is a
distinct set of three edges on a shortest string; $A(6)$ is the
coefficient of $q^3$ in the string-sector block failure probability
under maximum-likelihood decoding.}
\label{tab:matched}
\begin{tabular}{lrrr}\toprule
Layout & Shortest strings & Distinct halves & $A(6)$\\\midrule
Cubic & 108 & 2\,160 & 1\,080\\
Tilted & 3240 & 34\,992 & 16\,200\\\bottomrule
\end{tabular}
\end{table}

A vertex check sees the endpoints of an error chain. Exchanging two
adjacent steps of a staircase changes its route by a plaquette
stabilizer, so every shortest tilted string has the same logical
action. Write $\bar Z_i$ for a logical string of winding $b_i$, and let
$\bar X_1,\bar X_2,\bar X_3$ be the basis of logical $X$ observables dual
to the periods in Eq.~\eqref{eq:matched-basis}: $\bar Z_i$ flips
$\bar X_i$ and neither of the other two. A winding
$c_1b_1+c_2b_2+c_3b_3$ with integer coefficients $c_i$ then flips
$\bar X_i$ exactly when $c_i$ is odd. The shortest tilted strings
therefore flip only $\bar X_1$. For the cubic layout the periods are
$b_i=d e_i$.

\begin{corollary}[Distance for retained observables]\label{cor:matched-observables}
Choose $r$ independent logical $X$ observables, allowing products of the
three basis observables. Let $\dZ^{(r)}$ be the largest, over these choices,
of the minimum weight of an undetectable data-$Z$ error that flips at
least one chosen observable; retaining all three gives $\dZ^{(3)}=\dZ$.
For the layouts of Proposition~\ref{prop:matched},
\begin{equation}\label{eq:retained-distances}
 \bigl(\dZ^{(3)},\dZ^{(2)},\dZ^{(1)}\bigr)=
 \begin{cases}
 (d,d,d) & \text{cubic},\\
 (d,5d/3,2d) & \text{tilted}.
 \end{cases}
\end{equation}
On the tilted torus, keep $(\bar X_2,\bar X_3)$ for the pair and
$\bar X_2$ for the single observable. With perfect checks and independent
data-$Z$ flips, the optimal probability of a wrong retained result is
$\Theta_d\bigl(q^{\dZ^{(r)}/2}\bigr)$ for these choices at fixed $d$ as $q\to 0^+$.
\end{corollary}

At $d=6$ the tilted pair first fails at order $q^5$, whereas every
cubic pair fails at order $q^3$. Thus the layout with the larger block
failure coefficient is better for these two measurements at sufficiently
low noise. There is also an interpretation for stored qubits: if the
unused basis qubits are treated as gauge, $\dZ^{(r)}$ is the exact dressed
Pauli distance for the stated choices. The noisy-check tests below
measure only the logical $X$ readout, not a full quantum memory.

The idea of protecting a subsystem rather than the whole block is not
new. Aasen \emph{et al.}\ give a four-dimensional example and use a
task-specific distance for Bell-pair preparation~\cite[Secs.~II\,C and
III\,C]{AHHW2025}; a two-dimensional toric-code experiment likewise treats
one logical qubit as gauge~\cite{AtomToric2026}. Here the comparison fixes
the qubit count and both code distances: changing which outcomes count
as failures reverses the ranking without changing the checks.

The same distinction applies to other lattices. For BCC in the basis
$(de_1,de_2,\frac d2(1,1,1))$, the three axis directions act as
$\bar Z_1,\bar Z_2,\bar Z_1\bar Z_2$ and span only two independent
logical directions. The third generator has a diagonal
representative of weight $3d/2$. FCC's axis strings all share one
class. The seven contact pairs of the Minkowski code realize all seven
nonzero classes. For example, in the basis of
Theorem~\ref{thm:minkowski}, $(2,-1,-3)=b_3-b_1$ acts as
$\bar Z_1\bar Z_3$.

All these codes retain check weights six and four.
Counting one ancilla per check multiplies their data-qubit cost by
$7/3$. Appendix~\ref{app:memory-circuits} gives the gate counts and
circuit depths for direct torus couplings.

\begin{figure}[tb]\centering
\includegraphics[width=\columnwidth]{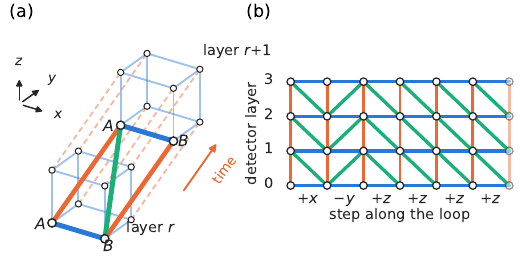}
\caption{\raggedright \textbf{Edges of the matching graph in space and time.}
Circles are detectors and edges are faults (Sec.~\ref{sec:circuit}).
Spatial (blue) and diagonal (green) edges are flips of a data qubit;
time edges (orange) are faults on a check's ancilla. All edges are read
from the detector error model of the direction-ordered circuit.
(a) One unit cube in detector layers $r$ and $r+1$; bold lines show the edges of
the qubit from $A$ to $B=A+e_x$. (b) The strip above a shortest loop of
the Minkowski code, of winding $(1,-1,4)$, for $R=3$ rounds; the last
column repeats the first vertex. On the step $-y$ the diagonal leans the
other way. Under the greedy schedule its direction varies from qubit to
qubit.}
\label{fig:edges}
\end{figure}

\section{A circuit-informed loop model}
\label{sec:entropy}

With independent data flips and perfect checks, a minimum-weight decoder
corrects every error of weight less than $\dZ/2$ in the string sector.
Distance alone does not fix the failure prefactor
(Proposition~\ref{prop:matched}). In a noisy syndrome circuit, a physical
fault need not be a single data flip. We count the ways a logical
string can wind around the torus. For each loop, we estimate how often
decoding fails across repeated measurement rounds, then correct for
errors counted on multiple loops of the same winding. This gives a
string-sector failure estimate without fitting to logical-error simulations.

\subsection{Syndrome circuit and matching graph}
\label{sec:circuit}

In an $X$-basis memory the logical $X$ outcomes are flipped by $Z$
strings, which the vertex checks detect. The vertex check at $A$ is the
product of $X$ over the six data qubits on the edges that meet at $A$.
Its ancilla is prepared in $|+\rangle$, controls one CNOT onto each of
these qubits, and is measured in the $X$ basis. A CNOT copies a $Z$ on
its target onto its control, so the ancilla registers a $Z$ on a data
qubit only if the $Z$ is already there when their CNOT acts.

The CNOTs of a round are applied in \emph{CNOT layers}, time steps in
which each qubit takes part in at most one CNOT. All vertex ancillas act
in parallel, one CNOT per CNOT layer. In the
\emph{direction-ordered} schedule the six vertex CNOT layers take the
directions $+x,+y,+z,-x,-y,-z$ in turn, so a qubit on an edge along
axis $i$ meets the check at its lower end in CNOT layer $i$ and the
check at its upper end in CNOT layer $i+3$. The \emph{greedy} schedule
puts each CNOT in the first CNOT layer in which both of its qubits are
free, so the order and the gap between a qubit's two vertex CNOTs vary
from qubit to qubit. Twelve plaquette CNOT layers complete a round, so
a round has $18$ CNOT layers, or $19$ with the usual seven greedy vertex
CNOT layers (Appendix~\ref{app:memory-circuits}). Circuit noise applies
two-qubit depolarization of strength $p$ after every CNOT, single-qubit
depolarization $p$ to each data qubit idle in a CNOT layer, and flips
with probability $p$ to every preparation and measurement.

A \emph{detector} compares the outcome of one vertex check in two
consecutive rounds, and the detectors of one round form a \emph{detector
layer}. Together with the initial state and the final data readout, $R$
rounds give $R+1$ detector layers. From here on, a layer without
qualification means a detector layer. A $Z$ on a data qubit persists, so
it fires each of its two checks once; a wrong outcome fires the same
check in two consecutive layers. Stim~\cite{Stim2021}
decomposes every fault into edges between the pairs of detectors that
it fires, and these edges form the matching graph (Fig.~\ref{fig:edges}).

Consider the qubit on the edge from $A$ to $B=A+e_x$ under the
direction-ordered schedule, so that it meets $A$ in CNOT layer $1$ and
$B$ in CNOT layer $4$ of every round. A $Z$ that arrives between its CNOT with $B$
in round $r-1$ and its CNOT with $A$ in round $r$ is first seen by both
checks in round $r$. It fires $(A,r)$ and $(B,r)$: a \emph{spatial}
edge in layer $r$. A $Z$ that arrives between the CNOTs with $A$ and $B$
in round $r$ misses the ancilla of $A$ in that round. It fires $(B,r)$
and $(A,r+1)$: a \emph{diagonal} edge. A fault on the ancilla of $A$
alone gives a wrong outcome of $A$ in round $r$ and leaves the data
intact. It fires $(A,r)$ and $(A,r+1)$: a \emph{time} edge. With no
other faults, $A$ reads $+1,-1,-1$ in rounds $r-1,r,r+1$ in the first
case, $+1,+1,-1$ in the second and $+1,-1,+1$ in the third.

Counting these faults gives the flip probabilities of the three edges.
Two facts suffice. An idle data qubit receives $X$, $Y$ or $Z$ with
probability $\frac p3$ each, so it acquires a $Z$ component with
probability $\frac23p$ per CNOT layer. After a CNOT, each of the fifteen
nonidentity two-qubit Paulis occurs with probability $\frac p{15}$, so
$Z$ or $Y$ on one qubit together with two given options on the other
has probability $\frac4{15}p$. In bulk rounds of the direction-ordered
schedule, to leading order in $p$, every qubit then has
\begin{equation}
\begin{aligned}
q_s&=10\cdot\tfrac23p+2\cdot\tfrac{4}{15}p+\tfrac{48}{15}p=\tfrac{52}{5}p,\\
q_d&=\tfrac{4}{15}p+2\cdot\tfrac23p+\tfrac{4}{15}p=\tfrac{28}{15}p,\\
q_t&=2p+6\cdot\tfrac{4}{15}p=\tfrac{18}{5}p,
\end{aligned}
\label{eq:edge-probabilities}
\end{equation}
that is, $10.4p$, $1.87p$ and $3.6p$. The time edge collects the
preparation and measurement flips of the ancilla of $A$, each with
probability $p$, and a $Z$ or $Y$ on that ancilla with $I$ or $X$ on
the data qubit after each of its six CNOTs. The diagonal collects a $Z$
or $Y$ on the qubit with $I$ or $X$ on the ancilla of $A$ after their
CNOT, idling in the two CNOT layers between the CNOTs with $A$ and $B$,
and a $Z$ or $Y$ on both the qubit and the ancilla of $B$ after their
CNOT, which acts as if it came just before that CNOT. The spatial edge
collects idling in the other ten CNOT layers in which the qubit idles,
two vertex and eight plaquette CNOT layers; two vertex-CNOT faults, a
$Z$ or $Y$ on the qubit with $I$ or $X$ on the ancilla of $B$ after
their CNOT and a $Z$ or $Y$ on both the qubit and the ancilla of $A$
after theirs; and faults during plaquette CNOTs.

The last term needs one more step. Label the four data qubits of a
plaquette $q_1,\dots,q_4$ in the order of their CNOTs; in our schedule
$q_1,q_2$ are opposite edges, and so are $q_3,q_4$. The data qubit is
the control of a plaquette CNOT, so a $Z$ on the plaquette ancilla is
copied onto every data qubit whose CNOT comes later. A $Z$ on all four
qubits is the plaquette stabilizer, which the vertex checks do not see.
After the CNOT with $q_k$, a $Z$ or $Y$ on $q_k$ alone, on the ancilla
alone, or on both has probability $\frac4{15}p$ each. Up to the
stabilizer, these three faults flip $q_1$, $q_1$ and no qubit for $k=1$;
$q_2$, $q_3q_4$ and $q_1$ for $k=2$; $q_3$, $q_4$ and $q_3q_4$ for
$k=3$; and $q_4$, no qubit and $q_4$ for $k=4$. The twelve qubit flips
give $12\cdot\frac4{15}p=\frac{48}{15}p$ per plaquette. A two-qubit
flip fires four detectors, and Stim splits it into two spatial edges:
the one for $k=2$, which equals $q_1q_2$ up to the stabilizer, into the
edges of $q_1$ and $q_2$, and the one for $k=3$ into those of $q_3$ and
$q_4$. The twelve flips of one plaquette then fall on
$(q_1,q_2,q_3,q_4)$ as $(4,2,2,4)$. In each of the two plaquette
orientations that contain a given data qubit, the qubit is $q_1$ of one
plaquette and $q_2$ of the next, or $q_3$ of one and $q_4$ of the next,
so it collects $2\cdot(4+2)=12$ flips: every data qubit receives exactly
$\frac{48}{15}p$. Stim's
decomposed detector error model reproduces every term of
Eq.~\eqref{eq:edge-probabilities} (Appendix~\ref{app:edge-probabilities}).

\subsection{Loops and their space-time strips}
\label{sec:model}

The model builds on circuit-derived fault enumeration~\cite{Fowler2013}
and loop sums with overlap corrections~\cite[Sec.~V, Eq.~(19)]{Beverland2019}.
Regev \emph{et al.}\ count distinct leading errors in planar
codes~\cite[Sec.~III, Theorem~2 and Algorithm~1]{Regev2026} and
model noisy measurements by majority
vote~\cite[Sec.~VI]{Regev2026}. Sethi \emph{et al.}\ use intersections to
remove duplicates in leading code-capacity counts~\cite[Theorem~8 and
Sec.~3.3]{Sethi2026}. Beverland \emph{et al.}\ enumerate the
minimum-weight failing configurations of syndrome circuits, sorted by
syndrome and logical action~\cite[Sec.~4.3]{Beverland2025}. Here we include all rounds of a finite memory
experiment by decoding a strip above each winding loop.

When minimum-weight decoding fails, the error and correction together
contain a cycle with nontrivial winding. On that cycle the correction
is no heavier than the error~\cite{DKLP2002}. The \emph{strip} above a length-$\ell$ loop
contains the $\ell(R+1)$ copies of its vertices in the detector layers
and the spatial, time and diagonal edges along the loop
(Fig.~\ref{fig:edges}(b)). Summing the strip failure
probabilities gives the model estimate per round,
\begin{equation}
\epsilon(\Lambda)=\frac1R\sum_{v\in\Lambda\setminus2\Lambda}\ \sum_{\ell\ge\|v\|_1}K_\ell(v)\,\frac{V\,W_\ell(v)}{2\ell}\,\Pi_\ell(v).
\label{eq:model}
\end{equation}
Here $V=\det\Lambda$ and $W_\ell(v)$ counts length-$\ell$ walks from
$0$ to $v$ in $\Z^3$ that never immediately reverse a step.
$\Pi_\ell(v)$ is the strip failure probability averaged over loops of
that length and winding. A failure means that the error and matching
correction together wind an odd number of times along the loop. The
factor $K_\ell(v)\le1$ accounts for failing sets shared by loops of the
same winding (Sec.~\ref{sec:shared}).

A closed walk of winding $v$ flips a logical $X$ outcome exactly when
$v\notin2\Lambda$.
The factor $V$ counts starting vertices, and division by $2\ell$
accounts for starting points and orientations of a loop. The sum
includes both shortest loops and \emph{detours} of length
$\ell=\|v\|_1+2j$ with $j\ge1$. For a shortest winding
$VW_\ell(v)/(2\ell)$ is the number of loops. For detours the walks also
include closed walks that are not loops, and the sharing factor gives
these zero weight (Appendix~\ref{app:detours}). We compute $W_\ell(v)$ exactly
through $\ell=68$.

The strips use the edge probabilities $q_s,q_d,q_t$ of
Sec.~\ref{sec:circuit}, with separate preparation and readout layers.
The model treats decomposed graph components as independent even when
one physical fault produces several of them. It also replaces whole-graph
decoding by strip decoding. These approximations, and overlaps between
windings, remain sources of discrepancy.
Appendix~\ref{app:model-details} gives the edge probabilities and the
independent-ring limit.

\subsection{Shared failing sets}
\label{sec:shared}

Beverland \emph{et al.}\ assign repeated half-error configurations to
canonical staircase paths on the rotated two-dimensional toric
code~\cite[Sec.~III\,C, Eqs.~(13)--(14)]{Beverland2019}. We count
distinct error sets for a fixed three-dimensional winding.

A \emph{half-set} consists of $w=\lceil\ell/2\rceil$ edges of a
length-$\ell$ loop. This is the fewest flips that can cause an isolated
ring to fail, with half weight for a tie on an even ring. A half-set
on a shortest straight loop determines the loop within its winding.
On a staircase, the same edges can be joined by different routes.
For $v=(2,2,0)$, the $x$ edge leaving $0$ and the $y$ edge entering
$v$ belong to both $xxyy$ and $xyxy$. To count each set once among
the shortest loops of winding $v$, we put $K_{\|v\|_1}(v)=\rho(v)$,
with
\begin{equation}
\rho(v)=\frac{\text{number of distinct half-sets}}{\text{number of (loop, half-set) pairs}}\le1.
\end{equation}
The next proposition gives an exact count under a condition on the
period lattice. Using it in the failure model also assumes that
isolated-loop failure and tie weights approximate those of the full
decoder.

\begin{proposition}[Shared half-sets]\label{prop:shared}
Let $v\in\Lambda\setminus0$, $\ell=\|v\|_1$, $w=\lceil\ell/2\rceil$ and $|v|=(|v_1|,|v_2|,|v_3|)$. With one formal variable $x_i$ per axis, let
\begin{equation}
E(x)=x_1+x_2+x_3,\qquad P(x)=\prod_{i=1}^3\frac1{1-x_i}
\end{equation}
be the generating functions of a single edge and of a monotone gap displacement of any length, respectively. A gap records the number of steps along each axis, not their order. Let $[x^a]f$ denote the coefficient of $x_1^{a_1}x_2^{a_2}x_3^{a_3}$ in a power series $f$. Then
\begin{equation}
\rho(v)=\frac{\ell\,[x^{|v|}]\,(EP)^{w}}{w\binom{\ell}{|v_1|,|v_2|,|v_3|}\binom{\ell}{w}},
\end{equation}
which does not depend on $\Lambda$. The formula is exact whenever no nonzero $v'\in\Lambda$ has $|v'_i|\le|v_i|$ for every $i$ and $|v'_i|<|v_i|$ for some $i$, in particular when $v$ is a shortest vector of $\Lambda$. A half-set that lies on shortest loops of several windings is counted in $\rho$ of each.

For the coefficient expression above, let $u=|v|/\ell$ be the direction of $v$, a probability vector on the three axes, with Shannon entropy $H(u)=-\sum_iu_i\log u_i$. As the integer $t\to\infty$,
\begin{equation}
\rho(tv)=e^{-\|tv\|_1\,\chi(u)\,(1+o(1))},
\end{equation}
with the decay rate per edge $\chi(u)=H(u)+\log2-h(u)$, where
\begin{equation}
\begin{split}
h(u)=\min_{\theta}\Big[&\tfrac12\log\sum_ie^{\theta_i}-\tfrac12\sum_i\log\big(1-e^{\theta_i}\big)\\
&-\sum_i\theta_iu_i\Big];
\end{split}
\end{equation}
here $i$ runs over the support of $u$ (the axes with $u_i>0$) and the minimum is over the real $\theta_i<0$, $i$ in the support. Finally, $\chi(u)=0$ for the three coordinate directions and $\chi(u)>0$ for every other direction.
\end{proposition}

The proof in Appendix~\ref{app:proof-shared} records each gap
by its displacement, so different routes through the gap give the same
half-set. For $v=(2,2,0)$ this leaves $7V$ distinct sets among $9V$
loop--half-set pairs, giving $\rho=7/9$. Even at large distances,
different shortest windings can still share half-sets.
The asymptotic statement concerns the coefficient expression. On a
fixed torus the exactness condition fails for $tv$ with $t\ge2$, because
$v$ itself is then a nonzero period inside the box of $tv$; the
condition can hold on the scaled torus $\Z^3/t\Lambda$.

Detours require a further correction: a half-set can already contain
half of a shorter loop. We use an inverse-multiplicity
estimate~\cite{KarpLubyMadras1989,OwenMaximovChertkov2019} for $K_\ell(v)$,
with exact dynamic programming inside that estimate and a specified
continuation for unsampled orders (Appendix~\ref{app:detours}).
Applying these one-layer corrections to repeated-round failures is
another approximation.

\subsection{Why failures span several rounds}
\label{sec:rounds}

A logical failure can come from flips spread across several rounds,
even when no single round contains enough flips to cause a failure on
its own.
Consider first an isolated nontrivial loop in one round with perfect
syndromes and equal edge weights. The decoder sees where the flipped
segments end. The unflipped segments have exactly the same endpoints,
so they are a candidate correction too. Choosing them completes the logical loop
instead of undoing the error. For example, three flips on a five-edge
loop leave a two-edge alternative. Matching chooses those two edges and
fails. On a six-edge loop, the same three flips leave three edges:
the two corrections tie, and symmetric tie breaking fails half the time.

With repeated checks, the same ambiguity can involve flips in different
rounds. Picture the strip as a stack of copies of the spatial loop, one
per detector layer (Fig.~\ref{fig:edges}(b)). A \emph{lift} is a path in this stack that winds once
around the loop and returns to its starting vertex and layer. A spatial
edge follows the loop within one layer; a time edge changes layers at
the same check; a diagonal does both. As in Sec.~\ref{sec:circuit}, if
a data qubit meets check $A$ before check $B$, a flip between their
CNOTs in round $r$ is seen by $B$ that round but by $A$ only in the next
(Fig.~\ref{fig:edges}(a)).

Matching compares only corrections that explain all the observed
events and chooses one with the lowest total weight. An edge with flip
probability $q_e$ contributes $\omega_e=\log[(1-q_e)/q_e]$.
If the only events are $(B,r)$ and $(A,r+1)$, the diagonal connects
them directly. The spatial edge from $(B,r)$ costs less, but it
reaches $(A,r)$, one round too early. That route still needs a time
edge of weight $\omega_t$ at $A$ to reach $(A,r+1)$. Matching favours
the diagonal over this two-edge path if $\omega_d<\omega_s+\omega_t$.

Two events in the same round can also be joined by a correction that
visits the next round and returns. A local detour preserves the logical
state. A correction that takes the other way around the spatial loop
completes a logical winding with the actual error and causes a failure.

To see what a change of round costs on a lift, consider just one
crossing from $B$ to $A$, without fixing the round of arrival. Both
the spatial edge and the diagonal cross the same data qubit in one
step. If neither edge flipped, the competing correction pays $\omega_d$
instead of $\omega_s$ to arrive in the next round. The extra cost for
that crossing is $\omega_d-\omega_s\approx\log(q_s/q_d)$, or
$1.3$--$3.1$ in our circuits. The lift must also return to its
starting round. At low noise, both
flip probabilities are proportional to $p$ to leading order, so $p$
cancels in their ratio. On the lift, the competing correction uses
the edges that did not flip. One more spatial flip therefore adds
$\omega_s$ to the actual error's weight and removes $\omega_s$ from
the competing correction's weight. That shift of
$2\omega_s\approx2\log(1/q_s)$ can outweigh the cost of joining
flipped segments in different rounds.

The number of lifts grows exponentially with the loop length $\ell$,
and a failure need not stay within two rounds: a model that keeps only
two consecutive rounds has a ratio to simulation that still falls with
distance (Appendix~\ref{app:comparison-details}). We therefore keep all
rounds of each experiment. Even
with half the lift flipped, matching can find a lighter route elsewhere
in the strip. We favour flips on a randomly chosen lift, reweight
each sample to undo the bias, and decide failure by matching on the
whole strip. Paired weight perturbations approximate symmetric tie
breaking (Appendix~\ref{app:strip}).

The five- and six-edge examples show why repeated rounds need not
increase failure rates by the same factor at odd and even distances.
In one round, loops of lengths $2k+1$ and $2k+2$ both need at least
$k+1$ flips before matching can fail. At that count the wrong correction
on the odd loop starts lighter by one spatial-edge weight $\omega_s$;
it can pay for unflipped diagonals and still be lighter. On the even
loop it starts tied, so the same cost can tip the decision in favour
of the correct recovery. Appendix~\ref{app:rings} gives the
one-round calculation and Appendix~\ref{app:rounds-checks} the
numerical parity checks.

\subsection{The free-loop norm}
\label{sec:norm}

As we enlarge the torus at fixed noise, how fast does the total weight
of walks of a given winding decay? From the origin there is one shortest
walk to $(2t,0,0)$. To reach $(t,t,0)$ in the same $2t$ steps, we can
order the $t$ steps along $x$ and the $t$ steps along $y$ in
$\binom{2t}{t}$ ways. Both endpoints also admit longer walks.
Proposition~\ref{prop:kappa} gives the decay rate when each walk is
counted separately.

Let $B_\ell$ count the flipped edges on an isolated ring of length
$\ell$ in one round. If each edge flips independently with probability
$0<q<1/2$, then $B_\ell\sim\mathrm{Binomial}(\ell,q)$. The decoder
fails when more than half the edges flip and half the time on a tie,
so the ring's failure probability is
\begin{equation}
T_\ell(q)=\Pr(B_\ell>\ell/2)+\tfrac12\Pr(B_\ell=\ell/2).
\label{eq:ring}
\end{equation}
At fixed $q$, $T_\ell(q)$ equals $\mu^\ell$ up to polynomial factors in $\ell$, where
$\mu=2\sqrt{q(1-q)}$ (Appendix~\ref{app:rings}). To find the
exponential rate, we therefore study the \emph{free-loop sum}
\begin{equation}
G_\mu(v)=\sum_{\ell\ge\|v\|_1} W_\ell(v)\,\mu^\ell .
\end{equation}
Here $W_\ell(v)$ counts length-$\ell$ non-backtracking walks from $0$
to $v$; for a period $v$, each closes on the torus. ``Free'' means
that each walk contributes separately, without correcting for shared
errors. For example, the detour $(+x,+y,-x)$ and the single step $+y$
reach the same endpoint but contribute separate weights $\mu^3$ and $\mu$.
Only the immediate reverse is forbidden, so each nonempty walk has five
extensions: the total weight over all endpoints changes by a factor of
$5\mu$ per step, and the sum over lengths converges for $\mu<1/5$.

Recall from Proposition~\ref{prop:shared} that the directional entropy is
$H(u_v)=-\sum_{i:\,u_i>0}u_i\log u_i$, where $u_i=|v_i|/\|v\|_1$
is the fraction of shortest-walk steps along axis $i$.

\begin{proposition}[Directional decay of the free-loop sum]\label{prop:kappa}
For $0<\mu<1/5$ and nonzero $v\in\Z^3$, the exponential decay rate
of the free-loop sum along integer multiples of $v$ is
\begin{equation}
\begin{aligned}
\kappa_\mu(v)&=\lim_{t\to\infty}-\frac{1}{t}\log G_\mu(tv)\\
&=\max_\theta\Big\{\theta\cdot v:2\sum_{i=1}^3\cosh\theta_i
                      \le\mu^{-1}+5\mu\Big\}.
\end{aligned}
\end{equation}
The maximum, taken over $\theta\in\R^3$, defines a norm on all of
$\R^3$. Write $u_v=|v|/\|v\|_1$ for the fractions of shortest-walk
steps along the three axes, and $H(u_v)$ for their Shannon entropy.
For fixed $v\ne0$:
\begin{enumerate}
\item[(a)] As $\mu\to0$, $\kappa_\mu(v)=\|v\|_1[\log(1/\mu)-H(u_v)]+O(\mu\|v\|_1)$.
\item[(b)] As $\mu\rightarrow1/5^-$, $\kappa_\mu(v)=\sqrt{(1-\mu)(1-5\mu)/\mu}\;\|v\|_2\allowbreak\,(1+o(1))$.
\end{enumerate}
\end{proposition}

At fixed $\mu$, Proposition~\ref{prop:kappa} gives
$G_\mu(tv)=\exp[-t\kappa_\mu(v)+o(t)]$ as $t\to\infty$.
This is the decay rate of the sum over all walks to $tv$, including
detours. A smaller $\kappa_\mu(v)$ means slower decay. In the opening example,
$(2,0,0)$ and $(1,1,0)$ have the same $\ell_1$ length but directional
entropies $0$ and $\log 2$, respectively. Part~(a) shows why the
free-loop sum decays more slowly for the staircase at low noise.
At the other end of the convergence range, part~(b) gives Euclidean
length after division by a common vanishing
factor. Beverland \emph{et al.}\ describe this emergence of approximately
Euclidean geometry from Manhattan path counts~\cite[Sec.~V\,B,
Eq.~(29)]{Beverland2019}. Appendix~\ref{app:proof-kappa} takes the rate
from the known directional decay of the lattice Green
function~\cite[Theorem~1.3(i)]{MichtaSlade2022} and derives both limits
from its variational form.

Because several walks in $G_\mu(v)$ can contain the same set of flipped
edges, the failure model in Eq.~\eqref{eq:model} weights shortest loops
by $\rho(v)$ and detours by a length-dependent overlap factor $K_\ell(v)$
(Sec.~\ref{sec:shared}).

\begin{figure*}[tp]\centering
\includegraphics[width=0.92\textwidth]{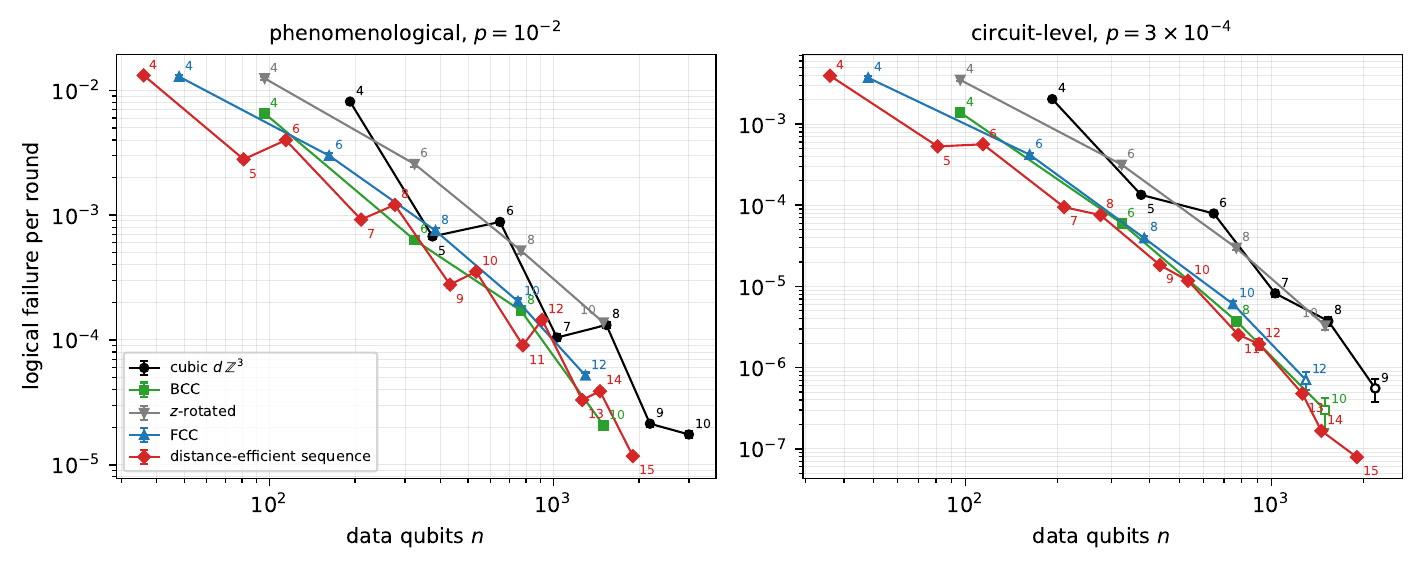}
\caption{\raggedright
\textbf{String-sector failure per round versus data qubits.}
Each code stores three logical qubits for $\dZ$ rounds (labels give $\dZ$).
For each $d=4,\ldots,15$, the red diamond gives the measured rate for a
cubic-lattice torus $\Z^3/\Lambda$ minimizing the data-qubit count subject
to $\dZ\ge d$. Cubic distances can be odd. BCC, FCC and $z$-rotated
distances are $2m$ at integer scale $m$.
The panels use phenomenological noise ($p=10^{-2}$, left) and circuit
noise with the greedy vertex schedule ($p=3\times10^{-4}$, right).
Bars show one approximate standard deviation. Open markers denote fewer
than $30$ failures against a target of $300$.
Appendices~\ref{app:simulation-inventory} and~\ref{app:sampling} list the
period lattices and shot caps.}
\label{fig:twisted}
\end{figure*}

\section{Memory tests and limits of the prediction}
\label{sec:sim}

\subsection{Memory experiment and observed geometry effects}

We simulate $X$-basis memory for $R=\dZ$ noisy rounds.
Stim~\cite{Stim2021} generates the circuit detector error model;
PyMatching~\cite{SparseBlossom2025} decodes the string sector.
The full readout fails if any of the three logical $X$ outcomes is wrong.
For shot failure probability $P_L$, we report
$\epsilon=1-(1-P_L)^{1/R}$. This reporting convention does not
assume independent logical failures across rounds.

Phenomenological noise depolarizes each data qubit with strength $p$
per round and flips each check outcome with probability $p$.
Circuit noise and the two vertex schedules are those of
Sec.~\ref{sec:circuit}. The circuit data in Fig.~\ref{fig:twisted}
(right) and in the fixed-size comparison of
Sec.~\ref{sec:fixedn} use the greedy schedule, which needs
seven vertex CNOT layers per round, except for the cyclic lattice with
$\dZ=14$, which needs six. The matched-pair experiment
(Sec.~\ref{sec:matched-circuit}) and the frozen-forecast test
(Appendix~\ref{app:frozen-forecasts}) use the direction-ordered schedule. The decomposed detector-graph distance
equals $\dZ$ for the $29$ tori checked with the greedy schedule.
A single physical fault can produce several graph edges, so counting
edges alone cannot certify a physical fault distance of $\dZ$.

For each target distance $d=4,\ldots,15$, the horizontal position of
the red diamond in Fig.~\ref{fig:twisted} gives the fewest data qubits
needed to encode three logical qubits with $\dZ\ge d$ on a
three-dimensional cubic-lattice torus. We call these twelve codes
the \emph{distance-efficient sequence}. Cools and Lyness give these
lattices up to signed coordinate permutations in their restricted
search~\cite[Tables~1 and 3]{CoolsLyness2001}.
Appendix~\ref{app:simulation-inventory} verifies each minimum over all
full-rank integer period lattices $\Lambda\subseteq\Z^3$. The height of
the diamond is the measured string-sector failure rate per round for
that code and noise model. The red diamonds at $d=5,7,8$ correspond
to cyclic lattices with $27,70,92$ vertices, respectively.
Appendix~\ref{app:frozen-forecasts} includes these three geometries in
a test of frozen model forecasts.

At equal string distance, layouts with shortest staircases fail more
often than cubic or BCC in Fig.~\ref{fig:twisted}. At equal qubit count,
tilting can give enough extra distance to compensate. BCC keeps its
shortest strings straight and, at distance eight in the circuit panel,
matches the cubic failure rate with half the data qubits. In the cubic
$5/6$ and $7/8$ pairs, the odd member has the lower measured rate
under phenomenological noise and the higher rate under circuit noise.
For cubic $9/10$, the phenomenological rate ratio is $1.22$
(Appendix~\ref{app:rounds-checks}).

On the smallest member of each family, membrane-sector failures were
much rarer than string-sector failures at the tested noise levels.
Our circuit comparisons therefore focus on the string-sector readout
(Appendix~\ref{app:membrane-sector}).

\begin{figure*}[tp]\centering
\begin{minipage}[c]{0.48\textwidth}
\includegraphics[width=\linewidth]{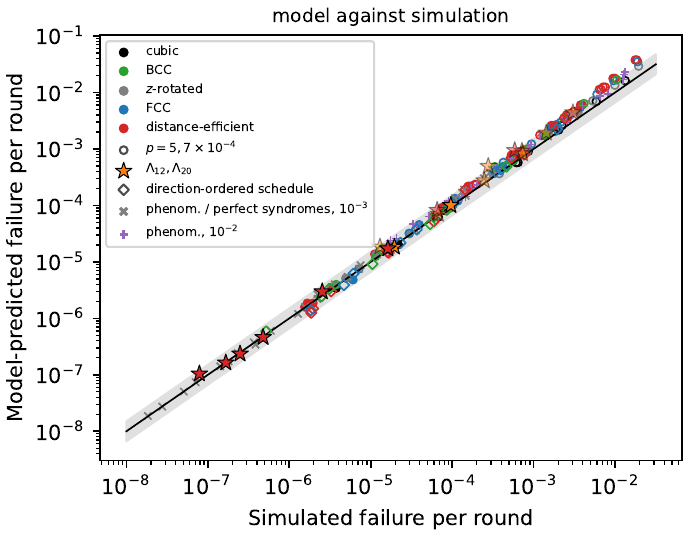}
\end{minipage}\hfill
\begin{minipage}[c]{0.48\textwidth}
\caption{\raggedright \textbf{Predicted (vertical) and simulated
(horizontal) failure per round.}
For the greedy vertex schedule, circles are filled at $p=2\times10^{-4}$
and $3\times10^{-4}$ and open at $5\times10^{-4}$ and $7\times10^{-4}$.
Stars mark additional lattices on that schedule: in red, cyclic
codes with $\dZ=9,11,13,14,15$; in orange, the
multiples by two and three of
$\Lambda_{12}=\langle(4,0,0),(1,3,0),(2,1,1)\rangle$ and
$\Lambda_{20}=\langle(4,0,0),(2,5,0),(0,3,1)\rangle$. Diamonds use the
direction-ordered schedule. Crosses mark phenomenological and
perfect-syndrome runs at $10^{-3}$. Plus signs mark phenomenological
runs at $10^{-2}$. The grey band shows agreement within a factor of $1.5$.
Appendices~\ref{app:comparison-details} and~\ref{app:forecast-error}
give comparison ranges and sampling limits.}
\label{fig:entropy}
\end{minipage}
\end{figure*}

\subsection{Noisy-check test of the matched pair}
\label{sec:matched-circuit}

Under perfect checks and maximum-likelihood decoding,
Proposition~\ref{prop:matched} gives the tilted layout a larger leading
block-failure coefficient than the cubic layout.
The six-round simulations at $d=6$ showed the same ordering at both
noise levels (Table~\ref{tab:matched-winding}). At $p=3\times10^{-4}$,
the tilted layout gave at least one wrong logical $X$ outcome $8.6$
times as often as the cubic layout. The loop model of
Sec.~\ref{sec:entropy} had predicted a ratio of $10.05$ before
sampling, inside the $95\%$ confidence interval $[6.97,10.65]$ of the
observed ratio (Appendix~\ref{app:matched-runs}).

On the tilted torus, every shortest logical $Z$ string winds along $b_1$
and flips only $\bar X_1$ (Corollary~\ref{cor:matched-observables}).
Most tilted failures in the simulations affected $\bar X_1$ alone.
When we rescored the lower-noise shots for $\bar X_2$ and $\bar X_3$
alone, the tilted layout had fewer failures. In the independent run at
$p=5\times10^{-4}$, with this pair chosen in advance, the tilted layout
had $7.1\%$ as many pair failures as the cubic layout.
At fixed qubit count and check schedule, tilting the torus lowered the
observed error rate of the $\bar X_2$ readout and raised that of the
$\bar X_1$ readout.

A search of the two circuits' detector error models gives the same
ordering of distances. Among faults that each trigger at most two
detectors, the fewest that flip a retained outcome without detection
number $(6,6,6)$ for the cubic layout and $(6,10,12)$ for the tilted
one, in the row order of Table~\ref{tab:matched-winding}. The tilted
values equal $(d,5d/3,2d)$ of Corollary~\ref{cor:matched-observables}
at $d=6$. The search omits faults that trigger four detectors, so
these numbers are upper bounds on the circuit-level fault distances
(Appendix~\ref{app:matched-runs}).

\begin{table}[tb]\centering
\caption{\raggedright \textbf{Failed shots in the matched-pair circuits.}
Each layout ran 500\,000 six-round shots at each physical error
rate $p$. Each row counts a shot once if at least
one listed logical $X$ outcome is wrong after decoding. The rows score
the same shots. For example, a shot with only $\bar X_3$ wrong counts
in the first two rows. Zero denotes no observed failures for that row.}
\label{tab:matched-winding}
\setlength{\tabcolsep}{6pt}
\begin{tabular}{lrr@{\qquad}rr}\toprule
 & \multicolumn{2}{c@{\qquad}}{$p=3\times10^{-4}$}
 & \multicolumn{2}{@{}c}{$p=5\times10^{-4}$}\\
Wrong outcome & Cubic & Tilted & Cubic & Tilted\\\midrule
$\bar X_1$ or $\bar X_2$ or $\bar X_3$ & 207 & 1778 & 1301 & 8708\\
$\bar X_2$ or $\bar X_3$ & 142 & 3 & 822 & 58\\
$\bar X_2$ & 77 & 0 & 410 & 11\\\bottomrule
\end{tabular}
\end{table}

\begin{figure*}[tp]\centering
\includegraphics[width=0.92\textwidth]{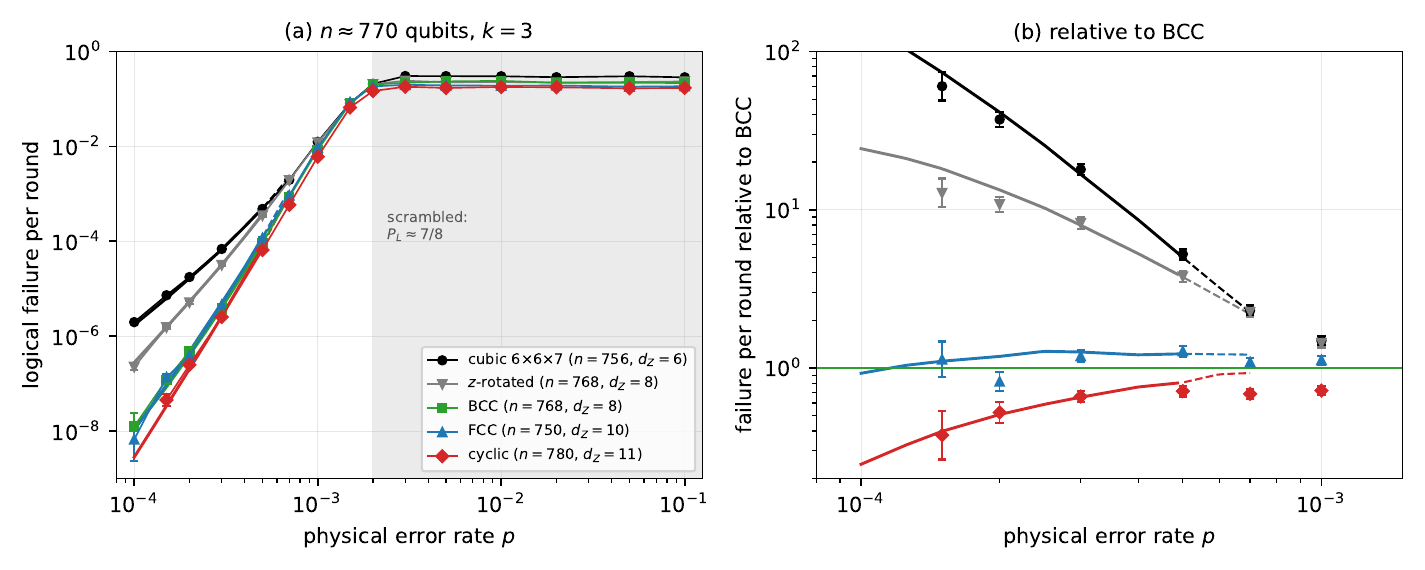}
\caption{\raggedright \textbf{Failure per round versus physical error
rate at fixed size.} The five codes have $k=3$ and $750$--$780$ data
qubits. Each uses $\dZ$ rounds of the greedy vertex schedule, circuit
noise and minimum-weight matching. (a) Points show runs with at least
one failure, stopped at $300$ failures or $3\times10^7$ shots. Lines
show Eq.~\eqref{eq:model}, solid through $p=5\times10^{-4}$ and dashed
up to $7\times10^{-4}$. Shading marks the random-readout regime.
(b) Rates relative to BCC where both codes have at least ten failures.
Error bars are $68\%$ Poisson intervals on failure counts.
Four points in (a) rest on fewer than $30$ failures and are
indicative: FCC and BCC at $p=10^{-4}$ ($2$ and $3$ failures), and the
cyclic code and BCC at $1.5\times10^{-4}$ ($15$ and $29$).}
\label{fig:fixedn}
\end{figure*}

\begin{table*}[tp]\centering\small\setlength{\tabcolsep}{7pt}
\begin{tabular}{lrrrrrr}
\toprule
$p$ & cubic & $z$-rot. & BCC & FCC & Mink. & checked min.\\\midrule
$2\times10^{-5}$ & 1029 & 768 & 768 & 384 & 912 & 336\\
$5\times10^{-5}$ & 1536 & 1500 & 768 & 750 & 912 & 567\\
$10^{-4}$ & 2187 & 2592 & 1500 & 1296 & 3078 & 1215\\
$1.5\times10^{-4}$ & 3000 & 4116 & 1500 & 2058 & 3078 & 1500\\
$2\times10^{-4}$ & 3993 & 6144 & 2592 & 3072 & 3078 & 2592\\
$3\times10^{-4}$ & 8232 & 8748 & 6144 & 4374 & 7296 & 4374\\
$5\times10^{-4}$ & 20577 & 26364 & 20736 & 16464 & 24624 & 16200\\
\bottomrule
\end{tabular}
\caption{\raggedright \textbf{Estimated data-qubit requirements for $10^{-9}$ logical failures per round.} The memory experiment has $R=\dZ$ rounds and direction-ordered vertex checks. For each family, we compare the tested integer scales and orientations and report the smallest code whose central loop-model prediction meets the target. ``Checked min.'' is the minimum over all evaluated reference and candidate codes. The full circuit may require a different number of data qubits to reach the target.}
\label{tab:entropy}
\end{table*}

\subsection{Prediction accuracy and remaining discrepancy}
\label{sec:compare}

Equation~\eqref{eq:model} uses each circuit's preparation, bulk and
readout layers without fitting to its failure rate. For the greedy
schedule and matching decoder at $p=(1\text{--}3)\times10^{-4}$,
the predictions fall within a factor of $1.5$ of the simulated
rates in all $57$ runs with $\dZ\ge5$ and at least $30$ failures
(Fig.~\ref{fig:entropy} and Appendix~\ref{app:comparison-details}).
These runs cover $44$ distinct code--noise settings on $24$ codes,
because $13$ settings were simulated twice, and only five of the $57$
runs lie below $p=2\times10^{-4}$.
Omitting the time and
diagonal edges underestimates failure at larger distances. Keeping
only two consecutive rounds still misses how the rate changes with
size. At higher noise the model can overestimate failure. Shared
errors among longer loops may contribute (Appendix~\ref{app:model-checks}).

With direction-ordered checks, the model increasingly underestimates
failure on larger staircase layouts. For the distance-18 Minkowski
code at $p=5\times10^{-4}$ it predicts $68\%$ of the simulated rate.
The cause of this drift remains open. We report point estimates in
Fig.~\ref{fig:entropy} because their strip sampling errors omit
correlations within a loop. The resource forecasts include those
correlations (Appendix~\ref{app:forecast-error}).

\subsection{Fixed qubit number}
\label{sec:fixedn}

At nearly equal size, the five layouts in Fig.~\ref{fig:fixedn} range
from a $6\times6\times7$ cubic box to a cyclic code with $\dZ=11$.
The cyclic code attains the minimum determinant at distance eleven
and has the lowest measured failure rate per round. At
$p=2\times10^{-4}$ to $10^{-3}$, it fails $0.5$ to $0.7$ times
as often per round as BCC. FCC gains two units of distance over BCC
with little change in failure rate. Thus the value of extra distance
depends on which strings the layout introduces.

Above $p\approx2\times10^{-3}$, all five readouts approach random
outcomes. The three logical bits then give $P_L\approx7/8$ per shot,
and the reported per-round rates differ because each code runs for
$\dZ$ rounds. Appendix~\ref{app:fixed-size-check} compares the model
with the low-noise data.

\subsection{Integer resource forecasts}
\label{sec:design}

Table~\ref{tab:entropy} gives the smallest evaluated integer codes
predicted to reach $10^{-9}$ failures per round with $R=\dZ$, using
the equal-error resource criterion of
Refs.~\cite{WatsonBarrett2014,ORourkeDevitt2025,TricycleJMB2025}.
It counts data qubits. Including one ancilla per check multiplies
each cost by $7/3$.
A torus with $V$ vertices has $3V$ edge qubits, $V$ vertex checks
and $3V$ plaquette checks, giving $7V$ qubits in total. Thus, when we consider the syndrome extraction, the total qubit count increases by a factor of $7/3$.

Integer scales can reverse a resource comparison. At $p=2\times10^{-5}$
the smallest Minkowski code misses the model target, and its next scale
uses $912$ data qubits. For a different period lattice, the central
model predicts that $336$ data qubits suffice, with a predicted
rate of $9.97\times10^{-10}$ that lies below the target by less than
its sampling error of $0.35\times10^{-10}$; over the tested
sensitivity scenarios the smallest qualifying code has $288$ to $384$
data qubits. The search is finite
(Appendix~\ref{app:resource-sensitivity}), and the size-dependent model
error in Sec.~\ref{sec:compare} limits this extrapolation.

\begin{figure*}[tp]\centering
\includegraphics[width=0.88\textwidth]{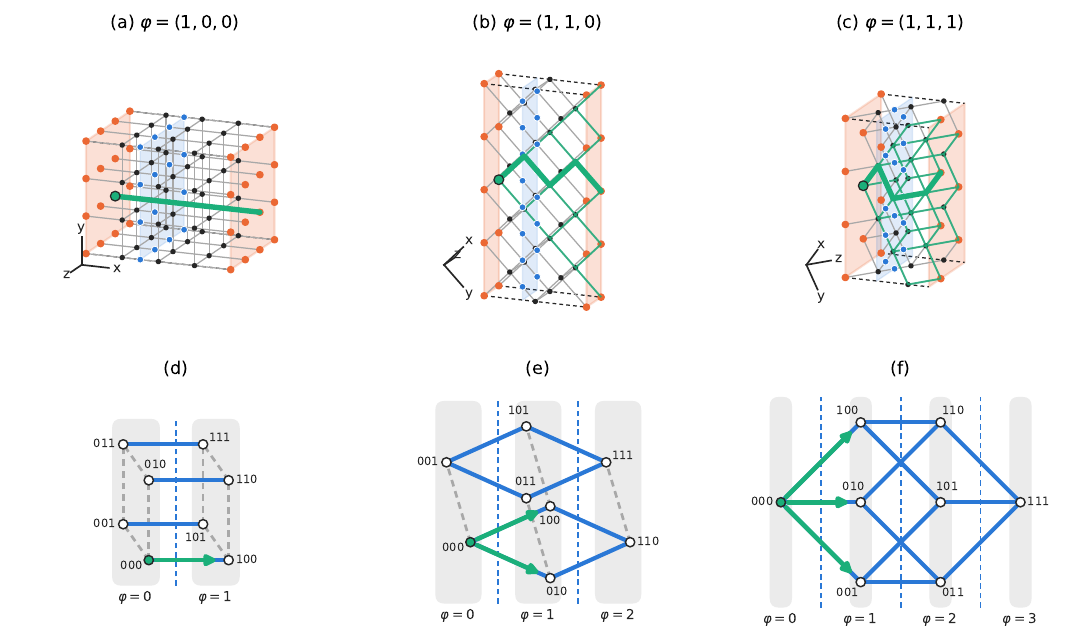}
\caption{\raggedright \textbf{Geometry of the three boundary orientations.}
(a--c) At $(\dZ,\dX)=(4,16)$ the axis-aligned, $(1,1,0)$ and $(1,1,1)$
volumes contain $136$, $77$ and $73$ data qubits (grey edges).
The normal $\varphi$ points right. Rough faces (orange) lie at
$\varphi(x)=0,\dZ$, and dashed sides are smooth. Green paths show all
shortest strings from the marked vertex ($1$, $16$ and $52$,
respectively), one in bold per panel.
Blue dots mark the $16$ qubits of a minimum membrane at $\varphi(x)=3/2$.
(d--f) Unit-cube vertices $xyz$ are grouped by level. Blue edges cross
a half-integer membrane plane (dashed). Dashed grey edges stay within a level.
Green arrows mark the one, two or three possible first steps from
$000$ along a shortest string. The crossing fractions $1/3$, $2/3$ and
$1$ give the large-size limits of $\dZ w/n$ for a flat membrane of
weight $w$. Because $\dX\le w$, the ratio $\dZ\dX/n$ cannot exceed them
in that limit; at finite size it can
(Sec.~\ref{sec:patch-orientations}).}
\label{fig:patchgeom}
\end{figure*}

These forecasts lie far below any failure rate that we can sample. At
measurable rates, we compared forecasts fixed before the
confirmation samples were drawn with simulations of 18 layouts, each run
for $R=6$ rounds of the direction-ordered schedule at $p=10^{-4}$ and
$2\times10^{-4}$ (Appendix~\ref{app:frozen-forecasts}). In the $23$ of
the $36$ layout--noise cells with at least $30$ observed failures, all
with $\dZ\le8$, the frozen forecast was $0.85$ to $1.37$ times the
observed shot failure probability, with median $1.05$.
We computed this summary after sampling. The question registered in
advance was whether the model picks the cheapest layout that meets a
target, and the data did not settle it; the remaining $13$ cells had too
few failures for a ratio (Appendix~\ref{app:frozen-forecasts}).

\begin{figure*}[tp]\centering
\includegraphics[width=\textwidth]{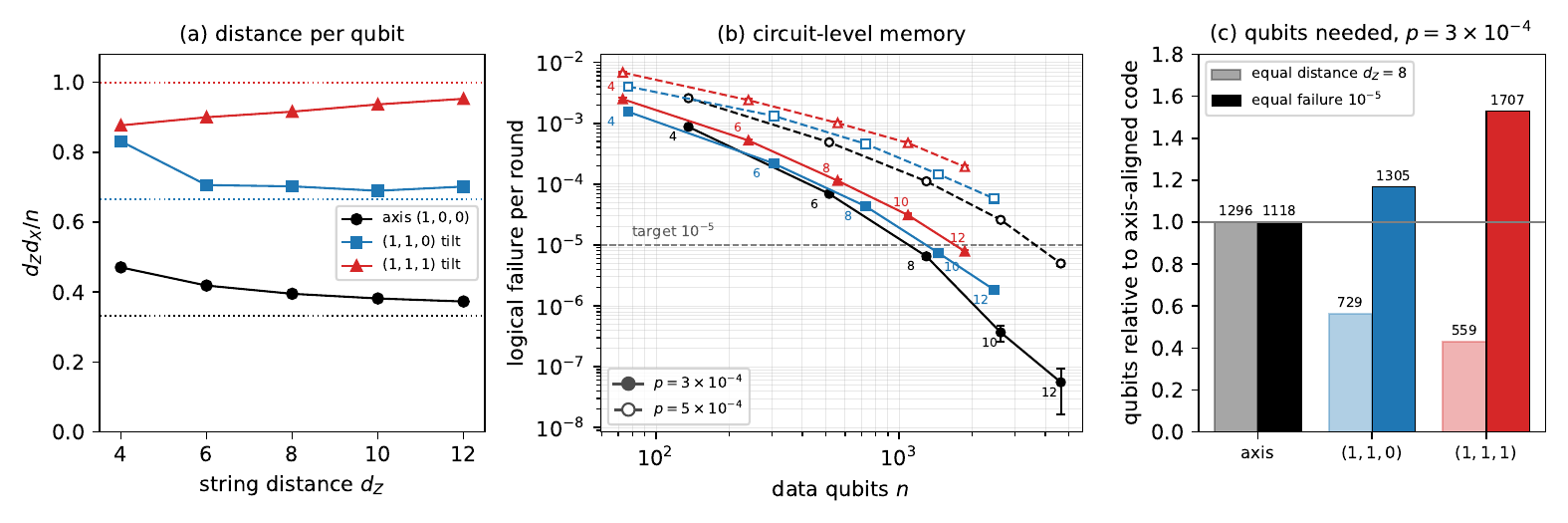}
\caption{\raggedright \textbf{Distance and memory cost of tilted volumes.}
Each code encodes one logical qubit in $n$ data qubits, with $\dX\ge\dZ^2$.
(a) $\dZ\dX/n$ at $\dZ=4,6,8,10,12$. The dotted lines at
$1/3$, $2/3$ and $1$ are the large-size values for flat membranes
in the axis-aligned, $(1,1,0)$ and $(1,1,1)$
volumes, respectively.
(b) Circuit-level string-sector failure per round at $p=3\times10^{-4}$
(filled) and $5\times10^{-4}$ (open). Runs use $\dZ$ rounds (point
labels) and minimum-weight matching, targeting $300$ failures in at
most $3\times10^6$ shots. Bars show approximate binomial standard errors.
At the lower $p$, the $\dZ=12$ runs reached the cap with $2$ failures
for the axis-aligned code, $66$ for $(1,1,0)$ and $283$ for $(1,1,1)$.
The axis-aligned estimates at $\dZ=10$ ($11$ failures) and $12$ have
large relative uncertainties.
(c) Relative to the axis-aligned code, both tilts save qubits at
$\dZ=8$ (light bars) and cost more at $10^{-5}$ failures per round
with $p=3\times10^{-4}$ (dark bars, interpolated before rounding up
to an available code size).}
\label{fig:patches}
\end{figure*}

\section{Open boundaries: tilted volumes}
\label{sec:patches}

Boundary orientation provides another way to change distance per data
qubit. Consider a three-dimensional surface code on a finite, simply
connected volume of the cubic lattice. With two rough boundary
components, the code encodes one logical qubit~\cite{ZouLo2025}. Strings join
the rough components, and membranes end on the smooth remainder.
The code is defined relative to the rough boundary: its vertices
carry no $X$ checks, and its internal edges carry no
qubits~\cite{3DToric_code2,ZouLo2025}.

\subsection{The length--width bound}

For these volumes, the counterpart of the periodic packing bound is
the classical network length--width inequality~\cite[Theorem~3]{MooreShannon1956}.

\begin{proposition}[Length--width bound]\label{prop:menger}
For a surface code on a simply connected volume of the cubic lattice
with two rough boundary components, $\dZ\dX\le n$.
\end{proposition}

\begin{proof}
A membrane in the nontrivial class is the coboundary of a set of vertices containing one rough component and not the other, that is, the set of edges with exactly one endpoint in it, an edge cut separating the two components. By Menger's theorem, the edge form of max-flow/min-cut, the minimum cut $\dX$ equals the maximum number of edge-disjoint strings joining the two components. Each string has at least $\dZ$ edges, so $\dZ\dX\le n$. This is the length--width inequality of Moore and Shannon~\cite{MooreShannon1956}; in two dimensions the rotated surface code attains it.
\end{proof}

\subsection{Three boundary orientations}
\label{sec:patch-orientations}

The rough boundary levels of the three families in Fig.~\ref{fig:patchgeom}
have integer normals $\varphi=(1,0,0),(1,1,0)$ and $(1,1,1)$, respectively.
Writing $\varphi(x)=\varphi\cdot x$, the data-qubit region lies
between levels $0$ and $\dZ$. The underlying cubical complex also
includes the rough levels $\varphi(x)\in\{-1,0\}$ and
$\{\dZ,\dZ+1\}$; two levels make each rough component connected
before passing to the relative code.
The transverse coordinate pairs are $(x_2,x_3)$,
$(x_1-x_2,x_3)$ and $(x_1-x_2,x_1+x_2-2x_3)$, respectively.
For each family and each $\dZ$, a scan over the two transverse
widths selects, among the scanned shapes, the volume with the fewest
data qubits whose membrane
distance, computed exactly as a minimum edge cut, satisfies
$\dX\ge\dZ^2$; Appendix~\ref{app:volume-shapes} gives the scanned
shapes, the fifteen selected volumes and examples of more elongated
shapes that need fewer qubits. The axis-aligned code is
standard, and the $(1,1,0)$
layout is the rotation of Ref.~\cite[Sec.~V\,A]{Huang2023}.

A shortest string advances one level per edge, and a flat membrane
cuts the edges crossing one half-integer level. Tilting puts more of
the bulk edges across these levels (Fig.~\ref{fig:patchgeom}).
In the bulk, an edge along axis $i$ joins two levels that differ by
$\varphi_i\in\{0,1\}$, so the edges along the $\|\varphi\|_1$ axes with
$\varphi_i=1$ each cross exactly one of the $\dZ$ half-integer levels
and the other edges cross none. To leading order, a fraction
$\|\varphi\|_1/3$ of the edges is thus shared equally among $\dZ$ flat
membranes, and a flat membrane of weight $w$ has
$\dZ w/n\to\|\varphi\|_1/3$ when $\dZ$ and both transverse widths grow,
that is, $1/3$, $2/3$ and $1$ for the three families. Since $\dX\le w$
in every volume, $\dZ\dX/n$ cannot exceed these values in that limit.
They are not bounds at finite size. The rough levels carry no qubits
within a level, so a finite volume has relatively more crossing edges
than the bulk: the selected axis-aligned volumes have
$n=\dZ^3+2\dZ(\dZ-1)^2$ and $\dZ\dX/n$ falls from $0.47$ at $\dZ=4$ to
$0.37$ at $\dZ=12$, and the $(1,1,0)$ volumes have $0.83$ at $\dZ=4$
and $0.69$ to $0.71$ at $\dZ=6$ to $12$. Only the value $1$ for
$(1,1,1)$ holds at every size, as the length--width bound. We have not
proved that the $(1,1,1)$ family attains that bound asymptotically. In
the fifteen selected volumes, the flat membrane is a minimum cut for the
axis-aligned and $(1,1,0)$ families, whereas the $(1,1,1)$ volumes with
$\dZ\ge6$ have minimum cuts of weight $36,64,102,148$, lighter than
their lightest flat membranes of weight $37,65,104,151$. For the
$(1,1,1)$ family $\dZ\dX/n$ rises from $0.88$ at $\dZ=4$ to $0.95$ at
$\dZ=12$.
Already at $\dZ=8$ the $(1,1,1)$ volume uses only $43\%$ as many data qubits as the
axis-aligned code at the same membrane distance (Fig.~\ref{fig:patches}(a,c)).

\subsection{Circuit-memory cost}

In the axis-aligned code, each shortest string goes straight. Tilting gives
two or three choices per bulk step, so the number of shortest
staircases grows roughly as $2^{\dZ}$ or $3^{\dZ}$
(Fig.~\ref{fig:patchgeom}(d--f)). We tested the resulting memory cost
by measuring one logical $X$ outcome after $\dZ$ rounds with the
circuit noise and matching decoder of Sec.~\ref{sec:sim}.

At $p=3\times10^{-4}$, the axis-aligned code has the lowest estimated
data-qubit cost at $10^{-5}$ failures per round. The $(1,1,1)$ code
needs about $1.5$ times as many data qubits (Fig.~\ref{fig:patches}(c)).
We interpolate $\log\epsilon$ linearly in $n^{1/3}$ between the two
simulated sizes that bracket the target. Their string distances are
$6$ and $8$ for the axis-aligned code, $8$ and $10$ for $(1,1,0)$,
and $10$ and $12$ for $(1,1,1)$.

At tested distances $4,6$, Stim's graphlike
search~\cite{Stim2021} finds a logical fault set of weight $\dZ$
for each family. This supplies an upper bound on the full
circuit-level distance, since the search omits faults that flip
more than two detectors.

\begin{figure}[tbp]\centering
\includegraphics[width=\columnwidth]{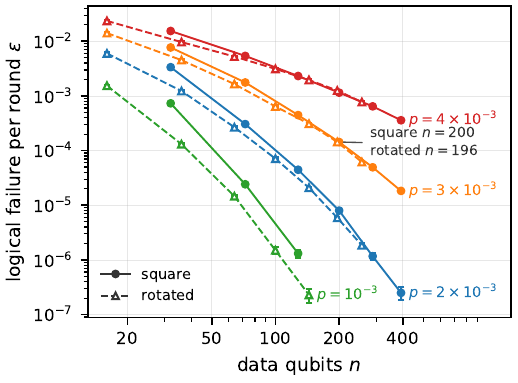}
\caption{\raggedright \textbf{Circuit memory on two-dimensional tori.}
Filled circles and solid lines show square tori. Open triangles and
dashed lines show $45^\circ$ rotated tori. For a fixed physical error
rate $p$ (colour), read horizontally at the desired logical failure
probability per round $\epsilon$. The curve farther left uses fewer
data qubits. Points show runs with at least $10$ failures, and bars
give approximate binomial standard errors.}
\label{fig:2d}
\end{figure}

\subsection{Comparison with two-dimensional codes}
\label{sec:2d-comparison}

The packing advantage weakens when translated into distance at
fixed size. Within a uniformly scaled family, write
$n=C\dZ^D$, where $C$ depends on its shape. Improving the packing
by a factor $\gamma$, so that $C$ is divided by $\gamma$, increases
$\dZ$ at fixed $n$ by only $\gamma^{1/D}$; within each such family
$\dX\propto\dZ^{D-1}$. Staircase multiplicity meanwhile grows
exponentially in string length. This competition occurs already
in two dimensions~\cite[Secs.~III--V]{Beverland2019}, and helps
explain why a geometric improvement need not translate into the
same memory saving (see Fig.~\ref{fig:patches}(c) for the
three-dimensional volumes).

At string distance $d=10$, the rotated torus uses half as many data
qubits as the square torus ($100$ versus $200$). At physical error
rate $p=3\times10^{-3}$, the square torus with $d=10$ and the rotated
torus with $d=14$ both give about $1.4\times10^{-4}$ logical failures
per round (Fig.~\ref{fig:2d}). The rotated torus then uses $196$ data
qubits, a $2\%$ saving.

We interpolate $\log\epsilon$ linearly in $\sqrt n$ for each layout
to estimate its qubit count at a target $\epsilon$. Before rounding
to available sizes, the estimated savings are $9\%$ and $2\%$ at
$p=2\times10^{-3}$ for $\epsilon=10^{-5}$ and $10^{-6}$, and $3\%$ and
$11\%$ at $p=3\times10^{-3}$ for $\epsilon=10^{-4}$ and $10^{-5}$. The
second extrapolates beyond the largest rotated torus, and the fourth
beyond the largest torus of both layouts. At $p=10^{-3}$ rotation saves about $20\%$ at
$\epsilon=10^{-5}$. At $p=4\times10^{-3}$ the two estimates differ
by $2\%$ in favour of the square torus at $\epsilon=10^{-3}$. These
estimates carry no error bars. The smallest saving uses a
square-torus point with $14$ failures, and we do not regard differences
of a few per cent as resolved.

For planar surface codes, O'Rourke and Devitt find savings of about a
quarter at $10^{-12}$ failures per $\dZ$ rounds and
$p=10^{-3}$~\cite{ORourkeDevitt2025}.
At that target our planar circuit extrapolations give a rotated-to-square
ratio of $68$--$72\%$ in total qubits, including check ancillas, for
$p=(1\text{--}4)\times10^{-3}$. These extrapolations are rough. The
runs use Stim's generated rotated and unrotated memory circuits, whose
noise model differs from that of Sec.~\ref{sec:circuit}, with $\dZ$
rounds and matching, and their records have no seeds. For each layout
we fit a line to the logarithm of the shot failure probability against
distance, using distances of at least six and points with at least $30$
failures, and extend it to the target. The fitted points reach no lower
than $6\times10^{-7}$, so the extension covers more than five decades,
and we did not estimate its uncertainty. A boundary-to-boundary walk
model with two parameters fitted to
the square circuits predicts savings of $13.4\%$ at $p=10^{-3}$ and
$9.5\%$ at $2\times10^{-3}$. It misses the difference in circuit path
multiplicities.

\section{Discussion}
\label{sec:discussion}

The classical lattice optimum at string distance
$6m$~\cite[Eqs.~(1.4)--(1.5) and Sec.~4.2]{CoolsLyness2001}
gives the minimum data-qubit count $114m^3$ for a periodic cubic-lattice
toric code. Our membrane certificate gives $\dX=19m^2$ on the same
family (Theorem~\ref{thm:minkowski}). With perfect
checks and maximum-likelihood decoding, two layouts can have the same
qubit count and both distances yet have leading failure coefficients
whose ratio grows at least exponentially with distance
(Proposition~\ref{prop:matched}).
Both results hold independently of the loop model and the distance
conjectures.

Which layout to use depends on the intended measurement. Many shortest
strings on the tilted torus flip the same observable. A longer winding
is needed to disturb either of the other two. This explains why the
cubic layout can better protect the full readout while the tilted one
better protects the retained pair. The noisy circuits show this reversal
at distance six. Its scaling and dependence on the schedule remain open.

Because distance does not fix the memory cost, we developed a loop
model that estimates the string-sector failure rate of a noisy memory
circuit from the circuit itself (Sec.~\ref{sec:entropy}). It sums over
the winding loops of the torus, counted by non-backtracking walks, and
removes the failing sets shared by loops of one winding through a count
that is exact under the condition of Proposition~\ref{prop:shared}.
Its edge probabilities come from the circuit's detector error model.
For each loop it decodes the whole space-time strip with the matching
decoder, so it includes failures whose flips are spread over several
rounds. No parameter is fitted to logical-error simulations. For the
greedy schedule and matching decoder at
$p=(1\text{--}3)\times10^{-4}$, its predictions lie within a factor
of $1.5$ of the simulated rates in $57$ runs, which cover $44$
code--noise settings on $24$ codes with $5\le\dZ\le15$
(Table~\ref{tab:model-ranges}). For
18 layouts from the same families, forecasts fixed before
the confirmation sampling stayed
within a factor of $1.4$ wherever at least $30$ failures were
observed, all at $\dZ\le8$; this summary was made after sampling,
and the selection question registered in advance was not settled
(Appendix~\ref{app:frozen-forecasts}). The
model also accounts for an effect of parity. Under phenomenological
noise, the distance-efficient codes of odd distance fail less often
than their even neighbours. Under circuit noise, the odd member fails
more often in three of the five pairs tested, because diagonal edges
let the lighter wrong correction on an odd loop pay for a change of
round (Sec.~\ref{sec:rounds}). For the pair $13/14$ the model gives an
odd-to-even ratio of $2.82$, against $2.87$ observed
(Appendix~\ref{app:rounds-checks}).

The model needs no sampled logical failures, so it can forecast rates
far below those that simulation reaches. At $10^{-9}$ failures
per round (Table~\ref{tab:entropy}), the cubic code needs more data
qubits than BCC at every $p\le3\times10^{-4}$, by $1.3$ to $2.0$ times
in the central values and $1.1$ to $2.7$ times over the tested
sensitivity scenarios; FCC needs fewer than BCC at the three lowest of
the seven noise levels and more at the next two, while the order at the
two highest is not resolved (Sec.~\ref{sec:design}); and integer
scales can reverse a comparison. These numbers are extrapolations.
Under the direction-ordered schedule the model increasingly
underestimates failure on larger staircase layouts, down to $68\%$ of
the simulated rate for the distance-18 Minkowski code at
$p=5\times10^{-4}$ (Sec.~\ref{sec:compare}). Errors shared across
windings, correlated faults and strip-by-strip decoding are further
approximations whose effect at large distance remains open.

We also studied codes with boundaries (Sec.~\ref{sec:patches}). A
three-dimensional surface code on a simply connected volume with two
rough boundary components encodes one logical qubit and obeys the
length--width bound $\dZ\dX\le n$ (Proposition~\ref{prop:menger}), the
open-boundary counterpart of the packing bound. Tilting the rough
boundaries to the normals $(1,1,0)$ and $(1,1,1)$ puts more edges
across each membrane level. The $(1,1,1)$ volumes that we built
reach $\dZ\dX/n=0.88$ to $0.95$ for $\dZ=4$ to $12$, against the bound
of one; we have not proved that the family attains the bound
asymptotically. At $\dZ=8$ the $(1,1,1)$ volume uses $43\%$ as many data qubits as
the axis-aligned code at the same membrane distance. The memory
circuits nevertheless favour the axis-aligned code. At
$p=3\times10^{-4}$ and $10^{-5}$ failures per round, interpolation
between simulated sizes gives both tilted volumes a higher data-qubit
cost, about $1.5$ times for $(1,1,1)$. Tilting multiplies the shortest
strings, roughly as $2^{\dZ}$ or $3^{\dZ}$, which is the same
competition between distance and multiplicity as on the torus. These
open-boundary results come from simulation; we have not yet applied
the loop model to them.

Three copies of a three-dimensional toric code support a
constant-depth CCZ circuit built from the cup product: Breuckmann
\emph{et al.}\ give it for the cubic
torus~\cite[Sec.~6.4.3]{CupsGates2026}, and Jacob \emph{et al.}\ give a
depth-two circuit for the twisted $114$-qubit
code~\cite[Sec.~V\,C]{TricycleJMB2025}. The cubical cup product is
defined on every twisted torus. For ten period lattices, among them
$\Lambda_{38}$, the cubic, BCC, FCC and $z$-rotated lattices and two
arbitrary ones, we checked numerically that the resulting circuit has
depth two and gives the expected phase on all $512$ triples of logical
classes; we give no proof for a general period lattice.
For logical classes $a,b,c$ the
expected phase is that of the triple cup product on the
three-torus, $(-1)^{\det[a,b,c]}$
in any period basis over $\F_2$, which extends by trilinearity the
coordinate-basis product of Ref.~\cite[Eqs.~(64)--(65)]{Zhu2025}.
Our resource counts cover string-sector memory with direct torus
couplings. An operational CCZ estimate would also require routing costs
and the correlated errors induced by the gate.

\begin{acknowledgments}
X.\,Z.\ is supported by the Croucher Foundation. The author thanks Prof. Hoi-Kwong Lo from the University of Toronto and The National University of Singapore for helpful discussions.
\end{acknowledgments}

\section*{AI Use Disclosure}
This work was carried out with the help of generative AI tools: Claude
(Anthropic; mainly Claude Opus~5.5, with Claude Fable~5.2 and Claude
Haiku~5.5 in some sessions), run through Claude Code. Under the
author's direction, the tools assisted with the literature search, with
drafting proofs and derivations, with writing and running the
simulation and analysis code, and with drafting and revising the text.
The author chose the research questions and the claims to include, and
reviewed the output. Numerical results were recomputed from saved
scripts and data, except for two exploratory checks that
Appendix~\ref{app:model-checks} marks as such, and proofs were checked by exact computation on
finite cases and in separate AI audit sessions. The author retains full
responsibility for the content and the scientific claims made in this
manuscript.

\section*{Data and code availability}
Code, numerical records and exact certificates are available from the corresponding author upon request.

\appendix
\section{Proofs}
\label{app:proofs}

The proofs follow the order of the main text.

\subsection{\texorpdfstring{Proposition~\ref{prop:dist}: distance bounds}{Distance bounds}}
\label{app:proof-dist}

\begin{proof}
For part~(a), note that a monotone path from $0$ to $v\in\Lambda$ closes into a loop of length $\|v\|_1$. When $v$ is a shortest vector, this loop is simple and $v\notin2\Lambda$, so the loop is a nontrivial cycle; the proof of Proposition~\ref{prop:nmin} in Appendix~\ref{app:proof-nmin} gives both facts by arguments that hold in any dimension. Hence $\dZ\le\lambda(\Lambda)$. Conversely, a nontrivial $\Z_2$-cycle splits into closed walks whose classes add up to its class. One of them therefore lifts to a path from $x$ to $x+v$ with $v\ne0$, of length at least $\|v\|_1\ge\lambda(\Lambda)$.

For part~(b), let $\Phi(x)=\lfloor\varphi\cdot x+s\rfloor$ on $\Z^D$, for a real offset $s$. Then $\Phi(x+v)=\Phi(x)+\varphi(v)$ for $v\in\Lambda$, so its coboundary $\delta\Phi$, the function $\Phi(y)-\Phi(x)$ on each edge from $x$ to $y$, descends to an integer cocycle of class $\varphi$ on $T_\Lambda$. Take $\varphi$ of least $\ell_1$ norm in its class, that is, among the vectors of $\varphi+2\Lambda^*$. Then $|\varphi_i|\le1$ for every $i$: since $e_i\in\Z^D\subseteq\Lambda^*$, the vector $2e_i$ lies in $2\Lambda^*$, and subtracting it when $\varphi_i>1$, or adding it when $\varphi_i<-1$, keeps the class and shortens $\varphi$. Every edge therefore carries $0$ or $\pm1$.

We now count the edges of direction $i$ that carry $\pm1$. Because $\varphi(\Lambda)\subseteq\Z$, the map $x\mapsto\varphi\cdot x\bmod1$ is a group homomorphism from the $V$ vertices of $\Z^D/\Lambda$ to $\R/\Z$. Its image is a finite cyclic group; write $\operatorname{ord}\varphi$ for its order. The image consists of the multiples of $1/\operatorname{ord}\varphi$, and each of them is the value at exactly $V/\operatorname{ord}\varphi$ vertices, because the fibres of a homomorphism have equal size. The coordinate $\varphi_i=\varphi\cdot e_i$ is itself such a multiple. Suppose $\varphi_i\ge0$. The edge from $x$ to $x+e_i$ carries $\lfloor\varphi\cdot x+s+\varphi_i\rfloor-\lfloor\varphi\cdot x+s\rfloor$, which is $1$ when the fractional part of $\varphi\cdot x+s$ lies in the half-open interval $[1-\varphi_i,1)$ and $0$ otherwise. As $x$ runs over the vertices, this fractional part takes $\operatorname{ord}\varphi$ equally spaced values, and an interval of length $\varphi_i$ contains exactly $\varphi_i\operatorname{ord}\varphi$ of them, whatever the offset $s$. Hence $V\varphi_i$ edges of direction $i$ carry $1$. For $\varphi_i<0$ the same argument gives $V|\varphi_i|$ edges that carry $-1$. For example, $\Lambda=4\Z$ in one dimension and $\varphi=1/4$ give $\operatorname{ord}\varphi=4$, and exactly one of the four edges carries $1$.

Reducing mod 2 gives a membrane of weight at most $V\|\varphi\|_1$. A shortest $\varphi$ is primitive, so $\varphi\notin2\Lambda^*$ and its class is nonzero.

For part~(c), take, for every vertex $x$ and every $v$ in the support of $\nu$, all monotone lattice paths from $x$ to $x+v$. Give each path weight $\nu(v)/(N_v t_\nu)$, where $N_v$ is the number of orderings of $v$. The load of an edge is the total weight of the paths through it, counted once per traversal. By translation invariance, every direction-$i$ edge then carries load $\mathbb E_\nu|v_i|/t_\nu\le1$, and the total weight is $V/t_\nu$. Every path closes into a closed walk of class $v$, and $\alpha(v)=1$, so it traverses the edges of every membrane of class $\alpha$ an odd number of times in total, hence at least once. Summing the loads over the edges of the membrane gives the bound.
\end{proof}

Every nonzero membrane class pairs oddly with some basis vector $v$.
A point mass on $v$ in Proposition~\ref{prop:dist}(c) gives a lower
bound of $V/\|v\|_\infty$ on the weight of every membrane in that class.
For the $z$-rotated lattice in Table~\ref{tab:lattices}, $V=d^3/2$
and the displayed basis vectors have $\|v\|_\infty\le d$.
For FCC, $V=d^3/4$ and the face-diagonal basis vectors
$\tfrac d2(e_i+e_j)$, $1\le i<j\le3$, have $\|v\|_\infty=d/2$.
Thus $\dX\ge d^2/2$ for both families. The dual covectors $e_3/d$
and $2e_1/d$, respectively, give the matching upper bound by
Proposition~\ref{prop:dist}(b).

\subsection{\texorpdfstring{Theorem~\ref{thm:minkowski}: exact distances of Minkowski codes}{Exact distances of Minkowski codes}}
\label{app:proof-minkowski}

\begin{proof}
Write $b_1=(1,-1,4)$, $b_2=(2,2,2)$ and $b_3=(3,-2,1)$
for the basis in the theorem. Each $b_i$ satisfies
$x_1+7x_2+11x_3\equiv0\bmod38$, and their determinant has absolute
value $38$, the index of this congruence, so they generate
$\Lambda_{38}$. Throughout the proof a membrane class $\alpha$ is
labelled by the string $\alpha(b_1)\alpha(b_2)\alpha(b_3)$ of its
values on this basis, which is also the basis of the logical labels in
Sec.~\ref{sec:extremal}. A second basis connects $\Lambda_{38}$ with
Minkowski's packing. The packing basis of Betke and
Henk~\cite[Sec.~5]{BetkeHenk2000}, multiplied by three and rotated by
$(x,y,z)\mapsto(y,-x,z)$, is $(3,-2,1),(-2,1,3),(1,3,-2)$. It
satisfies the same congruence and has determinant $38$, so it
generates the same lattice.

Lattice rules supply the string distance. In the notation of Cools
and Lyness, the dual lattice of a rule is our period lattice: its
determinant is the number of quadrature points, and its shortest
$\ell_1$ length is the enhanced degree~\cite[Theorem~2.1 and
Eq.~(2.2)]{CoolsLyness2001}. Their rule with $38$ points and enhanced
degree six has a dual lattice with Hermite normal form rows $(1,0,7)$,
$(0,1,11)$, $(0,0,38)$~\cite[Table~3]{CoolsLyness2001}, that is, the
lattice $y_3\equiv7y_1+11y_2\bmod38$. The signed coordinate permutation
$(x_1,x_2,x_3)\mapsto(y_1,y_2,y_3)=(x_2,x_3,-x_1)$ takes $\Lambda_{38}$
onto it, because $x_1+7x_2+11x_3\equiv0$ reads $-y_3+7y_1+11y_2\equiv0$.
The distance can also be checked directly:
none of the $230$ nonzero integer vectors with $\|x\|_1\le5$ satisfies
the congruence, and $b_2$ has length six, so $\lambda(\Lambda_{38})=6$.
Scaling the periods by $m$ gives
$\dZ=6m$ and $n=3\cdot38m^3$ at equality in Eq.~\eqref{eq:3d}; in
lattice-rule language this is the $m$-copy of the
rule~\cite[Sec.~4.2]{CoolsLyness2001}.

For the upper bound on $\dX$, the dual vector $\varphi=(1,7,11)/38$ has $\|\varphi\|_1=\frac12$, so $\dX\le38m^3\cdot\frac{1}{2m}=19m^2$.

For the lower bound on $\dX$, we give, for each class $\alpha$, a distribution on vectors $v\in\Lambda_{38}$ with $\alpha(v)=1$, written in standard coordinates, that has $\mathbb E|v_i|=2$ for $i=1,2,3$:
\begin{itemize}
\item the four classes with $\alpha(b_2)=1$, namely $010$, $011$, $110$ and $111$: the point mass on $b_2=(2,2,2)$;
\item class $100$: $\tfrac14(-2,1,3)+\tfrac14(4,1,-1)+\tfrac12(-1,-3,2)$;
\item class $001$: $\tfrac12(-2,1,3)+\tfrac14(-1,4,1)+\tfrac14(3,-2,1)$;
\item class $101$: $\tfrac13(1,-1,4)+\tfrac16(-1,4,1)+\tfrac12(3,-2,1)$.
\end{itemize}
For example, the three vectors of class $100$ are $b_1-b_3$, $-b_1+b_2+b_3$ and $b_1-b_2$. Each has an odd coefficient of $b_1$, so $\alpha(v)=1$, and the first coordinate has mean $\tfrac14\cdot2+\tfrac14\cdot4+\tfrac12\cdot1=2$.
Scaling by $m$ gives $t_\nu=2m$, and Proposition~\ref{prop:dist}(c) gives $\dX\ge38m^3/(2m)=19m^2$.
\end{proof}

The proof uses no computer search beyond the check of the $230$
vectors. As a separate numerical check at $m=1$, mixed-integer
programming gave every per-class distance: strings $6$ in every class, and membranes $19$ for the classes $010$, $110$ and $111$, $20$ for $001$, $100$ and $101$, and $25$ for $011$.

\subsection{\texorpdfstring{Proposition~\ref{prop:bcc}: BCC and the classical lattice classification}{BCC and the classical lattice classification}}
\label{app:proof-bcc}

Proposition~\ref{prop:bcc} interprets Minkowski's classification of
free lattice cross-polytopes~\cite[Lemma~2.1]{BetkeHenk2000}.
Its two types give $\Z^3$ and $\Z^3+\Z(1,1,1)/2$ after normalization
by the string distance. The direct proof also supplies the membrane
certificate and string count.

\begin{proof}
We first prove the determinant bound. The hypothesis puts $d\Z^3$ inside $\Lambda$. Let $h\ne0$ be an element of the quotient group $\Lambda/d\Z^3$, and let $x\in(-\frac d2,\frac d2]^3$ be its representative. Then $x$ is a nonzero vector of $\Lambda$, and it is not $\pm de_i$, because its coordinates have magnitude at most $d/2$. The hypothesis therefore gives $\|x\|_1>d$. Order the coordinates so that $|x_1|\ge|x_2|\ge|x_3|$. Since $|x_1|\le d/2$, the other two magnitudes sum to more than $d/2$, so $|x_2|>d/4$ and hence also $|x_1|>d/4$.

Now consider $2h$. Its representative in the same cube has $\ell_1$ norm $\sum_if(|x_i|)$, where $f(y)=2y$ for $y\le d/4$ and $f(y)=d-2y$ for $d/4<y\le d/2$: doubling a coordinate of magnitude above $d/4$ takes it out of the cube, and subtracting $d$ brings it back. If $|x_3|\le d/4$, this norm is $2d-2(|x_1|+|x_2|-|x_3|)$. It is below $d$, because $|x_1|+|x_2|>d-|x_3|$ and $|x_3|\le d/4$ give $|x_1|+|x_2|-|x_3|>d/2$. Otherwise the norm is $3d-2\|x\|_1<d$. In either case the representative of $2h$ is a lattice vector shorter than $d$, and the hypothesis leaves only the zero vector. So $2h=0$, every coordinate of $x$ is $0$ or $d/2$, and $\|x\|_1>d$ forces $x=\frac d2(1,1,1)$. The quotient therefore has at most one nonzero element: $[\Lambda:d\Z^3]\le2$, i.e.\ $\det\Lambda\ge d^3/2$, with equality only for BCC. Conversely $\Lambda_{\rm BCC}$ satisfies the hypothesis: a vector outside $d\Z^3$ has three coordinates that are odd multiples of $d/2$, so its $\ell_1$ length is at least $3d/2$, and the only vectors of $d\Z^3$ of length at most $d$ are $0$ and $\pm de_i$.

It remains to find the distances and string count of BCC. Its dual $\Lambda^*$ is $\frac1d$ times the face-centred cubic lattice $\{x\in\Z^3:\sum_ix_i\ \text{even}\}$. Hence $\lambda(\Lambda^*)=2/d$, and Proposition~\ref{prop:dist}(b) gives $\dX\le d^2$. At $d=2$ every nonzero class is odd on at least one of the four lattice vectors $(\pm1,\pm1,1)$: a class that is even on all four is even on $(1,1,1)$ and on their differences $(2,0,0)$ and $(0,2,0)$, which together generate $\Lambda_{\rm BCC}$. All coordinates of such a vector have magnitude one, so the point mass on it is a certificate with $t_\nu=1$; scaling gives $\dX\ge\frac{d^3/2}{d/2}=d^2$. The minimum-weight strings are the $V/d$ straight lines in each of the three directions.
\end{proof}

\subsection{\texorpdfstring{Proposition~\ref{prop:nmin}: counting shortest strings}{Counting shortest strings}}
\label{app:proof-nmin}

The proof also fixes the normalization of the loop counts used in the
model of Sec.~\ref{sec:entropy}.

\begin{proof}
A minimum-weight nontrivial cycle is a single simple loop; otherwise a lighter nontrivial cycle could be split off. It lifts to a path of $\dZ$ edges from $x$ to $x+v$ with $v\in\Lambda\setminus0$, so $\dZ\le\|v\|_1\le\dZ$ and the path is monotone. Conversely, a monotone path from $x$ to $x+v$ closes into a loop of $\dZ$ edges. The loop is simple: any two vertices of the path other than its two ends are less than $\dZ$ apart in $\ell_1$, so they cannot differ by a nonzero lattice vector. It is nontrivial, because $v\notin2\Lambda$; otherwise $v/2$ would be a shorter lattice vector. There are $V$ starting vertices and $\binom{\dZ}{|v_1|,|v_2|,|v_3|}$ orderings of the steps, and for a fixed sign of $v$ each loop arises once from each of its $\dZ$ vertices. The pair $\pm v$ therefore contributes $V\binom{\dZ}{|v_1|,|v_2|,|v_3|}/\dZ$ loops, and summing over the pairs gives Eq.~\eqref{eq:nmin}.
\end{proof}

Table~\ref{tab:lattices} lists the contacts of the five reference
families. At even distance $d$, each of the four staircase contacts of
the $z$-rotated lattice has $\binom d{d/2}$ step
orderings, the count for the rotated two-dimensional
code~\cite[Sec.~III]{Beverland2019}. The fourteen contacts of $\Lambda_{38}$
have $30$, $60$ or $90$ orderings.

\subsection{\texorpdfstring{Proposition~\ref{prop:matched}: matched distances, different failure coefficients}{Matched distances, different failure coefficients}}
\label{app:proof-matched}

\begin{proof}
We first establish the code parameters. The three tilted periods in
Eq.~\eqref{eq:matched-basis} have absolute determinant $V=d^3$. A
lattice vector $x=c_1b_1+c_2b_2+c_3b_3$, with integer coefficients
$c_i$, has coordinates satisfying
\begin{equation}\label{eq:matched-coordinates}
\begin{aligned}
 x_1+x_2+x_3&=dc_1,\\
 x_1-x_2&=2dc_2,\\
 2x_3-x_1-x_2&=3dc_3.
\end{aligned}
\end{equation}
If $\|x\|_1\le d$, the last two left-hand sides have absolute values
at most $d$ and $2d$. Thus $c_2=c_3=0$, and the only contacts are $\pm b_1$.
For the membrane upper bound, $(1,-1,0)/(2d)$ takes values $(0,1,0)$
on the basis $(b_1,b_2,b_3)$ and has $\ell_1$ norm $1/d$. For the lower
bound, every nonzero membrane class is odd on at least one basis row,
all of whose coordinates have magnitude at most $d$. Using that row
alone in Proposition~\ref{prop:dist}(c) gives $\dX\ge V/d$; the dual
covector gives the reverse inequality. The same argument on the cubic
basis uses $e_1/d$ for the upper bound. Thus both codes have
$n=3d^3$, $k=3$, $\dZ=d$ and $\dX=d^2$. Equation~\eqref{eq:nmin} gives
$3d^2$ shortest supports for cubic and
$d^2\binom{d}{d/3,d/3,d/3}$ for tilted.

Next we identify the errors that contribute at leading order. Two errors with the same
syndrome but different logical classes differ by a nontrivial cycle of
weight at least $d$. If both have weight at most $d/2$, they must therefore
be disjoint halves of a shortest support. Conversely, complementary
halves of any shortest support give exactly such an ambiguity. No
error of smaller weight can have this syndrome: it would differ in
class from at least one half and their weights would sum to less than
$d$. For the same reason every error of weight $d/2$ with this
syndrome is itself a half of a shortest support: it differs in class
from at least one of the two halves, and the two then form a nontrivial
cycle of weight at most $d$, hence exactly $d$, which makes them
disjoint halves of a shortest support. For a syndrome $s$ with an ambiguity, let $\mathcal E_j(s)$ be the
set of \emph{distinct} errors of weight $d/2$ that have this syndrome
and lie in logical class $j$ relative to a fixed recovery, so that
$|\mathcal E_j(s)|$ counts them. Each has probability
$q^{d/2}(1-q)^{3d^3-d/2}$. For fixed $d$ at small $q$, maximum
likelihood selects a class with the largest leading multiplicity
$|\mathcal E_j(s)|$. Hence
\begin{equation}\label{eq:matched-syndrome}
 A(d)=\sum_s\Bigl(\sum_j|\mathcal E_j(s)|-\max_j|\mathcal E_j(s)|\Bigr).
\end{equation}
Ties leave the coefficient unchanged. Syndromes without such a pair
cannot contribute at order $q^{d/2}$. This counts errors rather than
loop--error incidences. Sorting the minimum-weight errors by
syndrome and logical class and keeping the largest class is the
counting rule of the max-class minimum-weight decoder of Beverland
\emph{et al.}~\cite[Sec.~4.3]{Beverland2025}, which they apply to
syndrome circuits; here it is applied to data errors with perfect
checks. See also the general leading-order counting
methods in Refs.~\cite{Beverland2019,Sethi2026}.

For the cubic layout, consider a straight ring. A nonempty proper edge
subset of it has at least two boundary vertices. Distinct straight rings on the
cubic torus meet in at most one vertex. The syndrome of a half
therefore determines its ring, and on that ring a given boundary has
exactly two preimages, the half and its complement. One fails. Each of
the $3d^2$ rings contributes $\binom{d}{d/2}/2$ to Eq.~\eqref{eq:matched-syndrome}.

For the tilted layout, orient each shortest loop toward $b_1$. Height
$x_1+x_2+x_3\bmod d$ is well defined on the torus and increases by one
on every step. Pair the edges starting at even heights with their
complement at odd heights. Either half determines its loop: order the
known edges by cyclic height and fill each one-step gap. There is at
most one positive unit edge joining the required endpoints, since two
different choices would differ by a nonzero period of norm at most two.
The resulting pairs are disjoint as sets of error configurations. Both
halves have the same syndrome and differ by a logical string, so no
decoder can correct both. Every decoder, maximum likelihood included,
therefore fails on at least one error of weight $d/2$ per shortest
loop, and these errors are distinct for distinct loops. With
$d^2\binom{d}{d/3,d/3,d/3}$ loops this gives the tilted lower bound in
Eq.~\eqref{eq:matched-coeff}. Stirling's formula gives
$\binom{d}{d/3,d/3,d/3}\sim3^{d+3/2}/(2\pi d)$ and
$\binom{d}{d/2}\sim2^d\sqrt{2/(\pi d)}$, so the lower bound divided by
$A_{\rm cubic}(d)$ is asymptotic to $(3/2)^d\sqrt{3/(2\pi d)}$. This is
the stated ratio.
\end{proof}

The entries of Table~\ref{tab:matched} come from an exact count at
$d=6$. From each of the 216 vertices, enumerate
the three straight six-step words for cubic, or the 90 permutations of
two steps along each positive axis for tilted. Identify a loop by its
unordered edge support, and deduplicate all three-edge subsets of these
supports. Bin each subset by its boundary and by its three parities
against the dual-basis floor membranes of Proposition~\ref{prop:dist}.
For a fixed boundary these parities distinguish the logical classes.
Applying Eq.~\eqref{eq:matched-syndrome} to the 1080 ambiguous cubic
syndromes and the 11\,124 tilted ones gives the coefficients $A(6)$.

\begin{proof}[Proof of Corollary~\ref{cor:matched-observables}]
Let $S(\ell)$ be the span in $\Lambda/2\Lambda$ of windings of length
at most $\ell$ in the $\ell_1$ norm. This is also the span generated by
closed error chains of weight at most $\ell$: any such chain splits into
edge-disjoint closed walks, and a monotone walk realizes each period
with at most its $\ell_1$ length in edges. Retaining $r$ independent
observables leaves a kernel of dimension $3-r$, and any such kernel is
realizable. All windings shorter than a given length can therefore be
ignored precisely when their span fits inside this kernel. Equivalently,
\begin{equation}\label{eq:retained-span}
 \dZ^{(r)}=\min\{\ell:\dim S(\ell)>3-r\}.
\end{equation}
The three cubic axes enter together at $d$. For the tilted lattice,
Eq.~\eqref{eq:matched-coordinates} shows that
$x=c_1b_1+c_2b_2+c_3b_3$ satisfies $\|x\|_1\ge |x_1-x_2|=2d|c_2|$.
If $c_2=0$, then $x_1=x_2=d(c_1/3-c_3/2)$ and $x_3=d(c_1/3+c_3)$, so
\begin{equation}\label{eq:retained-length}
 \frac{\|x\|_1}{d}=2\left|\frac{c_1}{3}-\frac{c_3}{2}\right|
   +\left|\frac{c_1}{3}+c_3\right|\ge\frac{3|c_3|}{2}.
\end{equation}
For $c_3=1$ the real minimum is at $c_1=3/2$; the adjacent integers $1,2$
give $5/3,2$, respectively. Negating $x$ handles $c_3=-1$, and
$|c_3|\ge2$ costs at least $3d$. Consequently the first three independent
classes are witnessed by $b_1$, $b_1+b_3$ and $b_2$, at $d,5d/3,2d$.
Their parities give the claimed optimal pair and single observable.

For the subsystem interpretation, take the unused logical Pauli pairs as
gauge generators along with the original stabilizers. An undetectable
Pauli acts nontrivially on a retained qubit only if its $Z$ component
flips a retained $X$ (weight at least $\dZ^{(r)}$) or its $X$ component is a
nontrivial membrane (weight at least $d^2$). Its support contains that of
each component. Since $\dZ^{(r)}\le 2d\le d^2$ and a pure $Z$ string attains
$\dZ^{(r)}$, the minimum weight outside the gauge group is exactly $\dZ^{(r)}$
among Paulis commuting with the stabilizers.

Fix $r$ and the stated choice of observables. A minimum-weight syndrome
correction is right on all retained observables whenever the error has
weight less than $\dZ^{(r)}/2$: otherwise error plus correction would be
an undetectable flip of weight less than $\dZ^{(r)}$. Optimal decoding
for the retained result can do no worse on average. Conversely, split a
shortest retained flip into two halves (all distances here are even).
They give the same syndrome and different retained results, so at least
one must be decoded incorrectly. Each has probability
$q^{\dZ^{(r)}/2}(1-q)^{3d^3-\dZ^{(r)}/2}$, which gives the matching lower
order at fixed $d$.
\end{proof}

\subsection{\texorpdfstring{Proposition~\ref{prop:shared}: shared half-sets}{Shared half-sets}}
\label{app:proof-shared}
\hypertarget{app-halfsets}{}

\begin{proof}
We first identify distinct half-sets with cyclic sequences of edges and
unordered gap displacements. An exponential change of measure then
reduces their asymptotic count to a lattice local limit theorem. We
verify its hypotheses, including the rounding for odd lengths, before
explaining the limits of the counting formula.

We start by counting loops and loop--half-set pairs.
Reflect axes so that $v\ge0$, and let $I=\{i:v_i>0\}$.
A length-$\ell=\|v\|_1$ walk of winding $v$ is monotone.
Call a nonzero period $v'\in\Lambda$ \emph{short for $v$} if
$|v'_i|\le v_i$ for every $i$, with at least one strict inequality.
The hypothesis excludes such periods; a shortest $v$ always satisfies
it. Consequently every monotone walk of winding $v$ is simple:
a return before its last step would give a short period.
There are $\binom{\ell}{v_1,v_2,v_3}$ step words from each of the $V$
vertices, and a simple loop has $\ell$ choices of starting vertex.
Thus there are $V\binom{\ell}{v_1,v_2,v_3}/\ell$ loops and
$V\binom{\ell}{v_1,v_2,v_3}\binom\ell w/\ell$ loop--half-set pairs.

Choose an edge of a half-set $S$ as its \emph{anchor}. Reading a loop
through $S$ from that edge gives $w$ blocks $(i_j,g_j)$, $j=1,\dots,w$:
the $j$th chosen edge, along axis $i_j\in I$, followed by the nonnegative
displacement $g_j$ to the next chosen edge. A gap records step counts,
not their order.
The block displacements sum to $v$.
Each variable $x_i$ marks one step along axis $i$, and
\begin{equation}
\begin{aligned}
 P&=\prod_i(1+x_i+x_i^2+\cdots)=\sum_{g\ge0}x^g,\\
 N&=[x^v](EP)^w .
\end{aligned}
\end{equation}
Choosing a term from each successive edge factor $E$ and gap factor
$P$ therefore counts exactly the block sequences of displacement $v$.

Conversely, a block sequence and an anchor tail $y$ place the tail of
the $j$th chosen edge at the vertex
\begin{equation}
 y_j=y+\sum_{k<j}(e_{i_k}+g_k),
\end{equation}
so the anchor has tail $y_1=y$. After all blocks the position is $y+v$, identified with $y$ on the
torus. Fill each gap with any ordering of its steps. This gives a
monotone loop containing the placed edges. A gap $g$ has
$\binom{\|g\|_1}{g}$ fillings; only mixed-axis gaps have more than one.

It remains to show that each anchored half-set arises from exactly one
block sequence.
Lift a loop from an anchor tail $y_0$ to $y_0+v$. Every chosen edge
has its lift in this box. Two possible lifts of a direction-$i$ edge
differ by a period $v'$ with $|v'_j|\le v_j$ for every $j$
and $|v'_i|\le v_i-1$. The hypothesis forces $v'=0$.
These unique lifts fix the order of the chosen edges by their
coordinate sums and hence fix every gap. This proves that every loop
through the anchored half-set is one of the gap fillings just
constructed. It also ensures that the placed edges are distinct.
Thus placements of block sequences are in bijection with anchored
half-sets: there are $VN/w$ distinct half-sets. Dividing by the pair
count proves the stated formula for $\rho$.

For comparison, $P_{\rm ord}=1/(1-E)$ records an ordered gap.
A sequence counted by $[x^v](EP_{\rm ord})^w$ is a step word with its
first letter and $w-1$ further letters marked. Cutting before each
mark gives the inverse correspondence. Hence
\begin{equation}
\begin{aligned}
 \relax[x^v](EP_{\rm ord})^w
   &=\binom{\ell}{v_1,v_2,v_3}\binom{\ell-1}{w-1},\\
 \rho(v)&=\frac{[x^v](EP)^w}{[x^v](EP_{\rm ord})^w}.
\end{aligned}
\label{eq:rhoratio}
\end{equation}
The numerator forgets each gap's route; the denominator retains the
whole loop. Since $P\le P_{\rm ord}$ coefficientwise, $\rho\le1$.
On one active axis $P=P_{\rm ord}$ and $\rho=1$ exactly.
Only variables in $I$ contribute to $[x^v]$; all subsequent products
and vectors are restricted to those axes.

As an example, take $v=(2,2,0)$, so that $\ell=4$ and $w=2$.
The six step words give $3V/2$ loops and $9V$ pairs.
The $x$ edge leaving $0$ and the $y$ edge entering $v$ give blocks
$(1,(1,1,0),2,0)$. Their monomial is
$x_1(x_1x_2)x_2=x^v$. The mixed gap can pass through $(2,0,0)$ or
$(1,1,0)$, producing $xxyy$ or $xyxy$ through the same half-set.

The edge pairs $(1,1),(1,2),(2,1),(2,2)$ leave total gaps
$(0,2,0),(1,1,0),(1,1,0),(2,0,0)$, respectively.
Two steps on one axis split between the two gaps in three ways;
a mixed pair splits in four ways. Therefore
\begin{equation}
\begin{aligned}
 N&=[x_1^2x_2^2]\frac{(x_1+x_2)^2}{(1-x_1)^2(1-x_2)^2}\\
  &=3+4+4+3=14,\\
 \rho(2,2,0)&=\frac{14V/2}{9V}=\frac79 .
\end{aligned}
\end{equation}
Per vertex, five sets have one filling and two sets have two fillings.
Exact enumeration on $\Z^3/\langle(2,2,0),(0,20,0),(0,0,20)\rangle$
gives the same ratio.

We now turn to the asymptotics.
From here on $\rho$ denotes the coefficient expression, whether or
not the lattice hypothesis holds. Fix $u=v/\ell$, put $L=t\ell$,
$w_t=\lceil L/2\rceil$ and $N_t=[x^{tv}]F^{w_t}$, with $F=EP$ the
generating function of one block, so that $N_t$ counts the sequences
of $w_t$ blocks with total displacement $tv$.
For the denominator write $M_t=\binom L{tv_1,tv_2,tv_3}$ and
$B_t=\binom L{w_t}$. The type-class bound~\cite[Theorem~11.1.3]{CoverThomas2006}
and the fact that $B_t$ is a largest binomial coefficient give
\begin{equation}
\begin{aligned}
 (L+1)^{-|I|}e^{LH(u)}&\le M_t\le e^{LH(u)},\\
 2^L/(L+1)&\le B_t\le2^L,\\
 \rho(tv)&=\frac{L N_t}{w_t M_t B_t}.
\end{aligned}
\end{equation}
It remains to prove $\log N_t=Lh(u)+O(\log t)$.

Assign the formal variables positive numerical weights
$x_i=e^{\theta_i}$, where every $\theta_i<0$ so the geometric series
converge. If $f_b=[x^b]F$ counts blocks of displacement $b$, then
$F(e^\theta)=\sum_b f_b e^{\theta\cdot b}$ is their total weight and
\begin{equation}
 \log F(e^\theta)=\log\sum_i e^{\theta_i}
                    -\sum_i\log(1-e^{\theta_i}).
\end{equation}

Normalize these weights to independent block displacements $Y_j$
with $\Pr_\theta(Y_j=b)=f_be^{\theta\cdot b}/F(e^\theta)$.
The edge and gap are independent: the edge chooses axis $i$ with
probability $e^{\theta_i}/\sum_j e^{\theta_j}$, and the gap coordinates
are geometric with ratios $e^{\theta_i}$. Expanding $F^J$, every
sequence of total displacement $a$ receives the same factor
$e^{\theta\cdot a}/F(e^\theta)^J$, so
\begin{equation}
 [x^a]F^J=F(e^\theta)^J e^{-\theta\cdot a}
          \Pr_\theta(Y_1+\cdots+Y_J=a).
\label{eq:tilt}
\end{equation}
This is the exponential change of measure used in
Cram\'er's theorem~\cite[Sec.~2.2]{DemboZeitouni1998}.
Termwise differentiation inside the convergence domain gives
\begin{equation}
\begin{aligned}
 \partial_{\theta_i}\log F(e^\theta)
   &=\frac{\sum_b b_i f_b e^{\theta\cdot b}}{F(e^\theta)}
     =\mathbb E_\theta Y_{1,i},\\
 \nabla^2\log F(e^\theta)&=\operatorname{Cov}_\theta Y_1 .
\end{aligned}
\end{equation}
The support is every nonzero nonnegative integer vector, so it lies
in no affine hyperplane. The covariance is positive definite and
$\Psi(\theta)=\frac12\log F(e^\theta)-\theta\cdot u$ is strictly convex.

Next we show that $\Psi$ has a unique minimizer.
Set $m_\theta=\min_i|\theta_i|$, $M_\theta=\max_i|\theta_i|$ and
$u_{\min}=\min_{i\in I}u_i>0$. The edge sum is at least
$e^{-m_\theta}$. A coordinate attaining $M_\theta$ also gives
$-\theta\cdot u\ge m_\theta+u_{\min}(M_\theta-m_\theta)$.
Every negative log gap term is positive; retaining any one, indexed
by $j$, therefore yields
\begin{equation}
\begin{aligned}
 \Psi(\theta)&\ge-\tfrac12\log(1-e^{\theta_j})
            +\tfrac12m_\theta\\
 &\quad+u_{\min}(M_\theta-m_\theta)\\
 &\ge-\tfrac12\log(1-e^{\theta_j})
            +\min(\tfrac12,u_{\min})M_\theta .
\end{aligned}
\end{equation}
The first term diverges as $\theta_j\to0^-$ and the second as any
coordinate tends to $-\infty$. Sublevel sets are compact inside the
open convergence domain. Thus $\Psi$ has a unique minimizer $\theta^*$,
with $h(u)=\Psi(\theta^*)$ and $\mathbb E_{\theta^*}Y_j=2u$.
The mean block length is two, as required for one block per two steps.
All following block probabilities use this tilt.

With this tilt we bound $N_t$ from above and below.
For the upper bound, bound the probability in Eq.~\eqref{eq:tilt} by one and write
$\delta_t=w_t-L/2\in\{0,1/2\}$. This gives the saddle-point
upper bound~\cite[Sec.~VIII.2]{FlajoletSedgewick2009}
\begin{equation}
\begin{aligned}
 \log N_t&\le w_t\log F(e^{\theta^*})-t\theta^*\cdot v\\
 &=Lh(u)+\delta_t\log F(e^{\theta^*})\\
 &\le Lh(u)+\tfrac12|\log F(e^{\theta^*})|.
\end{aligned}
\end{equation}
The last term is fixed as $t$ grows and handles either sign of the
logarithm.

For the lower bound, choose $c=1$ if $\ell$ is even and $c=2$
otherwise, and put $J_0=c\ell/2$. Write $t=c\tau+r$, $0\le r<c$.
Then $w_t=\tau J_0+\lceil r\ell/2\rceil$. Nonnegative coefficients
allow us to retain sequences whose two parts have displacements
$\tau cv$ and $rv$:
\begin{equation}
\begin{aligned}
 A_\tau&=[x^{\tau cv}]F^{\tau J_0},\\
 N_t&\ge A_\tau[x^{rv}]F^{\lceil r\ell/2\rceil}\ge A_\tau .
\end{aligned}
\end{equation}
The remainder coefficient is one for $r=0$ and positive for $r=1$:
split any monotone word of displacement $v$ into
$\lceil\ell/2\rceil$ nonempty pieces and take each piece's first
edge and remaining gap. For example, $\ell=3,t=5$ gives two groups
of three blocks and a remainder of two blocks.

Group $J_0$ independent blocks and subtract their integral mean:
\begin{equation}
\begin{aligned}
 Z_j&=\sum_{k=(j-1)J_0+1}^{jJ_0}Y_k-cv,\\
 \mathbb E Z_j&=2J_0u-cv=0,\\
 \Sigma:=\operatorname{Cov}Z_j&=J_0\operatorname{Cov}Y_1,\\
 A_\tau&=e^{\tau c\ell h(u)}p_\tau .
\end{aligned}
\end{equation}
Here $p_\tau=\Pr(Z_1+\cdots+Z_\tau=0)$.
The last identity is Eq.~\eqref{eq:tilt} with $J=\tau J_0$ and
$a=\tau cv$. The $Z_j$ are identically distributed and independent,
have all moments because their gaps are geometric, and have positive
definite covariance $\Sigma$.

Their support is
$\{y-cv:y\in\Z^I_{\ge0},\ \|y\|_1\ge J_0\}$.
Every sum of $J_0$ blocks belongs to this set; conversely, partition
any word of such a displacement into $J_0$ nonempty pieces to realize
it. For every support point $z$, each $z+e_i$ is also in the support.
Their differences generate $\Z^I$ after any translation of the
support. This verifies strong aperiodicity, excluding missing return
times. The lattice local central limit theorem~\cite[P7.9]{Spitzer1976}
therefore gives
\begin{equation}
\begin{aligned}
 p_\tau&\sim\frac{1}{(2\pi\tau)^{|I|/2}\sqrt{\det\Sigma}},\\
 \log N_t&\ge Lh(u)-\tfrac{|I|}{2}\log t-O(1).
\end{aligned}
\end{equation}
For the second conclusion use $N_t\ge A_\tau$,
$\tau c\ell=L-r\ell$ and $\log\tau=\log t+O(1)$; the bounded
remainder changes only the constant. Together with the upper bound,
this proves $\log N_t=Lh(u)+O(\log t)$ for both parities.

Substitution in the exact ratio, using $1\le L/w_t\le2$ and the
denominator bounds above, gives
\begin{equation}
\begin{aligned}
 \log\rho(tv)&=Lh(u)-LH(u)-L\log2+O(\log t)\\
             &=-L\chi(u)+O(\log t).
\end{aligned}
\end{equation}
When $|I|\ge2$, the positivity proved below makes the error $o(L\chi)$,
giving the stated exponential form. On one axis $\rho=1$ exactly,
$H=0$, $h=\log2$ and $\chi=0$, without division by a zero rate.
If $t\ell$ is even, $r=0$ and $N_t=A_\tau$ exactly. The local limit
factor $t^{-|I|/2}$ then cancels the Stirling factors
$t^{-(|I|-1)/2}t^{-1/2}$ of $M_tB_t$, leaving a positive constant
prefactor for $\rho(tv)e^{L\chi(u)}$.

It remains to show that $\chi(u)>0$ off the coordinate axes.
For ordered gaps set $s_i=e^{\theta_i}$ and $\sigma=\sum_i s_i<1$.
The corresponding objective is
$\Psi_{\rm ord}=\frac12\log[\sigma/(1-\sigma)]-\sum_i u_i\log s_i$.
At fixed $\sigma$, the information inequality~\cite[Theorem~2.6.3]{CoverThomas2006}
gives $-\sum_i u_i\log s_i\ge H(u)-\log\sigma$, with equality at
$s_i=\sigma u_i$. Minimizing
$-\frac12\log[\sigma(1-\sigma)]$ then gives $\sigma=1/2$ and
$\min\Psi_{\rm ord}=H(u)+\log2$. Evaluate $\Psi$, whose minimum is $h$,
at this same point. Since $P_{\rm ord}=2$, this proves
\begin{equation}
 \chi(u)\ge\tfrac12\log\left[2\prod_{i\in I}(1-u_i/2)\right]>0
 \qquad (|I|\ge2).
\end{equation}
The strict inequality uses
$\prod_i(1-u_i/2)>1-\sum_i u_i/2=1/2$ when at least two components
are positive. For a uniform direction on $k$ axes, the stationary
condition gives $s=1/(k+1)$ and
\begin{equation}
 \chi(u)=\log\frac{2k}{k+1}
                 +\frac{k-1}{2}\log\frac{k}{k+1}.
\end{equation}
The values for $k=1,2,3$ are $0$, $\log(4\sqrt6/9)=0.0849$ and
$\log(9/8)=0.1178$. The two-axis exponent agrees with the growth
rate of the upper bound on failing staircase configurations in
Ref.~\cite[Appendix~B, Eq.~(B5)]{Beverland2019}, which its authors
believe to be tight.
For $v=(1,1,0)$ the tilt $s=1/3$ makes the marked step uniform on
the two axes, and each independent gap coordinate has variance $3/4$.
The block therefore has variance $1$ along each axis and covariance
$-1/4$ between them, giving covariance determinant $15/16$. Since
$F(s,s)/s^2=27/2$, Eq.~\eqref{eq:tilt} and the local limit formula give
$N_t\sim(27/2)^t/(2\pi t\sqrt{15/16})$. The corresponding denominator
is $M_tB_t/2=\tfrac12\binom{2t}{t}^2\sim16^t/(2\pi t)$ by Stirling's
formula. Their ratio gives
$\rho(t,t,0)\sim(4/\sqrt{15})(27/32)^t$. At $t=5$, $20$ and $80$ the
exact ratio is $0.4489$, $0.03466$ and $1.2927\times10^{-6}$, and the
asymptotic formula gives $0.4417$, $0.03454$ and $1.2916\times10^{-6}$.

Without the hypothesis on $\Lambda$ the count can fail. A period short for $v$ can put two lifts
of the same edge in the box from $y_0$ to $y_0+v$. The same half-set
can then give two block sequences. For example, take
$\Lambda=\{x:x_1\equiv9x_2+7x_3\bmod12\}$, $\dZ=4$ and $v=(-2,3,1)$.
With $x$ denoting a step $-e_1$, both $xxyyyz$ and $xzyyyx$ start at
$0$ with the same $x$ edge. The $y$ edges with tails at $(-2,0,0)$
and $(-2,2,0)$ on the first loop have lifts at $(-1,2,1)$ and
$(-1,0,1)$ on the second. The shifts $(1,2,1)$ and $(1,-2,1)$ are
periods, so these are the same two edges, visited in reverse order.
The formula gives $23/40$ for $\rho(v)$ and direct counting gives
$57/100$.

We enumerated every loop and half-set for $\|v\|_1\le9$ and
$\|v\|_1<2\dZ$ on $24$ cyclic tori with $11\le V\le35$ and
$3\le\dZ\le5$: the tori $(12;9,7)$ and $(27;23,17)$ of
Appendix~\ref{app:simulation-inventory} and $22$ kept from $40$
cyclic lattices drawn at random with a fixed seed, namely those with
$\dZ\ge3$.
The formula was exact for all $46$ shortest
windings, as proved above. It overcounted four of the $142$ longer
windings, by at most $0.9\%$.

This check stops at $\|v\|_1<2\dZ$ for a reason. If a period $v'$ short
for $v$ lies in the same orthant as $v$, then $v'$ and $v-v'$ are both
nonzero periods, so $\|v\|_1=\|v'\|_1+\|v-v'\|_1\ge2\dZ$. Below that
length every monotone walk is therefore simple, and only the two-lift
mechanism above can fail. From $\|v\|_1=2\dZ$ onwards a monotone walk
can also revisit a vertex. Then neither the loop count nor the
coefficient expression need be exact, and Eq.~\eqref{eq:model} uses
the expression for such windings as an approximation whose error we
have not bounded.

Different shortest windings can share half-sets even when each
within-winding count is exact. Write
$\Lambda_{\rm FCC}=m\{x\in\Z^3:\sum_i x_i\equiv0\bmod2\}$ for the FCC
period lattice with $\dZ=2m$.
The first $m$ consecutive $x$ edges form a half-set of both shortest
loops $x^m y^m$ and $x^m(-y)^m$ from the origin. Their windings differ
by $(0,2m,0)\notin2\Lambda_{\rm FCC}$, since $(0,1,0)$ has an odd
coordinate sum. Thus the shared
half-set belongs to different logical classes of loops at every scale.
At $m=3$, $162$ sets occur in five windings each and the rest in one,
giving $4\cdot162=648$ excess counts. Computing a failure coefficient
also requires deciding between logical classes for each syndrome.
\end{proof}

\subsection{\texorpdfstring{Proposition~\ref{prop:kappa}: directional decay of the free-loop sum}{Directional decay of the free-loop sum}}
\label{app:proof-kappa}

\begin{proof}
Fix $0<\mu<1/5$ and a nonzero $v\in\Z^3$. At each endpoint $tv$,
$G_\mu(tv)$ sums over all walk lengths $\ell$, with weight $\mu^\ell$
per walk. We let the integer $t$ grow along this ray and use the
known decay of a lattice Green function to find the rate.

We first express $G_\mu$ through this Green function.
There are $6\cdot5^{\ell-1}$ non-backtracking walks of length
$\ell\ge1$, since only the reverse of the last step is forbidden.
Thus $\sum_x G_\mu(x)$ is finite for $5\mu<1$, and the Fourier transform
$\widehat G_\mu(k)=\sum_{x\in\Z^3}G_\mu(x)e^{-ik\cdot x}$ converges.
Averaging $e^{-ik\cdot x}$ over the six possible single steps gives
$\widehat D(k)=\tfrac13\sum_{i=1}^3\cos k_i$. Write $c=\mu^{-1}+5\mu$.
Since $c-6=(1-\mu)(1-5\mu)/\mu>0$, we can set $a=\sqrt{c/6-1}>0$.
With the empty walk at $0$, the walk identity~\cite[Sec.~2.2,
Eqs.~(2.20)--(2.21)]{FitznerHofstad2013} gives
\begin{equation}
\begin{aligned}
 \widehat G_\mu(k)&=\frac{1-\mu^2}{1+5\mu^2-6\mu\widehat D(k)}
 =\frac{1-\mu^2}{6\mu}\,\widehat C_a(k),\\
 \widehat C_a(k)&=\frac{1}{1+a^2-\widehat D(k)}.
\end{aligned}
\end{equation}
The hat denotes the Fourier transform
$\widehat C_a(k)=\sum_{x\in\Z^3}e^{-ik\cdot x}C_a(x)$ at wavevector $k$.
At lattice site $x$, the inverse transform gives the Green
function~\cite[Eq.~(1.1)]{MichtaSlade2022},
\begin{equation}
 C_a(x)=\int_{[-\pi,\pi]^3}\frac{e^{ik\cdot x}}{1+a^2-\widehat D(k)}
        \,\frac{d^3k}{(2\pi)^3}.
\end{equation}
The prefactor is independent of $k$, so
$G_\mu(x)=[(1-\mu^2)/(6\mu)]C_a(x)$. In the geometric series for
$\widehat C_a(k)$, $\widehat D(k)^n$ transforms back to the probability
of reaching $x$ in $n$ independent uniform steps. Its weight is
$(1+a^2)^{-(n+1)}$.

This representation gives the decay rate at fixed $\mu$.
For fixed $a>0$, Michta and Slade~\cite[Definition~1.1 and
Theorem~1.3(i)]{MichtaSlade2022} give
$G_\mu(tv)=(A_{\mu,v}/t)e^{-tr}(1+o(1))$ with $A_{\mu,v}>0$ and
$r=m_a|v|_a$. The mass $m_a=\operatorname{arcosh}(1+3a^2)$ is the
decay rate along a coordinate axis. The anisotropic norm $|v|_a$ measures
the rate along $v$ in units of this axial rate. With $\tau=u_a(v)$ and
$c=6(1+a^2)$, their definition gives
\begin{equation}
\label{eq:kappa-direction}
\begin{aligned}
 2\sum_{i=1}^3\sqrt{1+\tau^2v_i^2}&=c,\qquad \tau>0,\\
 |v|_a&=\frac{1}{m_a}\sum_{i=1}^3 v_i\operatorname{arsinh}(\tau v_i).
\end{aligned}
\end{equation}
For $v\ne0$ the first sum increases strictly from $6$ to infinity,
so $c>6$ fixes $\tau$ uniquely. At $e_1=(1,0,0)$ it gives
$\sqrt{1+\tau^2}=\cosh m_a$, hence $\tau=\sinh m_a$ and $|e_1|_a=1$.
Taking logarithms of the asymptotic formula gives
$-t^{-1}\log G_\mu(tv)=r+(\log t-\log A_{\mu,v}+o(1))/t\to r$.
Thus $\kappa_\mu(v)=m_a|v|_a$.

Next we show that this rate is the maximum stated in the proposition.
Setting $k=i\theta$ in the Fourier denominator gives
$\mu[c-S(\theta)]$, where $S(\theta)=2\sum_i\cosh\theta_i$.
To maximize $\theta\cdot v$ subject to $S(\theta)\le c$, take
$\theta_i^*=\operatorname{arsinh}(\tau v_i)$. Then
$\cosh\theta_i^*=\sqrt{1+\tau^2v_i^2}$, so Eq.~\eqref{eq:kappa-direction}
gives $S(\theta^*)=c$ and $\theta^*\cdot v=r$. The outward normal
there is $\nabla S(\theta^*)=2\tau v$. Since each $\cosh\theta_i$ is
convex, $S$ lies above its tangent plane at $\theta^*$. For every
feasible $\theta$ this gives
\begin{equation}
 2\tau v\cdot(\theta-\theta^*)\le S(\theta)-S(\theta^*)\le0.
\end{equation}
Thus $\theta^*$ attains the claimed maximum. For real $v$ the same
maximum is finite by compactness and positive for $v\ne0$, since
$S(0)=6<c$ puts a ball about zero inside the feasible set. Scaling
$v$ scales the objective, and symmetry under $\theta\mapsto-\theta$
allows a negative factor to be replaced by its absolute value.
For the triangle inequality, maximize
$\theta\cdot(v+w)=\theta\cdot v+\theta\cdot w$ and bound each term
by its own maximum.

For the staircase $v=(1,1,0)$, the maximizer is $(s,s,0)$ with
$4\cosh s+2=c$. Hence $\kappa_\mu(v)=2s=2\operatorname{arcosh}\bigl((c-2)/4\bigr)$.
At $\mu=1/10$, $\cosh s=17/8=\cosh(\log 4)$, so the rate is $\log16$
and $G_\mu((t+1)v)/G_\mu(tv)\to1/16$.

For the low-noise limit in part~(a),
keep $v$ fixed and write $L=\|v\|_1$ and $u_i=|v_i|/L$. As $\mu\to0$,
$c$ grows and so does $\tau$. Applying
$x\le\sqrt{1+x^2}\le x+1$ to $x=\tau|v_i|$ in
Eq.~\eqref{eq:kappa-direction} gives $c-6\le2\tau L\le c$.
Hence $2\tau L=\mu^{-1}(1+O(\mu))$. For $u_i>0$, substitute
$2\tau|v_i|=(2\tau L)u_i$ into the large-$x$ formula
$\operatorname{arsinh}x=\log(2x)+O(x^{-2})$ to obtain
\begin{equation}
 |\theta_i^*|=\log(1/\mu)+\log u_i+O(\mu).
\end{equation}
Each $\theta_i^*$ has the sign of $v_i$, so
$\kappa_\mu(v)=\sum_{i:\,u_i>0}Lu_i|\theta_i^*|$. The first term
contributes $L\log(1/\mu)$ and the second contributes
$L\sum_{i:\,u_i>0}u_i\log u_i=-L H(u_v)$, with total error $O(\mu L)$.

For the staircase, $u_1=u_2=1/2$, so $L H(u_v)=2\log 2$.
Its shortest walks give the same term. Choosing which $t$ of the
$2t$ steps run along the first axis gives $\binom{2t}{t}$ walks of
weight $\mu^{2t}$ each, and Stirling's formula gives
$t^{-1}\log\binom{2t}{t}\to2\log 2$.

For the limit at the convergence boundary in part~(b),
write $\eta=c-6=(1-\mu)(1-5\mu)/\mu$. As $\mu\rightarrow1/5^-$,
$\eta\rightarrow0^+$, so the first line of Eq.~\eqref{eq:kappa-direction}
forces $\tau\rightarrow0^+$. With $v$ fixed, expanding
$\sqrt{1+x^2}=1+x^2/2+O(x^4)$ in that line and
$\operatorname{arsinh}x=x+O(x^3)$ in the second gives
\begin{equation}
\begin{aligned}
 \eta&=\tau^2\|v\|_2^2+O(\tau^4),\\
 \kappa_\mu(v)&=\tau\|v\|_2^2+O(\tau^3).
\end{aligned}
\end{equation}
Taking the square root of the first line and dividing the second by
it gives $\kappa_\mu(v)/\sqrt{\eta}\to\|v\|_2$. For the staircase,
$\eta=4(\cosh s-1)\sim2s^2$, so its exact rate $2s$ has limiting
ratio $\sqrt2=\|(1,1,0)\|_2$.
\end{proof}
\section{Lattices, circuits and sampling}
\label{app:methods}

This appendix specifies the simulated objects: the period lattices of
the distance-efficient sequence, the memory circuits and their noise,
and the shot counts behind the figures. Most simulation code used in this work
was constructed and run with AI assistance, as described in the AI Use
Disclosure after the acknowledgments.

\subsection{Lattices of the distance-efficient sequence}
\label{app:simulation-inventory}

The distance-efficient sequence in Fig.~\ref{fig:twisted} uses the
following oriented cyclic lattices $x_1\equiv a_2x_2+a_3x_3\bmod N$.
\begin{center}\small
\begin{tabular}{@{}r@{\ }l@{\quad}r@{\ }l@{\quad}r@{\ }l@{}}
$d$ & $(N;a_2,a_3)$ & $d$ & $(N;a_2,a_3)$ & $d$ & $(N;a_2,a_3)$\\
\midrule
$4$ & $(12;9,7)$ & $8$ & $(92;83,53)$ & $13$ & $(421;182,16)$\\
$5$ & $(27;23,17)$ & $10$ & $(178;167,103)$ & $14$ & $(486;57,41)$\\
$7$ & $(70;54,45)$ & $11$ & $(260;94,40)$ & $15$ & $(635;274,146)$
\end{tabular}
\end{center}
At $d=6$ and $12$ it uses $\Lambda_{38}$ and $2\Lambda_{38}$. At $d=9$ it uses
$\langle(36,0,0),(8,4,0),(25,1,1)\rangle$, of index $144$, with
$432$ data qubits. Both panels also include the cubic periods
$d\mathbb Z^3$ at $d=5,7,9$, with $n=3d^3=375,1029,2187$ data qubits,
respectively.

For the frozen-forecast test in Appendix~\ref{app:frozen-forecasts}, we fixed one
orientation at each of $N=27,70,92$ by taking the lexicographically first
pair $(a_2,a_3)$ with $0\le a_2,a_3<N$ attaining the maximum string
distance. This gives $(4,10)$, $(16,25)$ and $(9,39)$, respectively.
Each pair is obtained from the corresponding entry above by negating
both coefficients modulo $N$, which reflects the first coordinate.

Cools and Lyness give an index and a basis at each plotted distance
from their restricted search for $\mathcal K$-optimal lattice rules of
enhanced degree $d$~\cite[Tables~1 and 3]{CoolsLyness2001}. At each
distance a signed permutation of the coordinate axes takes a tabulated
lattice to the one used here.
We specify the simulated orientations because a
reflection can change the ordered circuit. Aasen \emph{et al.}\ had
already established the unrestricted minima through distance
seven~\cite[Table~I and Sec.~X]{AHHW2025}. At $d=12$, the packing bound
and $2\Lambda_{38}$ give the minimum analytically.

For each $d=4,\ldots,15$, we verified the minimum index over all
full-rank integer period lattices with $\dZ\ge d$. Starting at the
packing bound $\lceil 19d^3/108\rceil$, we enumerate lower-triangular
row Hermite normal forms by increasing index. The diagonal is positive, and each entry below it runs from
zero to one less than its column's diagonal entry. We reject a form
if it contains a nonzero period of $\ell_1$ length below $d$.
Each layout in this sequence attains the first admissible index. For example,
at $d=9$ the enumeration excludes indices $129$ through $143$, and
the index-$144$ lattice above supplies the $432$-qubit circuit.
For the $276$- and $534$-qubit layouts at $d=8$ and $10$, respectively,
integer-rounded path-packing bounds meet the membrane upper bounds,
giving $\dX=34$ and $51$.

\subsection{Memory circuits and noise}
\label{app:memory-circuits}

In the torus $X$-basis memory circuit, we prepare the data in
$|+\rangle$, measure vertex then plaquette checks in each of $R$
rounds, and finally measure the data in $X$. Each vertex ancilla
starts in $|+\rangle$, controls one CNOT on each of its six incident
edges, and is measured in $X$. The greedy schedule considers these
CNOTs in order of vertex index and then data-qubit index. The
direction-ordered schedule assigns one CNOT layer to each of
$+x,+y,+z,-x,-y,-z$, in that order. In both circuits each plaquette
ancilla starts in $|0\rangle$, receives four CNOTs, and is measured
in $Z$. Plaquettes parallel to the $xy$, $xz$ and $yz$ planes use
four CNOT layers per orientation, for a total of twelve. With $V$ vertices,
one ancilla per check gives $V+3V=4V$ ancillas for $3V$ data qubits.
Each round uses $6V+12V=18V$ CNOTs, in $18$ or $19$ CNOT layers on
the greedy tori tested and $18$ on the direction-ordered tori.

Each CNOT is followed by two-qubit depolarization, with probability
$p/15$ for each nonidentity two-qubit Pauli. In each CNOT layer an
idle data qubit receives $X,Y,Z$ with probability $p/3$ each.
Preparations and measurements flip with probability $p$.
Faults at distinct circuit locations are independent; their
propagated detector components need not be.
A vertex detector compares consecutive check outcomes, using the
known initial value and the product of final data outcomes at the
temporal boundaries. The full readout fails if any of the three logical
$X$ outcomes is decoded incorrectly. Plaquette outcomes form no
detectors for this $X$-basis memory.
Stim~\cite{Stim2021} supplies the decomposed detector model and
PyMatching~\cite{SparseBlossom2025} decodes it.
For the $29$ tori checked with the greedy schedule, the shortest
cycle of the decomposed detector graph that flips a logical outcome
has $\dZ$ edges. A graph edge is a component
of a fault and not a fault, so this is a graph distance and not a
fault distance. The number of these shortest cycles depends on the
circuit, in contrast with the non-zigzagging diagonals
of the cyclic two-dimensional circuit in Ref.~\cite{Fowler2013cyclic}; related
perfect-syndrome and circuit counts appear in
Refs.~\cite{WatsonBarrett2014,Beverland2025}.

\subsection{Shot counts, error bars and seeds}
\label{app:sampling}

Every run in Fig.~\ref{fig:twisted} targets $300$ failures, and all
phenomenological points reach that target. Circuit points stop at a
cap of $2\times10^6$ shots, except the distance-efficient codes at
$\dZ=9,11,13,14,15$, which ran to the target in $1.82\times10^6$ to
$2.54\times10^8$ shots; the distance-nine runs had a cap of $10^9$
shots. Ten circuit points reach the cap first, with
$6$ to $235$ failures. Three of them, with $6$, $10$ and $17$
failures, are below $30$ and are indicative.
Figures~\ref{fig:twisted} and~\ref{fig:fixedn} use the same
distance-eleven circuit measurement, with $305$ failures in
$10\,927\,000$ shots ($\epsilon=2.54\times10^{-6}$).
A second, independent run of the same circuit at $p=3\times10^{-4}$
gave $302$ failures in $9\,200\,000$ shots
($\epsilon=2.98\times10^{-6}$). The two runs differ by $2.0$ standard
errors of their difference. The figures show the first run. Pooling the two gives $607$
failures in $20\,127\,000$ shots and $\epsilon=2.74\times10^{-6}$.
Appendices~\ref{app:rounds-checks} and~\ref{app:fixed-size-check} state
how the pooled value changes the ratios quoted for this point. With
the pooled value the point stays inside the ranges of
Sec.~\ref{sec:fixedn} and Table~\ref{tab:model-ranges}.
For a shot fraction
$\hat P$ from $N_{\rm shot}$ shots, the plotted error bar is
$\sqrt{\hat P(1-\hat P)/N_{\rm shot}}/R$, with $R=\dZ$.

For some simulations of codes with periodic or open boundaries in Figs.~\ref{fig:twisted}
and~\ref{fig:patches}, the records give shot and failure counts
without random seeds. The fixed-size simulations in Fig.~\ref{fig:fixedn}
record a seed for each point, as do the $\dZ=12$ runs for the
$(1,0,0)$ and $(1,1,0)$ volumes in Fig.~\ref{fig:patches}.
At $p=5\times10^{-4}$ the direction-ordered
BCC ($\dZ=14$), FCC ($\dZ=16$) and Minkowski ($\dZ=18$) points in
Fig.~\ref{fig:entropy} each pool independently seeded samples and
have more than $200$ failures.
Seeds are likewise missing from the records of the two-dimensional
tori in Fig.~\ref{fig:2d}, of the planar-code runs in
Sec.~\ref{sec:2d-comparison}, of the membrane-sector check below, and of
the greedy-schedule runs in Fig.~\ref{fig:entropy} other than the cyclic
codes with $\dZ\ge9$ and the fixed-size codes. These runs can be
repeated statistically but not shot for shot.

\subsection{Matched-pair runs}
\label{app:matched-runs}

The circuits in Table~\ref{tab:matched-winding} used the same
direction-ordered schedule and matching decoder, with 1512 qubits
per layout including ancillas. We fixed the shot counts in advance.
We obtained conservative 95\% confidence intervals for the tilted-to-cubic
failure ratio by dividing opposite endpoints of the
two layouts' two-sided 97.5\% exact binomial intervals. For the full
readout at $p=3\times10^{-4}$ the interval is $[6.97,10.65]$, and the
loop-model prediction of $10.05$ was made before sampling. For the
prespecified pair $(\bar X_2,\bar X_3)$ at $p=5\times10^{-4}$, the
interval is $[0.0475,0.1022]$.

The fault-distance search of Sec.~\ref{sec:matched-circuit} uses the
undecomposed detector models and excludes $3888$ four-detector
mechanisms in each layout.

\subsection{Membrane-sector check}
\label{app:membrane-sector}

For six small codes, namely the $\dZ=4$ cubic, BCC, $z$-rotated and
FCC codes, the $36$-qubit cyclic code $(12;9,7)$ and the $114$-qubit
Minkowski code, we decoded $Z$-basis
memory at $p=10^{-3}$ and $2\times10^{-3}$ by belief propagation with
ordered-statistics post-processing
(BP-OSD)~\cite{PanteleevKalachev2021,RoffeWhiteBurtonCampbell2020}.
It gives at most $10$ membrane-sector failures in $2\times10^4$ shots
(the largest counts, $8$ and $10$ at $p=2\times10^{-3}$, are for the
FCC and the cyclic code, whose membrane distance is eight),
compared with $2\times10^3$--$10^4$ string-sector failures. The decoder
uses the full detector model, at most $50$ iterations of minimum-sum
BP~\cite{FossorierMihaljevicImai1999}, and the combination sweep of
Ref.~\cite[Sec.~V]{RoffeWhiteBurtonCampbell2020} at order four, a greedy
variant of ordered-statistics decoding~\cite{FossorierLin1995}, as
implemented in the \texttt{ldpc} package~\cite{RoffeLDPC2022}. This
supports the string-sector focus only at
those tested points. No circuit-level membrane distance or
complete-memory performance is certified.

\subsection{Shapes of the open-boundary volumes}
\label{app:volume-shapes}

A volume of Sec.~\ref{sec:patches} contains the lattice points $x$
with $-1\le\varphi(x)\le\dZ+1$ whose two transverse coordinates lie in
given ranges. For the normal $(1,0,0)$ the ranges are
$0\le x_2\le a$ and $0\le x_3\le b$. For $(1,1,0)$ they are
$|x_1-x_2|\le a$ and $0\le x_3\le b$, and for $(1,1,1)$ they are
$|x_1-x_2|\le a$ and $|x_1+x_2-2x_3|\le b$.

For each family and each $\dZ$, a scan visits candidate pairs $(a,b)$ in
order of the product of the two transverse widths. For each candidate it
computes the membrane distance exactly, as the minimum edge cut between
the two rough components, and it keeps the candidate with the fewest
data qubits among those with $\dX\ge\dZ^2$. It stops when the product of
widths exceeds $1.6$ times that of the best candidate found. The
candidates are $b=a$ and $b=a+1$ for the axis-aligned family,
$\max(0,a-3)\le b\le2a+2$ for $(1,1,0)$ and $\max(1,a-3)\le b\le2a+3$
for $(1,1,1)$, with $1\le a<2\dZ$ in each case. The tilted families are
thus scanned over a bounded range of aspect ratios and the axis-aligned
family only over squares and near-squares. The selected volume is the
smallest within this scan. It is not the smallest over all shapes: the
examples after the table show more elongated shapes with fewer qubits.
The scan selects
$a=b=\dZ-1$ for every axis-aligned volume and the following pairs for
the tilted ones, with the data-qubit counts $n$ and membrane distances
$\dX$ shown.
\begin{center}\small\setlength{\tabcolsep}{4pt}
\begin{tabular}{@{}r@{\quad}rr@{\quad}crr@{\quad}crr@{}}
 & \multicolumn{2}{c@{\quad}}{$(1,0,0)$} & \multicolumn{3}{c@{\quad}}{$(1,1,0)$}
 & \multicolumn{3}{c@{}}{$(1,1,1)$}\\
$\dZ$ & $n$ & $\dX$ & $(a,b)$ & $n$ & $\dX$ & $(a,b)$ & $n$ & $\dX$\\
\midrule
$4$ & $136$ & $16$ & $(4,1)$ & $77$ & $16$ & $(3,3)$ & $73$ & $16$\\
$6$ & $516$ & $36$ & $(3,5)$ & $306$ & $36$ & $(4,5)$ & $240$ & $36$\\
$8$ & $1296$ & $64$ & $(4,7)$ & $729$ & $64$ & $(7,5)$ & $559$ & $64$\\
$10$ & $2620$ & $100$ & $(5,9)$ & $1450$ & $100$ & $(6,9)$ & $1089$ & $102$\\
$12$ & $4632$ & $144$ & $(9,7)$ & $2463$ & $144$ & $(7,11)$ & $1864$ & $148$
\end{tabular}
\end{center}

The limits on $a$ and $b$ keep elongated shapes out of the scan, and
some of these reach the same membrane distance with fewer data qubits.
For $(1,1,0)$ at $\dZ=6$ the two-layer slab $(a,b)=(9,1)$ has $264$
data qubits against $306$, and at $\dZ=12$ the shape $(18,3)$ has
$2337$ against $2463$. For $(1,1,1)$ the shape $(1,8)$ has $71$ against
$73$ at $\dZ=4$, and $(3,11)$ has $551$ against $559$ at $\dZ=8$. For
the axis-aligned family the $4\times9$ column has $511$ against $516$
at $\dZ=6$, and the $4\times16$ column $1268$ against $1296$ at
$\dZ=8$. The selected $(1,1,0)$ volume at $\dZ=4$ is itself a two-layer
slab, $(a,b)=(4,1)$, on the limit $b=a-3$; this is why its ratio
$\dZ\dX/n=0.83$ stands above the others of its family. In the extreme
case the single layer $x_3=0$ with $|x_1-x_2|\le8$, which lies outside
the scanned range of $a$, is a two-dimensional rotated surface code: it
has $(\dZ,\dX)=(4,16)$ with $64$ data qubits and $\dZ\dX/n=1$, fewer
qubits than the $77$ and $73$ of the tilted volumes in
Fig.~\ref{fig:patchgeom}. Section~\ref{sec:patches} therefore compares
the scanned shapes, whose two transverse widths are of the same order
except at $\dZ=4$, and its qubit counts are not minima over all
volumes.
\section{Computing the loop model}
\label{app:model-details}

Equation~\eqref{eq:model} needs four inputs: the edge probabilities
of the circuit, the walk counts $W_\ell(v)$, the sharing factors
$K_\ell(v)$ and the strip failure probabilities $\Pi_\ell(v)$. This
appendix says how each is computed, in that order, and then how the
resource forecasts are assembled. A dynamic program computes the walk
counts through $\ell_{\max}=68$.

\subsection{Edge probabilities}
\label{app:edge-probabilities}

The model takes $q_s,q_d,q_t$ (Fig.~\ref{fig:edges}) from the decomposed
detector error model of each circuit, with bulk values away from the
first and last rounds and separate values for the preparation and
readout layers, and treats the components of one fault as independent.
Phenomenological noise has $(q_s,q_d,q_t)=(2p/3,0,p)$. For the
direction-ordered schedule, keeping one noise source at a time in Stim
(idling, vertex CNOTs, plaquette CNOTs, preparation and measurement)
reproduces each term of Eq.~\eqref{eq:edge-probabilities} to within
$0.02\%$ at $p=10^{-5}$, with the same value for every data qubit of the
cubic torus $3\Z^3$ and of the $114$-qubit Minkowski code.

The greedy schedule meets a qubit's two vertex checks in either order,
with zero to four idle CNOT layers between them. Its $q_d$ is therefore
$\frac8{15}p$ plus $\frac23p$ per idle CNOT layer, from $0.53p$ to
$3.2p$. At distance six its bulk means on the reference families are
$q_s=(11.4\text{--}11.6)p$, $q_d=(1.35\text{--}1.5)p$ and $q_t=3.6p$;
idling contributes about $8.7p$ to $q_s+q_d\approx12.9p$, plaquette
CNOTs $3.2p$ and vertex CNOTs $1.1p$. The strips combine each qubit's own
$q_d$ with the layer mean $q_s$, so $\log(q_s/q_d)$ runs from $1.3$ to
$3.1$, the range quoted in Sec.~\ref{sec:rounds}. With each qubit's own
$q_s$, which falls as its $q_d$ rises because their sum is the same for
every qubit, the range is $1.1$ to $3.1$. The direction-ordered schedule
gives $\log(q_s/q_d)=1.7$ for every qubit.

\subsection{Independent rings}
\label{app:rings}

Removing time and diagonal edges separates a strip into independent
rings, one per detector layer. Each ring fails with the isolated-ring
probability $T_\ell(q)$ of Eq.~\eqref{eq:ring}, and the strip's parity
error is the parity of the ring failures. Its probability is
\begin{equation}
 \Pi^{\rm ring}_\ell=\tfrac12\left[1-\prod_r(1-2T_\ell(q_r))\right]
             \simeq\sum_rT_\ell(q_r).
\label{eq:ring-limit}
\end{equation}
Here $q_r$ is the spatial-edge probability in layer $r$. The final
sum is the rare-failure limit, which we call the
\emph{independent-ring sum} of the strip. The \emph{independent-ring
weight} of a term of Eq.~\eqref{eq:model} is its coefficient
$K_\ell VW_\ell/(2\ell)$ times this sum, and the
\emph{independent-ring estimate} is the total of these weights. It is
$R$ times Eq.~\eqref{eq:model} with every $\Pi_\ell$ replaced by its
independent-ring sum. Two rings of probability $t$ give
$2t(1-t)$, rather than $2t$, because two failures cancel in parity.
The strip probability $\Pi_\ell$ of Eq.~\eqref{eq:model} reduces to
$\Pi^{\rm ring}_\ell$ when the time and diagonal edges are removed
from the strip. With those edges the model estimate exceeded the
independent-ring estimate in every circuit comparison
(Appendix~\ref{app:rounds-checks}).
For large even $\ell$ and small $q$,
$T_\ell(q)\simeq\mu^\ell/\sqrt{2\pi\ell}$ with
$\mu=2\sqrt{q(1-q)}$. Odd $\ell$ first fails with $(\ell+1)/2$
flips. Perfect syndromes give one ring and $R=1$.

Rings of lengths $2k-1$ and $2k$ fail equally often:
$T_{2k-1}(q)=T_{2k}(q)$ for every $q$ and every integer $k\ge1$. To see
this, condition on the number $B_{2k-1}$ of flips among the first $2k-1$
edges of the longer ring. The two rings then differ in two cases only.
If $B_{2k-1}=k$ and the last edge does not flip, the longer ring ties
where the shorter one fails. If $B_{2k-1}=k-1$ and the last edge flips,
the longer ring ties where the shorter one succeeds. A tie counts one
half, so
\begin{equation}
\begin{aligned}
 T_{2k}-T_{2k-1}=\tfrac12\big[&q\Pr(B_{2k-1}=k-1)\\
                &-(1-q)\Pr(B_{2k-1}=k)\big]=0 ,
\end{aligned}
\end{equation}
because both terms equal $\binom{2k-1}{k}q^k(1-q)^k$. For example
$T_1(q)=T_2(q)=q$. In the independent-ring estimate, therefore, an odd
length and the next even length differ only through their loop counts
and sharing factors. Any further dependence on the parity of $\ell$
must come from time and diagonal edges (Sec.~\ref{sec:rounds}).

\subsection{Sharing factors for detours}
\label{app:detours}

For a uniformly drawn pair of a length-$\|v\|_1+2j$ detour and a
half-set $S$ on it, let $N(S)$ count same-length detours through $S$.
Let $\mathrm{new}(S)=1$ if $S$ contains no half-set of a shorter
loop of the same winding, and zero otherwise. The inverse-multiplicity
identity for unions~\cite{KarpLubyMadras1989,OwenMaximovChertkov2019}
gives the implemented factor
\begin{equation}
 K_{\|v\|_1+2j}(v)=\rho_j(v)
       =\mathbb E[\mathrm{new}(S)/N(S)].
\end{equation}
The weight $1/N(S)$ counts a half-set once however many detours pass
through it, and $\mathrm{new}(S)$ removes the sets already counted at a
shorter length. At order zero, $\rho_0=\rho$.

We obtain $\rho$ by repeated prefix sums of the coefficients of
$(EP)^w$. For $j\le3$, each sampled detour factor
uses $1000$ loop--half-set pairs. We call a non-backtracking walk from
$0$ to $v$ a \emph{bridge}. A bridge of the required length is drawn
uniformly, using backward walk counts. It closes into a loop on the
\emph{winding cylinder} $\Z^3/v\Z$, the quotient that identifies its
two ends. The closed walk is not a loop if its last step reverses its
first, which we call a \emph{tail}, or if it uses an edge twice; such
bridges receive zero weight. Exact dynamic programs on the cylinder
compute $\mathrm{new}(S)$ and $N(S)$. Exhaustive order-one checks for
five short windings agree pair by pair.

The sampled factors fall with detour order. By $j=3$ they are $0.11$
for $(8,0,0)$ and $0.011$ for $(2,2,2)$.
Unsampled orders continue with the median ratio $\rho_j/\rho_{j-1}$
among sampled windings of length at least eight with the same
number of active axes. Beyond order three the last ratio is reused;
shorter windings use their own $\rho_3/\rho_2$. The median ratios
at orders one to three
are $0.52$--$0.55$ for straight, $0.47$--$0.54$ for planar and
$0.41$--$0.43$ for three-axis windings; the ratios of individual
windings scatter around these medians. This continuation is an
approximation, varied in the sensitivity checks of
Appendix~\ref{app:forecast-error}.
These factors describe one layer and one winding. Applying them across
rounds is an approximation, and different windings can still share
errors (Appendix~\ref{app:proof-shared}).

\subsection{Strip failure probabilities}
\label{app:strip}

For every code and duration, strips use the actual spatial mean of
each layer, the time-edge mean between each pair of consecutive
layers, and the diagonal probability and orientation of each data
qubit, including preparation and readout. Each sample draws a fresh
loop: an independent bridge of the winding, rejected if it has a tail
or a repeated edge on the winding cylinder, and placed at a uniform
vertex.
The loop determines the orientation of its diagonals. A diagonal joins
one end of a data qubit in one layer to the other end in the next
layer, so the direction in which the loop crosses the qubit decides
whether the diagonal runs forwards or backwards along the loop. We
call the cyclic sequence of these orientations the loop's \emph{sign
word}; Fig.~\ref{fig:edges}(b) shows one with a single reversed step.
Detours keep the sign words of their sampled loops; no uniform
distribution over sign words is assumed.

The strip observable is parity across one cut. Each configuration is
decoded twice by PyMatching, with edge weights
$\log[(1-q_e)/q_e](1+10^{-4}\zeta_e)$ and
$\log[(1-q_e)/q_e](1-10^{-4}\zeta_e)$, where the $\zeta_e$ are
independent and uniform on $[-1,1]$. Its failure value is the average
of the two outcomes: $0$, $\tfrac12$ or $1$. Opposite perturbations
break a tie between two corrections in opposite ways, so such a tie
counts one half. We call this the \emph{paired perturbation rule}.
In an exhaustive length-four, two-layer example, its failure rate
differed from exact symmetric tie breaking by $0.056\%$. This check
used alternating diagonals, $p=3\times10^{-4}$,
$(q_s,q_t,q_d)=(10.4p,3.6p,1.87p)$ and perturbation seed zero.
This example is not the worst case. Over four such exhaustive
examples, each with four perturbation seeds, the discrepancy ranged
from zero to $4.0\%$, the largest in an example with ten tied
syndromes.
The physical simulations use ordinary PyMatching. The size of the
tie-rule discrepancy on larger strips is unknown.

\subsubsection{Importance proposal}

The failure probability of a strip is estimated by importance
sampling over lifts of the loop.
A Markov chain on detector layers proposes a lift around the loop,
conditioned to close by backward messages. Its transitions use time
and permitted diagonal edges. Edges on the lift receive enhanced
flip probabilities; other edges retain their physical probabilities.
If $Q(\mathcal L)$ is the probability of proposing lift $\mathcal L$
and $r_e$ is the proposed/physical ratio for the realized state of
edge $e$, the proposal/physical configuration density is
$\sum_{\mathcal L}Q(\mathcal L)\prod_{e\in\mathcal L}r_e$.
A transfer-matrix trace evaluates this full mixture, and its
reciprocal weights the decoder outcome. The proposal is a mixture of
four fixed components, which differ in their enhanced flip
probabilities and in their rates of layer changes. A pilot sample from
this mixture is followed by one cross-entropy
step~\cite{RubinsteinKroese2004}, which adds two components fitted to
the pilot's failing configurations. Production samples are drawn
independently from the adapted mixture; pilot samples are excluded
from the estimate.

\subsubsection{Sampled and extrapolated terms}

Let $j_{\max}$ be the largest detour order that is sampled. The
windings with $\|v\|_1\le\dZ+12$ that carry a stated fraction of the
independent-ring estimate are sampled at every order $j\le j_{\max}$.
The remaining terms are not sampled. Their contribution is estimated
as their independent-ring weight times the ratio of strip estimate to
independent-ring weight found at order $j_{\max}$. We call it the
\emph{extrapolated contribution}.

The circuit comparisons in Fig.~\ref{fig:entropy} use $1.2\times10^4$
configurations per winding and detour order, with $j_{\max}=4$ and
$99\%$ of the independent-ring estimate.
The $10^{-9}$ resource forecasts in Table~\ref{tab:entropy} use $6144$
configurations per term, including a $2048$-sample pilot, with
$j_{\max}=6$ and $99.5\%$.
Each production group contains $32$ configurations on one fresh loop.
The frozen forecasts of
Appendix~\ref{app:frozen-forecasts} used $4096$ pilot configurations
and $256$ independent production groups ($8192$ configurations) for
each distinct strip law, that is, for each set of terms whose strips
have identical statistics, also with $j_{\max}=6$ and $99.5\%$.

\subsection{Sampling error of the forecasts}
\label{app:forecast-error}

Figure~\ref{fig:entropy} reports point estimates. The forecasts in
Table~\ref{tab:entropy} and Appendix~\ref{app:frozen-forecasts} carry
sampling errors, estimated as follows. Configurations come in $n_{\rm g}$
independent groups of $n_{\rm c}=32$ that share one loop and one decoder
perturbation. With $Y_{bc}$ the importance-weighted decoder outcome of
configuration $c$ in group $b$ and $\bar Y_b$ the mean of group $b$, a
strip probability and its variance are estimated by
\begin{equation}
\widehat\Pi=\frac1{n_{\rm g}}\sum_{b=1}^{n_{\rm g}}\bar Y_b,\qquad
\widehat s_\Pi^{\,2}=\frac{\sum_b(\bar Y_b-\widehat\Pi)^2}{n_{\rm g}(n_{\rm g}-1)} .
\end{equation}
Using group means keeps the covariance between configurations of the
same loop; treating all $n_{\rm g}n_{\rm c}$ values as independent would
underestimate the variance. A zero sampled contribution with positive
model weight is treated as unresolved and cannot establish that a code
meets the resource target.

The extrapolated contribution is proportional to the sampled terms of
the highest order $j_{\max}$, so those terms enter twice. Let $a_i$ be
the loop-count and sharing coefficient of sampled term $i$, of detour
order $j_i$, and let $\Delta$ and $\mathcal T_{j_{\max}}$ be the
independent-ring weights of the unsampled terms and of the sampled terms
of order $j_{\max}$. The shot failure probability is estimated as
$\widehat F=\sum_i\tilde a_i\widehat\Pi_i$, with combined coefficients
\begin{equation}
\tilde a_i=a_i\Big(1+\mathbf 1_{\{j_i=j_{\max}\}}\,\Delta/\mathcal T_{j_{\max}}\Big).
\end{equation}
Strips of monotone loops with the same length and at most one step of
the minority sign have the same law, because all their sign words are
rotations of one another, so they share one estimate. If $\mathcal I$
runs over the sets of terms that share an estimate $\widehat\Pi_{\mathcal I}$,
the Monte Carlo variance per round is
$R^{-2}\sum_{\mathcal I}(\sum_{i\in\mathcal I}\tilde a_i)^2\widehat s_{\Pi_{\mathcal I}}^{\,2}$;
summing the coefficients before squaring keeps the covariance from the
common extrapolation and from a shared estimate.

The errors of the detour factors are propagated with $256$ draws that
recompute the independent-ring weights and the coefficients together;
they are approximate because the individual half-set weights were not
stored. The reported variance adds the spread of these draws to their
mean Monte Carlo variance. Separate sensitivity checks replace the
continuation ratios by their $10$th and $90$th percentiles, take the
extrapolation ratio from order $j_{\max}-1$, double the extrapolated
contribution, truncate the detour factors above order $3$, or reduce
$\ell_{\max}$ from $68$ to $60$. The extrapolated contribution reaches
$36.4\%$ of the central prediction among $245$ lattice/noise cases,
covering $179$ distinct integer lattices. The resulting sensitivity
range is not a confidence bound on the discrepancy between the loop
model and the correlated circuit.

\subsection{Integer screening}
\label{app:resource-sensitivity}

The resource library contains all $43\,981$ lower-triangular Hermite
normal forms with determinant at most $40$, reduced to $8427$ classes
under unsigned coordinate permutations. Coordinate reflections are
retained because they can change the ordered circuit. A lookup
computed for a nine-layer strip with the same probabilities in every
layer screens integer scales with $n\le2\times10^5$
and $3\le\dZ\le42$; shortlisted codes and references are reevaluated
with the actual finite experiment $R=\dZ$. The screening minimum is
not a certified library optimum.

Table~\ref{tab:entropy} lists central values. Each estimate also has
a sensitivity interval, which spans the scenario variations of
Appendix~\ref{app:forecast-error} and two statistical standard errors.
Selecting for each entry the smallest code whose lower end of the
interval meets the target, or the smallest whose upper end does,
changes fifteen of the $42$ entries. In data qubits, and in the column
order of the table, the resulting ranges are as follows. At
$p=2\times10^{-5}$: checked minimum $288$--$384$. At $5\times10^{-5}$:
checked minimum $567$--$648$. At $1.5\times10^{-4}$: cubic
$3000$--$3993$. At $2\times10^{-4}$: cubic $3993$--$5184$, $z$-rotated
$4116$--$6144$, checked minimum $2304$--$2592$. At $3\times10^{-4}$:
cubic $6591$--$8232$, BCC $4116$--$6144$, checked minimum
$4116$--$4374$. At $5\times10^{-4}$: cubic $17\,496$--$24\,000$,
$z$-rotated $26\,364$--$32\,928$, BCC $15\,972$--$20\,736$, FCC
$13\,182$--$16\,464$, Minkowski $14\,250$--$24\,624$, checked minimum
$13\,182$--$16\,464$. The other $27$ entries do not change. Several
entries are decided by less than their sampling error. For example, at
$p=3\times10^{-4}$ the $4116$-qubit BCC code has a predicted rate of
$(1.02\pm0.13)\times10^{-9}$ and misses the target by $2\%$, and at
$5\times10^{-4}$ the $15\,972$-qubit BCC code has
$(7.8\pm6.9)\times10^{-9}$. These ranges do not bound the difference
between the loop model and the full circuit.

Revealing all nonspatial faults to an ideal decoder leaves independent
rings with parity-failure probability $\Pi^{\rm ring}_\ell$ of
Eq.~\eqref{eq:ring-limit}. This cannot exceed the original
strip decoder's failure, so a positive weighted sum can exclude
undersized candidates within the loop model. It is not a lower
bound on the correlated physical circuit.
The screening lookup uses a strip with the same probabilities in every
layer. Increasing this strip from $9$ to $25$ layers changes
failure per layer by factors of $1.36\pm0.14$ at
$(p,\ell)=(3\times10^{-4},25)$ and $1.30\pm0.13$ at
$(5\times10^{-4},30)$. The final estimates use the actual preparation
and readout layers and the sampled sign words.
\begin{table}[tb]
\caption{\raggedright \textbf{Model/simulation ratio ranges.} PS denotes perfect
syndromes. The circuit rows use the greedy vertex schedule and require
$\dZ\ge5$ and at least $30$ observed failures. The direction-ordered
test and its unresolved size drift are reported in Sec.~\ref{sec:compare}.}
\label{tab:model-ranges}
\begin{tabular}{@{}lrr@{}}
\toprule
Comparison & Points & Ratio range\\
\midrule
Phenomenological $p=10^{-3}$; PS & $15+14+20$ & $0.89$--$1.50$\\
Same, excluding $\dZ=4$ & & $0.89$--$1.26$\\
Phenomenological $p=10^{-2}$ & $29$ & $0.99$--$1.76$\\
Circuit $p=(1\text{--}3)\times10^{-4}$ & $57$ & $0.69$--$1.46$\\
Circuit $p=5\times10^{-4}$ & $32$ & $0.93$--$1.58$\\
Circuit $p=7\times10^{-4}$ & $33$ & $0.90$--$1.79$\\
\bottomrule
\end{tabular}
\end{table}

\section{Checks of the loop model}
\label{app:model-checks}

This appendix compares Eq.~\eqref{eq:model} with simulation and tests
its approximations one at a time. All of these are finite numerical
checks at the stated parameters.

\subsection{Comparison ranges}
\label{app:comparison-details}

Table~\ref{tab:model-ranges} counts each simulation run once; two
independent runs of one code at one noise level count as two points.
Its first row combines $15$ phenomenological points at $p=10^{-3}$
with perfect-syndrome points at $q=10^{-3}$ and $3\times10^{-3}$, $14$
and $20$ of them.
At distance nine, the comparison in Fig.~\ref{fig:entropy}
uses the $435$-qubit cyclic lattice $(145;61,9)$, not the $432$-qubit
minimum-index code of Fig.~\ref{fig:twisted}.
For the lower-noise greedy circuits the $57$ runs are $2$, $3$,
$19$ and $33$ runs at $p=1$, $1.5$, $2$ and $3\times10^{-4}$. They
cover $44$ distinct code--noise settings, of which $13$ were simulated
twice against a single model value, on $24$ codes with
$5\le\dZ\le15$. The median predicted-to-simulated ratio is $1.06$ and the
rms factor is $1.17$, where the rms factor of ratios $x_i$ is
$\exp(\sqrt{\langle(\ln x_i)^2\rangle})$. These are point estimates;
Appendix~\ref{app:forecast-error} describes the sampling errors.

At circuit $p=10^{-3}$, $\mu\approx6.8\sqrt p>1/5$ and the
uncorrected non-backtracking sum diverges. The distance-four circuit
ratios, which the table excludes, are $1.1$--$1.6$ at
$p\le3\times10^{-4}$ and reach $2.1$ at $p=7\times10^{-4}$.

Simpler variants of the model do worse.
Omitting the strips leaves only $0.04$--$0.55$ of the observed rate at
$\dZ\ge10$. Keeping two rounds brings the estimates closer, although the
predicted-to-simulated ratio still falls with distance. A two-parameter
fit also leaves a size dependence and overestimates the
distance-fourteen rate by a factor of $2.2$. At each circuit noise
level $p$, this fit uses codes with $\dZ\ge5$ and at least $20$
failures, and adjusts the amplitude $A$ and fugacity $\mu$ in
$A\sum_{v\notin2\Lambda,\ell}V\mu^\ell W_\ell(v)/\ell$
by least squares on $\log\epsilon$.

\subsection{Repeated rounds and parity}
\label{app:rounds-checks}

\subsubsection{Size of the repeated-round effect}

Replacing every $\Pi_\ell$ in Eq.~\eqref{eq:model} by its
independent-ring sum, Eq.~\eqref{eq:ring-limit}, gives the model
without time and diagonal edges. Over the
$146$ circuit comparisons on the greedy schedule, namely the $122$
circuit points of Table~\ref{tab:model-ranges} and the $24$ runs at
$\dZ=4$, the ratio of the full estimate to this one
ranges from $1.08$ to $44.5$. It
is $1.3$--$1.8$ at $\dZ=6$, $p=3\times10^{-4}$. It exceeds
$20$ in five circuit comparisons, where it is $21$--$44$: BCC at
$\dZ=10$ and the lattice $\langle(12,0,0),(6,15,0),(0,9,3)\rangle$ at
$\dZ=12$, both at $p=5\times10^{-4}$ and $7\times10^{-4}$, and the cyclic $\dZ=15$ code at
$p=3\times10^{-4}$.

\subsubsection{Decoding across layers with spatial faults only}

To isolate the effect of decoding across layers, we removed the time
and diagonal fault components from the detector error model, kept all
edge weights, and decoded the samples in two ways: by matching on
the whole space-time graph, and by matching each layer on its own.
In seven such tests at $p=3\times10^{-4}$ and $5\times10^{-4}$, the ratio of the first
failure rate to the second was larger at odd distances $7,11$ than at
even distances $6,8$. In the two tests where both decoders saw the same
$10^6$ samples, at $p=5\times10^{-4}$, it was $1.99\pm0.25$ for the
cyclic distance-eleven code and $1.20\pm0.13$ for BCC at distance
eight. The model reproduces
this parity dependence. The other five tests come from an earlier
exploratory script that is no longer available: their failure and shot
counts are on record, but their seeds are not, nor whether the two
decoders saw the same samples.

\subsubsection{Odd and even distances}

Isolated rings of lengths $2k-1$ and $2k$ fail equally often
(Appendix~\ref{app:rings}), so a dependence on parity must come from
the loop counts or from time and diagonal edges.
Along the distance-efficient sequence in Fig.~\ref{fig:twisted},
the odd-distance codes from five through thirteen fail less often
than their next even-distance neighbours under phenomenological noise.
Under circuit noise the odd member fails more often in the pairs
$7/8$, $9/10$ and $13/14$, with odd-to-even ratios $1.25\pm0.10$,
$1.56\pm0.14$ and $2.87\pm0.23$. The pairs $5/6$ and $11/12$, at
$0.94\pm0.07$ and $1.30\pm0.20$, are not resolved. With the two
distance-eleven runs of Appendix~\ref{app:sampling} pooled, the $11/12$
ratio becomes $1.40\pm0.21$. For the pair $13/14$ the model gives $2.82$.
For cubic periods the observed ratios at $5/6$ and $7/8$ are,
respectively, $0.765$ and $0.793$ under phenomenological noise,
and $1.70$ and $2.19$ under circuit noise. The cubic $9/10$
phenomenological ratio is $1.22$, from $303$ failures in
$1.58\times10^6$ shots at distance nine and $304$ in
$1.74\times10^6$ at distance ten.
Section~\ref{sec:rounds} explains how diagonal edges can favour
failures at odd distance.

\subsection{Fixed-size codes}
\label{app:fixed-size-check}

Repeated rounds increase the distance-eleven code's failure rate more
than BCC's, yet the cyclic code retains its advantage
(Fig.~\ref{fig:fixedn}). At $p=3\times10^{-4}$ the predicted cyclic-to-BCC
ratio is $0.65$ and the observed ratio is $0.66$, or $0.71\pm0.05$
with the two distance-eleven runs of Appendix~\ref{app:sampling}
pooled. The agreement is looser than these two numbers suggest. The
curves of Fig.~\ref{fig:fixedn} come from a model evaluation separate
from the one behind Table~\ref{tab:model-ranges} and
Fig.~\ref{fig:entropy}. For the distance-eleven code at this $p$ the
two evaluations give $2.63\times10^{-6}$ and $2.98\times10^{-6}$. They
differ by $13\%$, about five times the stored sampling error of the
first, which omits correlations within a loop
(Sec.~\ref{sec:compare}). The second gives a predicted ratio of
$0.75$. The lower-noise ratio
at $p=1.5\times10^{-4}$ rests on just $15$ cyclic failures. Across all
five layouts the model/simulation ratios range from $0.69$ to $1.24$
at $p=(1\text{--}5)\times10^{-4}$ for the $22$ points with at least
ten failures; the two points with $2$ and $3$ failures give $1.59$ and
$0.92$.

\subsection{Forecasts frozen before sampling}
\label{app:frozen-forecasts}

For one further test, we fixed and recorded the pool, the forecasts and
the analysis before drawing any confirmation samples. Two sets of
circuit samples had been drawn earlier, during development: $5\times10^4$
shots on each of three registered cells at $p=2\times10^{-4}$, to time
the sampler, and the $9\times10^5$ diagnostic shots at
$p=5\times10^{-4}$ of Appendix~\ref{app:independence}, a third of
which sample the circuit itself. The model has no
parameter fitted to sampled failures, so neither set enters a
forecast. The pool has $18$
layouts, all from lattice families used elsewhere in this paper.
Relative to the greedy-schedule comparisons of
Table~\ref{tab:model-ranges}, what is new in this test is the schedule,
the duration and the samples. Fifteen come from the five families of
Table~\ref{tab:lattices}, in one orientation per scale: cubic at
$m=4,5,6,7$, $z$-rotated, BCC and FCC at $m=2,3,4$, and Minkowski at
$m=1,2$. The other three are the cyclic lattices with $V=27,70,92$ and
$\dZ=5,7,8$, in the orientations of
Appendix~\ref{app:simulation-inventory}. The string distances range
from $4$ to $12$. Each layout ran $R=6$ noisy rounds of the
direction-ordered schedule (Sec.~\ref{sec:circuit}) at $p=10^{-4}$ and
$2\times10^{-4}$, between preparation and final readout. A shot fails
if minimum-weight matching returns any of the three logical $X$
outcomes incorrectly. The frozen forecast of this shot failure
probability is $R\,\epsilon(\Lambda)$, the sum of strip failure
probabilities in Eq.~\eqref{eq:model}, computed for this circuit and
duration with the sampling of Appendix~\ref{app:strip}. Each
layout--noise cell received $2\times10^5$ shots, extended to
$2\times10^6$ when the first checkpoint did not settle the selection
question described below.

In the $23$ of the $36$ cells with at least $30$ failures ($39$ to
$1799$), the forecast divided by the observed fraction of failed shots
lies between $0.85$ and $1.37$, with median $1.05$. The two largest
ratios, $1.33$ and $1.37$, belong to the FCC layout with $\dZ=4$. All
$23$ cells have $\dZ\le8$, so the test does not reach the distances at
which Sec.~\ref{sec:compare} reports a drift. The other $13$ cells, six
of them with no failures, have too few failures for a ratio. We
computed this summary after sampling; it was not a registered endpoint.

The registered question was one of selection. At each noise level the
rule chose the cheapest layout whose forecast met $10^{-5}$ failures
per round, counting all $7V$ data and ancilla qubits. At $p=10^{-4}$ it
chose the cyclic layout with $V=70$, which uses $490$ qubits. One
failure in $2\times10^5$ shots puts an upper limit of
$5.33\times10^{-5}$ on its six-round failure probability, below the
target $1-(1-10^{-5})^6=6.0\times10^{-5}$. The limits are exact
binomial limits~\cite{ClopperPearson1934} with simultaneous coverage of
at least $0.96$ across the $36$ cells and their checkpoints. The
cheapest BCC and cubic layouts that met the target use $756$ and $875$
qubits. The chosen layout, however, is also the cheapest qualifying
layout among those of greatest string distance at each qubit budget,
so the test does not separate the model from a rule based on distance.
A cheaper $378$-qubit FCC layout remained unresolved, with $152$
failures in $2\times10^6$ shots and an interval containing the target.
The registered requirement that the choice cost at most $25\%$ more
than the cheapest feasible layout is therefore not confirmed, since
$490/378=1.30$. At $p=2\times10^{-4}$ the model's choice, BCC at $m=3$,
also remained unresolved, with $129$ failures in $2\times10^6$ shots.
We therefore use the study only as a test of forecast accuracy.

\subsection{Independence and sharing approximations}
\label{app:independence}

Three finite checks probe these approximations.
The model treats the components of one fault as independent
(Appendix~\ref{app:edge-probabilities}). A \emph{hook} is a fault
that the decomposed detector model represents by two edges. Splitting
each correlated plaquette hook into independent edges with the same
marginals changes
the simulated rate by $-6\%$ to $+6\%$ in eight codes with
$\dZ\le8$ and by $-12\%$ at distance eleven ($706$ failures in $10^6$
shots with split hooks against $806$ with correlated ones), all at
$p=5\times10^{-4}$.
For the distance-six BCC, FCC and Minkowski codes at $p=5\times10^{-4}$,
we compared three ways to sample noise with the same whole-graph decoder:
circuit faults, joint detector-error events, and the components of
those events drawn independently. With $10^5$ shots per sampler and code,
the three failure-rate intervals overlapped for each code. How much
strip decoding contributes to the drift at larger distances remains open.
We fitted no correction to these data.
A separate sharing diagnostic weights failing configurations by
$\mathrm{new}/M$, where $M$ counts same-winding failing strips, using
the union identity~\cite{KarpLubyMadras1989,OwenMaximovChertkov2019}.
On the nine-layer BCC, FCC and Minkowski strips tested, this raises
the estimated rates by $4\%$--$25\%$ in an exploratory aggregation
whose inputs are on record but whose script was not kept. The size drift persists. The
sampling errors for this check omit covariance within a loop.
These checks do not establish independence at other parameters.

\subsection{Leading-order half-set count}
\label{app:leading-order}

With perfect syndromes and low noise, failures come from half-sets on
shortest loops. The model counts them winding by winding: it sums
$\rho(v)$ times the number of (loop, half-set) pairs over the shortest
windings $v$, with a tie on an even loop weighted one half. A half-set
that lies on loops of two windings is then counted twice
(Appendix~\ref{app:proof-shared}). To measure
this excess we counted the distinct half-sets directly. We enumerated
every subset of size $w=\lceil\dZ/2\rceil$ of every monotone
loop of length $\dZ$ (and $\dZ+1$ for odd distance),
identifying sets up to translation. Each receives weight $V$ divided by
its translation stabilizer size.

On $24$ codes with $4\le\dZ\le12$, the winding-by-winding sum
equals the direct count for cubic and BCC and exceeds it by at most
$1.3\%$ for $\dZ\ge7$. At $\dZ=6$ the excess is
$4.4\%$ for the $z$-rotated code, $4.7\%$ for FCC and $8.8\%$ for the
Minkowski code. It is $9.2\%$ for the cyclic $\dZ=5$ code and
$28\%$--$52\%$ for the three staircase codes at $\dZ=4$.

The same enumeration was also decoded.
Matching is applied at every translate up to $3\times10^7$
decodings, and sampled translates beyond; a single fixed translate
can bias the result substantially, because matching ties depend on position.
For $\dZ\ge5$ the number of half-sets on which matching fails
is within $2.5\%$ of the count with ties weighted one half.

These
finite checks do not prove cross-winding exactness or an all-scale
error bound.

\bibliography{Tilting_3D_Toric_Code}
\end{document}